\documentclass[12pt]{Jucthesis}
\usepackage{graphicx}
\usepackage{version}
\usepackage{amssymb}
\usepackage{amsmath}

\def\gsim{\, \rlap{$>$}{\lower 1.1ex\hbox{$\sim$}}\,}
\def\lsim{\, \rlap{$<$}{\lower 1.1ex\hbox{$\sim$}}\,}

\newenvironment{vitaesectnodate}[1]{
\item {\bf #1}
\begin{list}{}{\leftmargin 0in \labelwidth 1.3in \labelsep .2in
\parsep 0in
}}{\end{list}}

\begin{document}


\title{Analytic Approaches to the Study of\\Small Scale Structure on Cosmic String Networks}
\author{Jorge V. Rocha}
\degreemonth{September} \degreeyear{2008} \degree{Doctor of Philosophy}
\approvalmonth{August} 
\chair{Professor Joseph Polchinski} 
\theorymember{Professor Donald Marolf} 
\experimentalmember{Professor Tommaso Treu} 
\numberofmembers{3} 

\field{Physics}
\campus{Santa Barbara}

\maketitle

\begin{frontmatter}

\thispagestyle{empty}\approvalpage

\copyrightpage

\begin{dedication}
\null\vfil
{\large
\begin{center}
To my parents, Carlos and Isabel
\end{center}}
\vfil\null
\end{dedication}

\begin{acknowledgements}

In writing these lines I think of the people without whom I would not be writing these lines.  To the person who taught me the inner workings of science, Joe Polchinski, I am deeply grateful.  Thank you, Joe, for sharing your knowledge and insight with me, for showing me the way whenever I got off track and for your powerful educated guesses.

Over these years I have also had the opportunity of collaborating with Florian Dubath.  For this and for his enthusiasm I am thankful.

I was fortunate enough to have Don Marolf and Tommaso Treu as committee members.  I learned much about physics from Don, and always with a good feeling about it.  I am grateful for your constructive criticism as well.  From Tommaso I appreciate him always keeping an open door for me and also showing so much interest.  I thank both of you for all that.

Having spent these years in UCSB I have benefited from the excellence of the local Physics Department, either through classes, gravity lunches or simple conversations.  I wish to thank Professors Ski Antonucci, David Berenstein, Matt Fisher, Steve Giddings, Sergei Gukov, Jim Hartle, Gary Horowitz, Peng Oh, Mark Srednicki, Bob Sugar and Tony Zee.

Now going back further in time, I am grateful to Professors Jo\~ao Pimentel Nunes and Jos\'e Mour\~ao, who advised me during the last years of undergraduate studies in Lisbon.  Thank you for introducing me to the wonders of string theory and for all your support that led me here.

I have also benefited from the presence of many former and fellow students of the high energy theory and gravity groups at UCSB, as well as postdocs.  I would like to thank those who taught me one way or another, either through journal club talks and discussions, interactions in classes (as classmates or TAs), or simply by sharing the office with me.  These people include Aaron Amsel, Curtis Asplund, Dan Balick, Charlie Beil, Johannes Br\"{o}del, Geoffrey Comp\`ere, Keith Copsey, Richard Eager, Henriette Elvang, Ted Erler, Scott Fraser, Mike Gary, Misato Hayashida, Sungwoo Hong, Matt Lippert, Anshuman Maharana, Nelia Mann, Ian Morrison, Jacopo Orgera, Jo\~ao Penedones, Matt Pillsbury, Sam Pinansky, Matt Roberts, Sam Vazquez, Amitabh Virmani and Sho Yaida.

My friends in Santa Barbara also helped making these years more enjoyable.  Among the ones I haven't already mentioned are Biliana, Carlos, Cec\'ilia, Coral, Daniel, Eduardo, German, Gokhan, James, Jo\~ao, Jonathan, Larisa, Lu, Melissa, Pedro, Ricardo, Rishi, Sabine, Scott, Sofia and Susana.  Furthermore, I would like to thank our neighbors at Family Student Housing who contributed to make it such a nice place to live at, and my soccer buddies for times well spent on the field.

On the financial side, I acknowledge support from {\it Funda\c{c}\~ao para a Ci\^encia e a Tecnologia} through grant SFRH/BD/12241/2003, which allowed me to focus more on research, rather than worrying about more mundane issues.

I saved for last the people I love and care for most deeply.  I would not have reached this point in my life if it had not been for the ones who brought me up and educated me.  Thank you Mother, thank you Father, for your constant support and encouragement.  I am also very happy (and proud, I admit) to say thank you to my daughter Sara, who keeps on showing me how joyful life can be, specially when she displays the most beautiful smile I have ever seen.  Finally, to you, Susana, I am more grateful than I can express.  Thanks for sharing my life over all these years, and for sharing your life with me.  Thank you for putting up with me through all the bad moments (helping me keep my mental sanity!) and for inflating the good ones.  Thank you for your contagious happiness.  And thank you for giving me Sara.

\end{acknowledgements}

\begin{vitae}
\ssp
\begin{vitaesection}{Education}
\item [2008] PhD in Physics, University of California, Santa Barbara
\item [2002] \textit{Licenciado} in Technological Physics Engineering, \textit{Instituto Superior T\'ecnico}, Lisbon, Portugal

\end{vitaesection}

\begin{vitaesectnodate}{Publications}


\item{``Analytic study of small scale structure on cosmic strings''},
  J.~Polchinski and J.~V.~Rocha,
  published in \texttt{Phys.Rev.D74:083504,2006},
  e-print: \texttt{hep-ph/0606205}. \\

\item{``Cosmic string structure at the gravitational radiation scale''},
  J.~Polchinski and J.~V.~Rocha,
  published in \texttt{Phys.Rev.D75:123503,2007},
  e-print: \texttt{gr-qc/0702055}. \\

\item{``Periodic gravitational waves from small cosmic string loops''},
  F.~Dubath and J.~V.~Rocha,
  published in \texttt{Phys.Rev.D76:024001,2007},
  e-print: \texttt{gr-qc/0703109}. \\

\item{``Scaling solution for small cosmic string loops''},
  J.~V.~Rocha,
  published in \texttt{Phys. Rev.Lett.100:071601,2008},
  e-print: \texttt{arXiv:0709.3284 [gr-qc]}. \\

\item{``Cosmic String Loops, Large and Small''},
  F.~Dubath, J.~Polchinski and J.~V. Rocha,
  published in \texttt{Phys.Rev.D77:123528,2008},\\
  e-print: \texttt{arXiv:0711.0994 [astro-ph]}. \\

\item{``Evaporation of large black holes in AdS: coupling to the evaporon''},
  J.~V.~Rocha,
  e-print: \texttt{arXiv:0804.0055 [hep-th]}. \\


\end{vitaesectnodate}

\begin{vitaesection}{Honors and Awards}

\item[2003]Broida fellowship award

\item[2003]Doctoral fellowship from \textit{Funda\c{c}\~ao para a Ci\^encia e
  a Tecnologia} (Portugal)

\end{vitaesection}

\end{vitae}

\begin{abstract}
We present an analytic model specifically designed to address the long standing issue of small scale structure on cosmic string networks.
The model is derived from the microscopic string equations, together with a few motivated assumptions.
The resulting form of the correlation between two points on a string is exploited to study smoothing by gravitational radiation, loop formation and lensing by cosmic strings.
In addition, the properties of the small loop population and the possibility of detecting gravitational waves generated by their lowest harmonics are investigated.
Whenever possible, we compare the predictions of the model to the most recent numerical simulations of cosmic string networks.
\end{abstract}

\addcontentsline{toc}{chapter}{Contents}
\tableofcontents

\listoffigures
\listoftables

\end{frontmatter}

\numberwithin{equation}{chapter}

\chapter{Introduction}

The subject of cosmic strings lies in the interface of many branches of Physics:
first proposed in the context of field theories and in close analogy with condensed matter systems, their existence would be relevant for cosmology and imply detectable gravitational effects.
More recently the connection with string theory has also been made, which resulted in the renewal of interest on the field.

\section{Cosmic strings: what are they?}

The idea that topological defects could form during a cosmological phase transition was first suggested by Kibble in 1976~\cite{Kibble:1976sj}.
The picture behind this proposal was that of a gauge theory that undergoes a symmetry breaking transition at some critical temperature.
In a hot Big Bang model, the temperature of the universe grows without bound as we go back in time.
At very high temperatures the gauge theory finds itself in a symmetry preserving configuration, but as the universe expands and consequently cools down, the critical temperature is reached and the theory enters a broken symmetry phase.

\begin{figure}[t]
\begin{center}
\ \includegraphics[width=30pc]{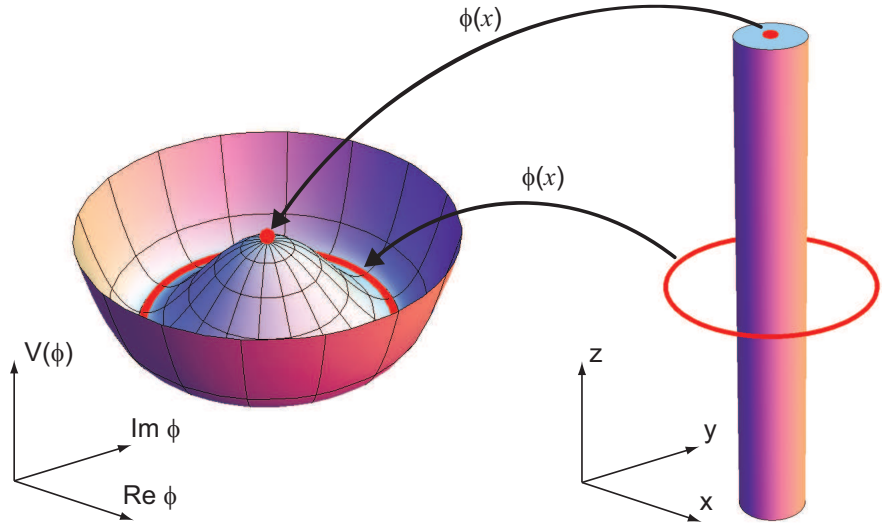}
\end{center}
\caption{The usual quartic ``Mexican hat'' potential for the complex Higgs field $\phi(x)$ is shown on the left.  A vortex string configuration maps spacetime points belonging to the core of the string (on the right side of the figure) to the local maximum of the potential $V(\phi)$ and large circles surrounding the string to the ring of degenerate minima of $V$.}
\label{vortex}
\end{figure}

In particular, whenever the field theory has a spontaneously broken $U(1)$ symmetry there are classical solutions which are localized in two spatial dimensions.
Such solitons are present for example in the Abelian Higgs model, where they are known as Nielsen-Olesen vortices~\cite{Nielsen:1973cs}.
This model couples an Abelian gauge field to a complex scalar field with a typical ``Mexican hat'' potential for which there is a ring of degenerate vacua (see Fig.~\ref{vortex}).
The scalar field configuration for the vortex maps circles in space (with sufficiently large radius) to the ring of the potential minima.
Continuity implies that somewhere in the middle the field sits at the local maximum of the potential.
Analogously, if we lay down a blanket in such a way that its rim encircles, say, a pillow, then some portion of the blanket must cover the pillow.
So the solution has a core where the energy density is localized.

One can now easily imagine extending this solution in one more dimension, thus obtaining a string.
The discussion above makes it evident that these defects are classified by their winding number, the (integer) number of times the field wraps around the ring of vacua as we trace out a closed path in physical space.
The stability (at least in the absence of sources) of these objects stems from this topological nature.
Summarizing, cosmic strings are linear threads of energy with cosmic-scale extensions that are produced in cosmological phase transitions.
Such objects can be very massive as their core retains the typical energy density of the universe before the phase transition.
Thus, cosmic strings represent remnants from the early universe.

Of course, other topological defects are possible.
The simplest one is the domain wall obtained as the `kink' solution for a real scalar field theory interpolating between the two minima of a potential with discrete ${\mathbb Z}_2$ symmetry, trivially extended in the two transverse directions.
Monopoles, which are point-like defects, also arise in theories for which the manifold of degenerate vacua is topologically a sphere.
All the possibilities are nicely classified by homotopy groups.

The crucial point is that these defects inevitably form in cosmological contexts by the Kibble mechanism.
As the temperature drops below the critical value, regions of space-time in the broken symmetry phase start nucleating.
The choice of vacuum by each region must be uncorrelated on distances larger than the cosmological horizon.
Typically, these nucleated bubbles expand with velocities approaching the speed of light and eventually start smashing into each other.
When that happens, occasionally strings are formed when the energy density in its core gets trapped by the non-trivial winding of the field around it.
The conservation of this winding number implies that these strings cannot end as this would require an excursion of the field over the top of the potential and such large fluctuations are prohibited once the temperature has dropped below the Ginzburg temperature.
So they either have infinite extension or they form closed loops.

The same {\it rationale} applies to defects of different dimensionality but domain walls and monopoles are both cosmologically dangerous: their energy density would dominate over the driving source of space-time expansion during part of the history of the universe, therefore ruining the cosmological evolution.
To prevent this catastrophe we postulate a period of accelerated expansion of the universe, known as inflation, thereby diluting all defects formed earlier.
Hence, cosmic strings will only be relevant if they form after or at the end of inflation.

\section{Cosmic strings in our universe?}

Following the seminal paper by Kibble there was a period during which the field was dormant but four years later Zel'dovich~\cite{Zeldovich:1980gh} and Vilenkin~\cite{Vilenkin:1981iu} brought an additional thrust to the subject, by pointing out that cosmic strings produced during a phase transition at a Grand Unification Theory (GUT) energy scale $\eta_{SB}$ ($\sim 10^{16} {\rm GeV}$) would give rise to density fluctuations of the right magnitude to explain galaxy formation.
Indeed, the tension $\mu$ of the string is proportional to the square of the symmetry breaking scale.
Multiplying by Newton's constant $G$ and the appropriate power of the speed of light $c$ we obtain the all-important dimensionless quantity $G\mu/c^2 \sim \eta_{SB}^2 / M_{Pl}^2$.
Here $M_{Pl} \approx 1.3 \times 10^{19} {\rm GeV}/c^2$ is the Planck mass, representing the scale at which gravitational and quantum effects become equally important.
One then concludes that GUT scale strings would generate effects of order $G\mu/c^2 \sim 10^{-6}$.
This combination is proportional to the fractional change in density that would be generated by cosmic strings and the value just given roughly corresponds to what is needed to seed galaxy formation.
Furthermore, it has been shown that the production of cosmic strings in supersymmetric GUT models is quite generic~\cite{Jeannerot:2003qv}.

Much work has been done since then in order to study properties and consequences of cosmic strings in the universe.\footnote{The interested reader is directed to Refs.~\cite{1994csot.book.....V,Hindmarsh:1994re} for nice detailed expositions.}
However, the interest in the subject started to fade away by the end of the nineties with the advent of experiments devoted to the study of the anisotropies in the Cosmic Microwave Background (CMB), namely COBE (Cosmic Background Explorer) and more recently WMAP (Wilkinson Microwave Anisotropy Probe).
The decline was partly caused by the very good agreement between the angular power spectrum of the CMB revealed by the measurements and the predictions of the competing theory of inflation.

In the inflationary paradigm, the density perturbations are generated by quantum fluctuations occurring during the period of accelerated expansion.
Modes with wavelengths greater that the Hubble distance $H(t)$, whose inverse gives the expansion rate of the universe, are frozen in.
When traced back in time, comoving distance scales observable today (and therefore smaller than $H$) cross the Hubble radius and remain frozen for some time before they re-enter.
Different scales stay frozen for different amounts of time in a very specific way and this in turn results in a distribution with characteristic peaks for the angular power spectrum of the density perturbations.

On the contrary, in the cosmic string scenario the density anisotropy is a direct consequence of the inhomogeneity of a universe filled with a network of strings.
The original proposal was that matter accreted around closed loops of cosmic strings.
At that time it was thought that the loop number density was very close to the density of galaxies, suggesting a real connection between the two entities.
However, simulations performed by the end of the eighties~\cite{Bennett:1989yp,Bennett:1989yq,Albrecht:1989mk,Allen:1990tv} showed that this was not the case: typically the string networks evolve into configurations with a much denser gas of loops, with sizes much shorter than the horizon scale and correspondingly increased velocities.
So the picture was replaced by the idea that structure formed in the wakes of moving long cosmic strings.
This is made possible by the gravitational properties of one-dimensional defects: a straight string produces a conical deficit in the geometry of the transverse directions so as it passes between two initially static objects it generates a relative attractive motion.
Still, this mechanism of generating density perturbations predicts a broad, nearly featureless, distribution for the CMB angular power spectrum and it became apparent that cosmic strings were disfavored already with the data collected by COBE~\cite{Albrecht:1997nt}.

So cosmic strings were dismissed as the main source of density perturbations responsible for structure formation.
Indeed, their contribution to the CMB anisotropy has been constrained to be no more then roughly $10\%$~\cite{Pogosian:2003mz}.
In a sense, it can be said that the inflationary theory had more chances of taking the lead as we have already mentioned that even in the case of a universe containing topological defects, we still need to appeal to inflation to avoid cosmological catastrophes.
Nevertheless, cosmic strings are not ruled out and in fact recent studies~\cite{Bevis:2007gh,Pogosian:2008am} indicate that a small contribution from these defects can actually provide better fits to the data, possibly explaining the excess of power in the anisotropy at small scales, for example.

\section{Cosmic superstrings}

More recently, the possibility of the emergence of cosmic strings from models of inflation in string theory, namely the brane inflation scenario~\cite{Dvali:1998pa}, has driven a revival of the field.\footnote{Reference~\cite{Polchinski:2004ia} gives an excellent introduction to the subject of cosmic superstrings.}
In this incarnation, topological defects may arise as fundamental strings (F-strings), D1-branes (D-strings), bound states of F- and D-strings or higher dimensional branes wrapping various cycles in the extra dimensions~\cite{Copeland:2003bj}.

When the possibility of superstrings taking the role of cosmic strings was considered in 1985 by Witten~\cite{Witten:1985fp} it did not seem promising: the fundamental strings, having tensions of order the Plank scale, would generate anisotropies in the CMB much larger than those observed.
In addition, these objects appeared to be unstable: in open string theories a long string would simply break into many small strings and closed strings would collapse due to the force exerted by the tension of a domain wall bounded by the string itself.

The realization that other (non-perturbative) objects exist in string theory and that compact extra dimensions can include throat regions in their geometry, which would suppress the tension of strings lying therein, provided a much more fertile ground for cosmic superstrings.
Indeed, in brane-inflationary models it is the inter-brane distance that drives inflation as two branes approach each other.
Eventually a tachyonic mode appears and inflation ends when the branes collide and annihilate.
The initial system contains U(1) gauge fields and this symmetry is broken at the end of the process.
So again one expects formation of lower dimensional defects.
The types of extended objects we are left with depend on the particular D-brane model but, interestingly, from our four dimensional  point of view one can (and even must) obtain only string-like defects~\cite{Jones:2002cv,Sarangi:2002yt,Jones:2003da,Dvali:2003zj}.
Therefore, the creation of cosmic superstrings at the end of inflation is expected whereas domain walls and monopoles are simply absent.

It should be noted that the properties and stability of these cosmic superstrings are model-dependent.
Depending on the D-brane geometry that produces inflation, the resulting string tension can be anywhere within the range $10^{-12} \lsim G\mu/c^2 \lsim 10^{-6}$~\cite{Sarangi:2002yt,Jones:2003da}.
The issue of stability has been reexamined in the light of the modern understanding of string theory in Ref.~\cite{Copeland:2003bj} where it was found that many models can accommodate cosmic superstrings that are at least metastable.

\section{Networks of cosmic strings}

When a network of cosmic strings is formed it is expected to resemble a tangle of random walks on scales greater than the correlation length of the Higgs field.
This is confirmed by numerical simulations~\cite{Vachaspati:1984dz}.
Some strings form closed loops but in an infinite universe strings of infinite extension (often called long strings) will also exist.
After formation the network will not remain static.
First of all, the strings have tension and so the segments will start moving around.
On the other hand, spacetime itself is evolving and so the network is initially in an out-of-equilibrium state.

The existence of cosmic strings populating our universe would generate effects we could detect, in principle.
Besides influencing the CMB angular power spectrum as already mentioned, other effects are also predicted:  the conical nature of spacetime around a string gives rise to very peculiar gravitational lensing properties~\cite{Vilenkin:1981zs} and this should also affect the CMB temperature pattern by introducing step-like discontinuities~\cite{Kaiser:1984iv};  their gravitational radiation properties leads to strong bursts of gravitational waves originating from special events (cusps and kinks) along the strings~\cite{Damour:2000wa};  also, the presence of a gravitational radiation stochastic background would distort the very precise time-periodicity of pulsars~\cite{Hogan:1984is};  finally, particle emission from annihilating strings, in particular gamma ray bursts~\cite{Berezinsky:2000vn}, is a possibility as well.

The lack of any such observation until present date translates into constraints on the cosmic string properties, typically an upper bound on the dimensionless string tension $G\mu/c^2$.
The most stringent current bounds come from matching the CMB power spectrum~\cite{Pogosian:2006hg}, which gives $G\mu/c^2 \lsim 2.7 \times 10^{-7}$ at $95\%$ confidence level.
The effect of cosmic strings on the angular power spectrum comes mostly from the long string population and depends primarily on the large scale properties of the network.
The reliability of the above bound stems from the fact that the large distance structure of cosmic strings is well understood.
A recent study of the gravitational lensing effect reports the constraint $G\mu/c^2 < 3.0 \times 10^{-7}$ at $95\%$ confidence level~\cite{Christiansen:2008vi} and this limit should be easily improved by an order of magnitude in the near future~\cite{Gasparini:2007jj}.
Measurements of pulsar timing exclude values of the dimensionless string tension above $10^{-9}$~\cite{DePies:2007bm} but these methods depend on more debatable network properties, most notably the typical size of loops formed by self-intersections of long strings and their velocities.

The previous paragraph illustrates the importance of understanding the evolution of cosmic string networks for establishing constraints on the parameters.
Another aspect is that knowledge about these systems provides a better guide to search for cosmic strings in our universe.
Finally, if one day such topological defects are discovered, we would hope to be able to interpret the data and obtain valuable information about the microscopic theory that supports those extended objects.
For example, one would wish to know if the detected strings (or rather their effects) were field theoretic cosmic strings or string theory counterparts.

Fortunately, there are some differentiating features.
In particular, when two segments of field theoretic vortex strings intersect the outcome is deterministic and for most of the range of parameters, namely the relative velocity, it results in a reconnection.
On the contrary, in string theory this intercommutation is a quantum process and the probability of reconnection can be highly suppressed.
Also the presence of bound states in string theory permits the existence of Y-junctions in the string network and such features are not possible for ordinary cosmic strings.

As it turns out, the evolution of cosmic string networks is a notoriously difficult problem.
The string equation of motion in flat spacetime is a simple wave equation but once we consider expanding spacetimes the equations of motion become non-linear.
In addition, the evolution of these systems is also influenced by intercommutation events (which generates kinks along the strings~\cite{Bennett:1987vf}) and gravitational radiation (which shortens the strings).
As we will discuss, gravitational radiation effects are naturally suppressed by (powers of) $G\mu/c^2$.
Given the current bounds on the dimensionless string tension one concludes that there are large ratios of length and time scales involved in the evolution of cosmic string networks, rendering a full numerical analysis impracticable.

The large scale properties, i.e. characteristics of the network on scales comparable to the horizon distance, have been fairly well understood from the earliest simulations~\cite{Albrecht:1989mk,Bennett:1989yp,Allen:1990tv}.
As long as there are a few long strings at formation of the network, the late time configuration approaches the so-called {\em scaling regime}, in which all length scales grow proportionally with time.
Independently of the initial conditions there will be a few dozen long strings crossing any horizon volume and on these large scales the strings trace out Brownian trajectories.
Given the importance of the scaling regime we dedicate the following subsection to describing its nature.

\subsection{The scaling regime}

The idea of a scaling regime for a cosmic string network was first introduced in~\cite{Kibble:1984hp}.
In addition to its obvious attractiveness, it was confirmed by the early simulations~\cite{Albrecht:1984xv,Bennett:1987vf,Albrecht:1989mk,Bennett:1989yp,Bennett:1989yq,Allen:1990tv}.
An important consequence of the scaling regime is that the energy density in strings remains a (small) fixed fraction of the background energy density in both radiation and matter cosmological eras as it redshifts with the same power of the scale factor in each epoch.
The by now consensual value for the contribution from long cosmic strings to the total energy density is roughly $60 G\mu/c^2$ ($10^3 G\mu/c^2$) times the matter (radiation) energy density in a matter (radiation) dominated universe~\cite{Albrecht:1989mk,Bennett:1989yp,Allen:1990tv,Martins:2005es,Ringeval:2005kr}.

Several scales can be defined for a cosmic string network.
Among them, the {\em characteristic length} of the network plays an important part.
It is defined as the length scale $L$ such that a typical volume $L^3$ of the network contains a length $L$ of long strings.
Another length scale is the {\em persistence length} $\xi$, i.e. the distance along the string beyond which points are uncorrelated.
In any case, we say that a quantity (with units of length) is scaling if it remains constant in units of the cosmological time $t$ as the network evolves.
Consequently, in a scaling regime the evolution is self-similar, in the sense that the network at a given time will resemble itself at a previous time after all lengths are scaled down appropriately.

Thus, in a scaling regime we have $L=\gamma c t$, for some constant $\gamma$, and the energy density in long strings becomes
\begin{equation}
\rho_\infty = \frac{\mu \, c^2 L}{L^3} = \frac{\mu}{\gamma^2 t^2}  \ .
\end{equation}
The consensual value for the proportionality constant is $\gamma^{-2} \sim 3$ in a matter-dominated universe and $\gamma^{-2} \sim 10$ in a radiation-dominated universe~\cite{Bennett:1989yp,Allen:1990tv,Martins:2005es,Ringeval:2005kr}.
Similarly, the persistence length $\xi$ is also observed to approach a scaling regime~\cite{Vincent:1996rb,Martins:2005es}.

A crucial feature of cosmic string networks is that the scaling regime is an attractor.
The intuitive picture behind this is the following.
If the long strings simply grew with the expansion of the universe their density would decrease as $a(t)^{-2}$, where $a(t) \propto t^{\nu}$ is the scale factor and $\nu = 1/2$ for a radiation-dominated universe and $\nu = 2/3$ in a matter-dominated era.
Note that such a density is growing with respect to a scaling density, which would decay as $t^{-2}$.
If we start out with too many long strings the probability of an encounter between two of them will be large, leading to a bigger loss of long strings into loops.
On the other hand, if we begin with a sparse density of infinite strings intercommutations will be rarer events and the long string density will increase relative to the scaling value and eventually reach it.
Therefore, the late time scaling behavior is independent of the initial conditions as long as there are a few long strings to begin with.\footnote{A notable exception occurs in superstring networks for which there are several attractors; which one is chosen at the outcome depends on the initial conditions in this case~\cite{Leblond:2007tf}.}

More recently the issue of scaling in cosmic superstring networks has also been studied numerically~\cite{Copeland:2005cy,Tye:2005fn,Leblond:2007tf,Urrestilla:2007yw}.
The existence of Y-junctions could possibly lead the network to freeze but the scaling regime seems to be a robust prediction in this case as well.
In fact, there are indications that the late time evolution of cosmic string and superstring networks might be very similar~\cite{Rivers:2008je}.
If this is the case, the results presented in this dissertation should also be applicable in the cosmic superstring scenario.

\subsection{Small scale structure}

By now, the evolution of structure on cosmic string networks on large scales is a well understood problem.
The situation gets more complicated and less clear once we turn to the short distances.
Indeed, the early simulations~\cite{Albrecht:1989mk,Bennett:1989yp,Allen:1990tv} showed evidence for the build-up of small scale structure along the strings, caused by kinks generated during intercommutations, and for the presence of a gas of tiny loops with sizes accumulating at the smallest possible scale -- the resolution scale.
To date, there have been many analytic and numerical studies of cosmic string networks, but no single work has yielded completely satisfactory results.
As mentioned above, the large ratios of length and time scales involved make a full numerical treatment impracticable.
On the other hand, analytic methods are difficult because of the highly non-linear nature of the system.
The hope is that a careful combination of analytic and numerical approaches will lead to a good understanding of these networks.

The properties of the small scale structure have been analyzed over the last decade and a half and its existence has deep implications for the networks and consequently for the cosmology.
Strings with a fair amount of short distance structure will self-intersect and produce small loops with high probability.
However, the typical size at which loops form has been an issue of much debate over the years.
Estimates range from the thickness of the string~\cite{Vincent:1996rb} at the lower end, to sizes just an order of magnitude below the horizon distance~\cite{Vanchurin:2005pa,Olum:2006ix}.
Less radically, Refs.~\cite{Ringeval:2005kr,Martins:2005es} suggest that loops form predominantly with sizes $10^{-3} - 10^{-4}$ times smaller that the horizon.
It should be noted that larger loops lead to enhanced signatures of several types, and consequently to tighter bounds on $G\mu/c^2$~\cite{Olum:2006at}.

Of course, after establishing the presence of small scale structure, one would like to understand its evolution.
Does the short distance structure approach a scaling regime as well?
Part of literature assumes that gravitational radiation is needed for scaling, and that it determines the size of loops.
However, the most recent simulations suggest that the late time properties evolve in a scaling fashion~\cite{Vanchurin:2005yb,Vanchurin:2005pa,Ringeval:2005kr,Martins:2005es,Olum:2006ix}, even though gravitational radiation is not included in their algorithms.

\subsection{Analytic approaches}

The present dissertation is based on analytic approaches to the study of scaling networks of cosmic strings. 
Several other studies have been previously conducted and we now briefly survey them.
This will serve as a platform for comparison with the work presented in this dissertation.

The one-scale model~\cite{Kibble:1984hp,Bennett:1985qt,Bennett:1986zn} nicely accommodates the scaling behavior of the network at late times as seen in simulations but only concerns the large scale properties.
A model with two scales was considered in~\cite{Kibble:1990ym}, one of which characterized the long-string density and the other one being the persistence length.
However, none of these explained the small scale structure observed in the simulations and so a third length scale was later added~\cite{Austin:1993rg} to specifically describe it.

The picture arising from the detailed studies of~\cite{Austin:1993rg} confirms that the problem of cosmic string network evolution is very complex: all the processes that occur during the evolution interact with each other.
Unfortunately, the large number of unknown parameters (and assumptions) somewhat reduce the appeal of this three-scale model.
The complexity of these systems was also revealed in attempts to use path-integral methods~\cite{Embacher:1992zk,Embacher:1992zm}.
There, even working in flat space and assuming the simplest possible string probability distribution computations quickly become quite involved.

An improvement over the original one-scale model came with the velocity-dependent one-scale model~\cite{Martins:1996jp,Martins:2000cs}, which allowed a varying averaged string velocity.
This approach can address the evolution in transient regimes and therefore in realistic cosmological scenarios but, once again, ignores the small-scale issues.

Along different lines, another model with a single scale was developed in~\cite{Allen:1990mp}.
The scale considered therein, determined by the number density of kinks, was found to approach very small values compared to the horizon distance but the model is in many ways too simplistic.

\section{Outline}

The materials presented in the following chapters relie on Refs.~\cite{Polchinski:2006ee,Polchinski:2007rg,Dubath:2007wu,Rocha:2007ni,Dubath:2007mf} and are mainly results of collaborations with Joseph Polchinski and Florian Dubath.

As we have alluded to before, the evolution of cosmic string networks is a notoriously difficult problem.
Nevertheless, the problem is well-posed~\cite{Polchinski:2007qc} and so our difficulties in studying such networks reflect their complexity as a whole and not a lack of understanding of the microscopics {\it per se}.
Therefore, it is not unreasonable to imagine that a good strategy to tackle the evolution of such networks starts by focusing on a microscopic description.
In Chapter~2 we set up an analytic model whose purpose is to describe the small scale structure on strings.
The model is derived from a few simplifying assumptions on the network evolution which are expected to hold over a large range of length scales.
This approach has the advantage of avoiding the nonlinearities of the system which dominate at the horizon scale.
The full non-linear string equations do not seem amenable to analytic techniques and at such large scales we must trust the numerical simulations.
The main conclusion is that the strings become smooth at short distances but there is a power law in this approach to straightness.
The exponent $\chi$ controlling the deviation from straightness is determined by cosmological parameters and large scale properties of the networks.

In Chapter~3 we use our model to investigate the issue of gravitational radiation from long cosmic strings.
The emission of gravitational radiation depends crucially on the spectrum of perturbations along the string and also leads to the smoothing of cosmic strings below a certain scale.
We find that this scale is smaller than the horizon distance by a power of $G\mu/c^2$ and the exponent is related to the spectrum of perturbations, namely $\chi$.

Chapter~4 is devoted to a study of loop formation from self-intersections of cosmic strings and takes the model of Chapter~2 as input.
The main surprise here is that without inserting an ultra-violet (UV) cutoff on the spectrum of fluctuations the rate of loop production diverges, despite the smoothness of the strings on small scales.
A careful treatment reveals that loops of all sizes form simultaneously (instead of cascading down from large loops to small ones) with a power law distribution determined by $\chi$.
This result is in good agreement with recent simulations~\cite{Ringeval:2005kr,Olum:2006ix}.
The role of the UV cutoff is taken by the gravitational radiation scale and it partly determines the normalization of the loop distribution.

We reserve Chapter~5 to a survey of the small loop population properties and in particular to the investigation of the appearance of a scaling regime in the loop distribution.
We point out that loops with sizes close to the gravitational radiation cutoff are formed with large Lorentz boosts.
Also, we show that the loop length distribution approaches a scaling regime at late times, both with and without the inclusion gravitational radiation.

In Chapter~6 we consider some consequences for observational signatures.
In a first part we calculate deviations from perfect lensing by cosmic strings and derive expectations for the alignment between several pairs of cosmic string lenses.
Subsequently, we investigate the detectability of (quasi-) periodic gravitational waves generated by loops, taking into account the large Lorentz factors characteristic of small loops.
We find that such mechanism of gravitational wave production can yield detectable signals for values of $G\mu/c^2$ of order $O(10^{-9})$, but this depends strongly on the direction of motion of the small loops relative to the detectors.
When integrated over the population of loops within the horizon, this gives an expected rate of detectability at Advanced LIGO equal to $10^{-4}$ events per year, for $G\mu/c^2 \sim 10^{-9}$.

Finally, we conclude in Chapter~7 with some final remarks.


In the remaining part of this dissertation we shall use units in which $c = 1$.

\chapter[Model for small scale structure]{Model for small scale structure on cosmic strings}

In this chapter we delineate a model designed to describe the small scale structure on cosmic strings.
The basic object we consider is the two-point function which describes how points along a single string are correlated with each other.
The full microscopic equations for the string network~\cite{Embacher:1992zk,Austin:1993rg} appear to be too complicated to solve, and so we model what we hope are the essential physical processes.
In order to make the problem more tractable we resort to simplifying assumptions and well-motivated approximations.

We begin by offering a preliminary Section~\ref{PRELIM} where the necessary ingredients from the cosmic string literature are introduced.
In Section~\ref{ASSUMPTIONS} we identify our assumptions, arguing that over a comprehensive range of scales the dominant effect in the evolution of cosmic string networks is the stretching due to the expansion of the universe.
Based on these assumptions, in Section~\ref{TWOPF} we are able to determine the form of the two-point function up to two parameters which are inferred by matching our solution to numerical simulations.
We will see that the string is actually rather smooth, in agreement with simulations~\cite{Martins:2005es}: its fractal dimension approaches one as we go to smaller scales.
There is, however, a nontrivial power law that reveals itself in the {\it approach} of the fractal dimension to one.
The critical exponent, which is determined by the mean string velocity, is related to the power spectrum of perturbations on the long string.
Section~\ref{DISCUSS} is devoted to discussion and comparison of our results for the two-point function with the above-mentioned simulations.

The model presented here was developed together with Joseph Polchinski in Ref.~\cite{Polchinski:2006ee} and follows that reference closely.


\section{Preliminaries: Basic ingredients\label{PRELIM}}

In Chapter~1 we saw that cosmic strings can arise both in field theory (namely in Grand Unification Theories) and string theory contexts.
Cosmic superstrings are honest one-dimensional objects but their field theoretic counterparts can also be regarded as one-dimensional defects, at least on scales much larger than their thickness.
Thus, it comes as no surprise that for the simplest class of ``vanilla'' cosmic strings their dynamics is described by the Nambu action:
\begin{equation}
S = - \mu \int \sqrt{-\gamma} \: d^2\sigma  \ .
\label{Nambu}
\end{equation}
A string evolved in time sweeps out a two-dimensional surface, the worldsheet.
The above action is simply proportional to the area swept by the string in spacetime.
The time-like and space-like coordinates parameterizing the worldsheet are denoted by $\sigma^0$ and $\sigma^1$, respectively, and the integral is taken over these coordinates.
The prefactor $\mu$ is nothing but the string mass per unit length and the quantity $\gamma$ is the determinant of the metric induced on the worldsheet from the embedding in four-dimensional spacetime.
Defining this embedding by the functions\footnote{Greek indices will be used for spacetime coordinates ($\mu = 0,\dots,3$) and lowercase Latin indices will represent worldsheet coordinates ($a = 0,1$).} $x^\mu(\sigma^a)$, the induced metric is given by
\begin{equation}
\gamma_{ab} = g_{\mu\nu} \partial_a x^\mu \partial_b x^\nu  \ .
\end{equation}
Expression~(\ref{Nambu}) represents an effective action for gauge strings which can be derived under rather general considerations and provides a good description whenever their radius of curvature is much larger than the thickness of the string~\cite{1994csot.book.....V}.

The action~(\ref{Nambu}) enjoys the property of invariance under a large set of transformations, namely reparametrizations of the worldsheet.
This freedom can be used to bring the worldsheet metric into diagonal form.
Furthermore, one can also identify the time-like coordinates of the worldsheet and of the ambient four-dimensional spacetime, $\sigma^0 = x^0 = t$.
Such a procedure still leaves a residual gauge freedom corresponding to time-independent reparametrizations of $\sigma^1$.\footnote{Since the time-like coordinate is already fixed in this gauge we shall drop the superscript from $\sigma^1$ in the rest of this dissertation.}
Therefore, in this so-called transverse gauge the spacetime coordinates of the string are fully specified by the spatial vector ${\bf x}(t,\sigma)$, where $\sigma$ parametrizes the string at fixed time $t$.
Note that the first gauge fixing condition amounts to the following constraint:
\begin{equation}
\partial_t {\bf x} \cdot \partial_\sigma {\bf x} = 0  \ .
\end{equation}
This states that at any given point on the string the velocity is orthogonal to the tangent vector, thus justifying the name of this particular gauge.

We are interested in cosmic strings evolving in cosmological spacetimes of the Friedmann-Robertson-Walker (FRW) type, for which the ambient metric $g_{\mu\nu}$ is defined by
\begin{equation}
ds^2 = g_{\mu\nu} dx^\mu dx^\nu
= - dt^2 + a(t)^2 d{\bf x} \cdot d{\bf x}
= a(\tau)^2 (- d\tau^2 + d{\bf x} \cdot d{\bf x})  \ .
\end{equation}
The conformal time $\tau$ has been introduced via $d\tau = a^{-1}(t) dt$ and the scale factor $a(t)$ determines the expansion of these spatially flat universes.
Therefore, two comoving observers (sitting at fixed ${\bf x}$ coordinates) separated by a physical distance $\ell$ experience a recessional velocity relative to each other equal to $v = H \ell$, where
\begin{equation}
H = \frac{1}{a} \frac{da}{dt}
\end{equation}
is the Hubble parameter, whose present value is about $73 \, {\rm Km} \, {\rm s}^{-1} {\rm MPc}^{-1}$.

We shall consider only the situation of universes dominated by radiation or by matter, in which case the scale factor grows in time as a simple power law,
\begin{equation}
a \propto t^\nu \propto \tau^{\nu'}\ ,\quad\quad \nu' = \nu/(1 - \nu)  \ ,
\end{equation}
where $\nu = 1/2$ and $\nu' = 1$ ($\nu = 2/3$ and $\nu' = 2$) in a radiation (matter) dominated era.
This implies that the Hubble parameter decreases with time as $t^{-1}$.
Therefore, a useful measure for timescales is provided by the Hubble time, $H^{-1}$.
This represents the time it would take for the universe to expand to twice its (linear) initial size assuming the expansion rate remained fixed at the initial value. 

An important feature of FWR spacetimes is the presence of a cosmological horizon: events can only be causally connected if they are space-separated by no more than the horizon distance
\begin{equation}
d_H(t)  =  a(t) \int_0^t \frac{dt'}{a(t')}  =  \frac{t}{ 1 - \nu }  \ .
\end{equation}
The second equality applies only to the case of a power law expansion.
In astrophysics the cosmological redshift $z$ is commonly used to specify distances; it contains the same information as the scale factor since the two are related by $1 + z(t) = a(t)^{-1}$, where conventionally the scale factor at present time is set to unity.

In an FRW background the equation of motion governing the evolution of a cosmic string, derived by varying the action~(\ref{Nambu}) with respect to the string embedding, is~\cite{Turok:1984db}
\begin{equation}
\ddot{\bf x} + 2\,\frac{\dot{a}}{a}\,(1-\dot{\bf x}^2)\, \dot{\bf x} = \frac{1}{\epsilon} \left( \frac{{\bf x}'}{\epsilon} \right)' \ .
\label{CS_EOM}
\end{equation}
Here $\epsilon$ is given by
\begin{equation}
\epsilon \equiv \left( \frac{{\bf x}'^2}{1-\dot{\bf x}^2} \right)^{1/2} \ .
\label{epsilon}
\end{equation}
These equations hold in the transverse gauge, now with the time-like worldsheet coordinate identified with the {\it conformal} time $\tau$.\footnote{From now on we will use dots and primes to refer to derivatives relative to the conformal time $\tau$ and the spatial parameter $\sigma$ along the string, respectively.}
It is the non-linear nature of these equations that undermine a complete analytic treatment.

The evolution of the parameter $\epsilon$ follows from equation (\ref{CS_EOM}),
\begin{equation}
\frac{\dot{\epsilon}}{\epsilon} = -2\,\frac{\dot{a}}{a} \, \dot{\bf x}^2  \ ,
\label{Eps_Evol}
\end{equation}
and it is related to the energy $dE$ carried by a segment of string with infinitesimal coordinate length $d\sigma$ in an expanding universe by $dE = \mu \, a(\tau) \epsilon(\tau,\sigma) d\sigma$.
The denominator in equation~(\ref{epsilon}) neatly accounts for the Lorentz factor expected for relativistic motions.

From the second derivative terms in equation~(\ref{CS_EOM}) it follows that signals on the string propagate to the right and left with $d\sigma = \pm d\tau / \epsilon$.
The equation of motion also includes a friction term which is controlled by the Hubble parameter $\dot{a}/a$.
Thus, in a flat spacetime the structure on a short piece of string at a given time is a superposition of left- and right-moving segments, and it is these that we follow in time in Section~\ref{TWOPF}.
In an expanding universe the left- and right-moving waves interact --- they are not free as in flat spacetime.

Let us now turn to the issue of evolution of cosmic string networks.
As we have discussed in the Introduction, a key property of such networks is that their energy density $\rho_s$ does not dominate over the radiation or matter energy densities in their respective epochs (for which $\rho_r \propto a(t)^{-4} \propto t^{-2}$ and $\rho_m \propto a(t)^{-3} \propto t^{-2}$).
If the only process acting on the strings were the conformal expansion imposed by the growth of the universe we would have $\rho_s \propto a(t)^{-2}$.
However, there are several other mechanisms that come into play and reduce $\rho_s$ down to a (small) constant fraction of the background energy density when the scaling regime is reached in each era.
Indeed, when scaling, the length of string within a horizon volume grows with $t$ (so the energy density contained in the cosmic string network is proportional to $\mu t^{-2}$) and the ratios $\rho_s/\rho_r$ and $\rho_s/\rho_m$ are proportional to the small parameter $G\mu$.
This fact is at the origin of the cosmological viability of cosmic strings.
We now discuss the individual processes mentioned above.

\subsubsection*{Stretching}
As the spacetime fabric expands, it tends to `drag' with it all objects sitting within.
As the FRW models are homogeneous and isotropic, one would na\"{i}vely expect that infinite strings would grow uniformly.
This is true on scales larger than the horizon size $d_H$ for which irregularities on a string are just conformally amplified.
However, considering small perturbations on an otherwise straight string and performing a linearized analysis~\cite{Vilenkin:1981kz} one finds that the modes with a wavelength shorter than $d_H$ do not grow in amplitude accordingly.
Thus, on smaller scales the string effectively straightens.
This effect is incorporated in the description of the string evolution by the Nambu action in curved spacetime.

\subsubsection*{Gravitational radiation smoothing}
Another source of smoothing comes from gravitational radiation.
Cosmic strings interact gravitationally through their energy-momentum tensor.
Straight strings do not generate any gravitational effect on surrounding matter but this is no longer true once we introduce oscillations.
In particular, any given point on the string interacts with the rest of the defect, resulting in power emitted as gravitational waves (GW).
This in turn steals energy from the cosmic string itself which then consequently straightens.
The strength of the GW emission is naturally controlled by the dimensionless string coupling $G\mu$, the product of Newton's constant and the string mass per unit length.
However, this smoothing by gravitational radiation becomes relevant only below a length scale proportional to $d_H$ and to a positive power of $G\mu$~\cite{Bennett:1989yq,Hindmarsh:1990xi,Quashnock:1990wv,Siemens:2002dj}.
We will have more to say about this in Chapter~3.

\subsubsection*{Intercommuting}
\begin{figure}[t]
\begin{center}
\ \includegraphics[width=23pc]{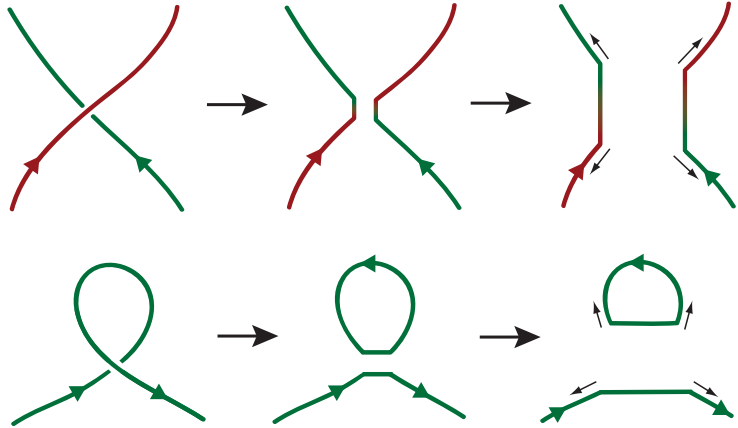}
\end{center}
\caption{The top panel shows an intercommutation event in which two segments of string exchange partners.  The bottom panel displays a single cosmic string self-intersecting and chopping off a loop.  In such processes pairs of kinks are formed on the intervening segments and travel in opposite directions along the string.}
\label{reconnect}
\end{figure}

As the cosmic string network evolves in time, often two segments of string intersect.
When this happens and there is just one type of string in the network only two outcomes are possible: either the segments reconnect (with probability $P$) or they pass through each other (with probability $1-P$).\footnote{These two classical processes dominate over situations resulting in entangled strings~\cite{Shellard:1987bv}.  Also, In more complicated networks one might have Y-junctions as well and therefore `bridges' between strings~\cite{Bevis:2008hg}.}
This process is shown in Fig.~\ref{reconnect}.
Field theory strings always reconnect, except when very extreme conditions are met by the collision parameters~\cite{Shellard:1987bv}, whereas for cosmic superstrings the reconnection is a quantum process whose strength depends on the value of the string coupling constant~\cite{Jackson:2004zg}.

An intercommutation event can result in the formation of closed loops.
This important process can happen when a string self-intersects, as illustrated in the bottom panel of Fig.~\ref{reconnect}.
A side effect of a reconnection is the appearance of pairs of kinks on each of the two intervening segments.
These kinks then start moving apart and open up (becoming less acute) as time elapses, again due to the expansion of the universe.

\subsubsection*{Decay of loops by emission of gravitational radiation}
Loops with sizes smaller than the horizon scale do not grow with the expansion of the universe.
These loops live in an approximately flat spacetime and so they oscillate almost freely, instead.
Consequently, they emit gravitational radiation and eventually decay completely, disappearing from the network.
Thus, the formation of small loops and their subsequent decay are essential to achieve the all-important scaling regime.

\section{Assumptions of the model\label{ASSUMPTIONS}}

We consider ``vanilla'' cosmic strings, a single species of local string without superconducting or other extra internal degrees of freedom.
In this dissertation we will assume for simplicity that every collision between two segments of string results in a reconnection, so that the intercommutation probability is $P = 1$.
The evolution of a network of such strings is dictated by three distinct processes:\footnote{We neglect damping forces arising from interactions with the surrounding radiation background. Such frictional terms enter the evolution equations in the same manner as the Hubble damping and become subdominant roughly after a time $(G\mu)^{-1}t_{\rm f}$, where $t_{\rm f}$ is the time of string formation~\cite{1994csot.book.....V}.}
\begin{itemize}
 \item{
  First, the expansion of the universe stretches the strings;
 }
 \item{
  Gravitational radiation also has the effect of straightening the strings, but it is significant only below a length scale proportional to $d_H$ and to a positive power of $G\mu$.
Since this is parametrically small at small $G\mu$, we will ignore this effect for the present purposes;
 }
 \item{
  Finally, intercommutations play an important role in reaching a scaling solution, in particular through the formation of closed loops of string.
At first we shall neglect this effect but we will be forced to return to this issue in Chapter~4.
 }
\end{itemize}

Let us consider the evolution of a small segment on a long string.
We take the segment to be very short compared to the horizon scale, but long compared to the scale at which gravitational radiation is relevant.
The scaling property of the network implies that the probability per Hubble time for this segment to be involved in a long string intercommutation event is proportional to its length divided by $d_H$, and so for short segments the intercommutation rate per Hubble time will be small.
Formation of a loop much larger than the segment might remove the entire segment from the long string, but this should have little correlation with the configuration of the segment itself, and so will not affect the probability distribution for the ensemble of short segments.
Formation of loops at the size of the segment and smaller could affect this distribution but this process takes place only in localized regions where the left- and right-moving tangents are approximately equal.\footnote{Nevertheless, the results of Chapter~4 indicate that the production of small loops is large.}
Thus, there is a regime where stretching is the only relevant process.

If we follow a segment forward in time, its length increases but certainly does so more slowly than the horizon scale $d_H$, which is proportional to the FRW time $t$.
Thus the length divided by $d_H$ decreases, and therefore so does the rate of intercommutation.
If we follow the segment backward in time, its length eventually begins to approach the horizon scale, and the probability becomes large that we encounter an intercommutation event.
Our strategy is therefore clear.
For the highly nonlinear processes near the horizon scale we must trust simulations.
At a somewhat lower scale we can read off the various correlators describing the behavior of the string, and then evolve them forward in time using the Nambu action until we reach the gravitational radiation scale.
The small probability of an intercommutation involving the short segment can be added as a perturbation.
This approach is in the spirit of the renormalization group, though with long and short distances reversed.

\section{Two-point functions at short distance\label{TWOPF}}

From Eq.~(\ref{Eps_Evol}) it follows that the time scale of variation of $\epsilon$ is the Hubble time, and so to good approximation we can replace $\dot{\bf x}^2$ with the time-averaged $\bar v^2$ (bars will always refer to root-mean-square (RMS) averages), giving $\epsilon \propto a^{-2\bar v^2}$ as a function of time only.\footnote{The transverse gauge choice leaves a gauge freedom of time-independent $\sigma$ reparameterizations.  A convenient choice is to take $\epsilon$ to be independent of $\sigma$ at the final time, and then $\epsilon$ will be $\sigma$-independent to good approximation on any horizon length scale in the past.}
The RMS velocities for points on long strings are taken from simulations~\cite{Martins:2005es}: 
\begin{eqnarray}
\mbox{radiation domination:}&&  \bar v^2 \approx 0.41 \ ,   \nonumber\\  [-10pt]
\mbox{matter domination:}&&  \bar v^2 \approx 0.35 \ .
\end{eqnarray}

From the definition of $\epsilon$ it follows that the energy of a segment of string of coordinate length $\sigma$ is
$E = \mu \, a(\tau) \epsilon(\tau) \sigma$.
For simplicity we will refer to $E/\mu$ as the length $l$ of a segment,
\begin{equation}
l = a(\tau) \epsilon(\tau) \sigma \ ,
\end{equation}
though this is literally true only in the rest frame.

As a consequence,
\begin{equation}
l \propto \tau^{\zeta'} \propto t^{\zeta} \ , \quad\quad  \zeta' = (1 - 2\bar v^2) \nu' \ , \quad\quad  \zeta = (1 - 2\bar v^2) \nu  \ .
\label{stretch}
\end{equation}
In the radiation era $\zeta_r \sim 0.1$ while in the matter era $\zeta_m \sim 0.2$.
Thus the physical length of the segment grows in time, but more slowly than the comoving length~\cite{Vilenkin:1981kz}, and much more slowly than the horizon length $d_H$.

For illustration, consider a segment of length $(10^{-6} \; {\rm to} \; 10^{-7}) d_H$, as would be relevant for lensing at a separation of a few seconds and a redshift of the order of $z \simeq 0.1$.
According to the discussion above, $l / d_H$ depends on time as $t^{\zeta - 1} \sim t^{-0.8}$ in the matter era.
Thus the length of the segment would have been around a hundredth of the horizon scale at the radiation-to-matter transition.
In other words, it is the nonlinear horizon scale dynamics in the radiation epoch that produces the short-distance structure that is relevant for lensing today, in this model.
This makes clear the limitation of simulations by themselves for studying the small scale structure on strings, as they are restricted to much smaller dynamical ranges.

For our purposes it is convenient to work not with the velocity and tangent vectors but instead with a linear combination of them, $\bf{p}_\pm \equiv \dot{\bf{x}} \pm \frac{1}{\epsilon}\bf{x}'$.
In the transverse gauge we are adopting, these are unit vectors as can easily be seen by employing the definition~(\ref{epsilon}).
Furthermore, in a flat spacetime ($\dot{a}=0$) the vector $\bf{p}_+$ ($\bf{p}_-$) would be a left-mover (right-mover) as it is annihilated by the differential operator $\partial_\tau - \frac{1}{\epsilon} \partial_\sigma$.
In terms of these left- and right-moving unit vectors the equation of motion~(\ref{CS_EOM}) can be written as~\cite{Albrecht:1989mk}
\begin{equation}
\dot{\bf p}_\pm \mp \frac{1}{\epsilon} {\bf p}'_\pm = - \frac{\dot{a}}{a} \left[ {\bf p}_\mp - ({\bf p}_+ \cdot {\bf p}_-)\, {\bf p}_\pm \right] \ .
\label{CS_EOM2}
\end{equation}
We will study the time evolution of the left-moving product ${\bf p}_+(\tau,\sigma) \cdot {\bf p}_+(\tau,\sigma')$.
For this it is useful to change variables from $(\tau,\sigma)$ to $(\tau,s)$ where $s$ is constant along the left-moving characteristics,
$\dot s - s'/\epsilon = 0$.
Then
\begin{eqnarray}
\partial_\tau ({\bf p}_+ (s,\tau) \cdot {\bf p}_+(s',\tau)) &=&
 - \frac{\dot a}{a} \biggl( {\bf p}_-(s,\tau)\cdot{\bf p}_+(s',\tau)  +  {\bf p}_+(s,\tau)\cdot{\bf p}_-(s',\tau)     \nonumber\\
&&\hspace{-100pt}  +  \alpha (s,\tau)\,{\bf p}_+(s,\tau)\cdot{\bf p}_+(s',\tau)  +  \alpha  (s',\tau)\,{\bf p}_+(s,\tau)\cdot{\bf p}_+(s',\tau) \biggr) \ ,
\label{bilinear}
\end{eqnarray}
where $\alpha = - {\bf p}_+ \cdot{\bf p}_- = 1 - 2v^2$.

The equations of motion~(\ref{CS_EOM2}, \ref{bilinear}) are nonlinear and do not admit an analytic solution, but they simplify when we focus on the small scale structure.
If ${\bf p}_+ (s,\tau)$ were a smooth function on the unit sphere, we would have $1 - {\bf p}_+ (s,\tau) \cdot {\bf p}_+(s',\tau) = O([s-s']^2)$ as $s'$ approaches $s$.
We are interested in any structure that is less smooth than this, meaning that it goes to zero more slowly than $[s-s']^2$.
For this purpose we can drop any term of order $[s-s']^2$ or higher in the equation of motion (smooth terms of order $s-s'$ cancel because the function is even).

Consider the product ${\bf p}_-(s,\tau) \cdot {\bf p}_+(s',\tau)$.
The right-moving characteristic through $(s,\tau)$ and the left-moving characteristic through $(s',\tau)$ meet at a point $(s,\tau-\delta)$ where $\delta$ is of order $s - s'$.\footnote{Explicitly, for given $\tau$ we could choose coordinates where $\epsilon(\tau) = 1$ and $s(\tau',\sigma) = \sigma + \tau' - \tau+ O( [\tau' - \tau]^2)$, and then $\delta = (s-s') /2$.} Eq.~(\ref{CS_EOM2}) states that ${\bf p}_+$ is slowly varying along left-moving characteristics (that is, the time scale of its variation is the FRW time $t$), and ${\bf p}_-$ is slowly varying along right-moving characteristics.
Thus we can approximate their product at nearby points by the local product where the two geodesics intersect,
\begin{equation}
{\bf p}_-(s,\tau) \cdot {\bf p}_+(s',\tau) = - \alpha(s',\tau - \delta) + O(s - s') \ .
\label{ppa}
\end{equation}
Then
\begin{eqnarray}
\partial_\tau ({\bf p}_+ (s,\tau) \cdot {\bf p}_+(s',\tau)) &=&
 \frac{\dot a}{a} \biggl( \alpha(s',\tau - \delta)  +  \alpha(s,\tau + \delta)     \nonumber\\
&& \hspace{-160pt}  -  \alpha (s,\tau)\,{\bf p}_+(s,\tau)\cdot{\bf p}_+(s',\tau)  -  \alpha(s',\tau)\,{\bf p}_+(s,\tau)\cdot{\bf p}_+(s',\tau) \biggr)  +  O(s - s') \ .
\end{eqnarray}
When we integrate over a scale of order of the Hubble time, the $\delta$ shifts in the arguments have a negligible effect $O(\delta)$ and so we ignore them.
Defining
\begin{equation}
h_{++}(s,s', \tau) = 1 - {\bf p}_+ (s,\tau) \cdot {\bf p}_+(s',\tau) \ ,
\end{equation}
we have
\begin{equation}
\partial_\tau h_{++}(s,s', \tau) =
 - \frac{\dot a}{a} h_{++}(s,s', \tau) [ \alpha(s',\tau ) +  \alpha(s,\tau) + O(s - s') ] \ .
\end{equation}
Thus
\begin{equation}
h_{++}(s,s', \tau_1) = h_{++}(s,s', \tau_0) \exp\Biggl\{ - \nu' \int_{\tau_0}^{\tau_1} \frac{d\tau}{\tau}
[ \alpha(s',\tau ) +  \alpha(s,\tau) + O(s - s') ]
\label{hppint}
 \Biggr\} \ .
\end{equation}

Averaging over an ensemble of segments, and integrating over many Hubble times (and therefore a rather large number of correlation times) the fluctuations in the exponent average out and we can replace $\alpha(s,\tau)$ with $\bar\alpha = 1 - 2\bar v^2$,
\begin{equation}
\langle h_{++}(s,s', \tau_1) \rangle \approx
\langle  h_{++}(s,s', \tau_0) \rangle (\tau_1/\tau_0)^{- 2 \nu' \bar\alpha} \ .
\label{avexp}
\end{equation}
Note that in contrast to previous equations the approximation here is less controlled.
We do not have a good means to estimate the error.
It depends on the correlation between the small scale and large scale structure (the latter determines the distribution of $\alpha$), and so would require an extension of our methods.
We do expect that the error is numerically small; note that if we were to consider instead $\langle \ln h_{++}(s,s', \tau) \rangle$ then the corresponding step would involve no approximation at all.
  
Averaging over a translationally invariant ensemble of solutions, we have
\begin{equation}
\langle h_{++}(\sigma - \sigma', \tau_1) \rangle \approx
(\tau_1/\tau_0)^{- 2 \nu' \bar\alpha}\langle h_{++}(\sigma - \sigma', \tau_0) \rangle \ .
\end{equation}
We have used the fact that to good approximation (once again in the sense of Eq.~(\ref{avexp})), $\epsilon$ is only a function of time, and so we can choose $\sigma = s - \int d\tau/\epsilon$ and $\sigma-\sigma' = s - s'$.
Equivalently,
\begin{equation}
\langle h_{++}(\sigma - \sigma', \tau) \rangle \approx \frac{ f(\sigma - \sigma') }{\tau^{2 \nu' \bar\alpha} } \ .
\label{soln}
\end{equation}

The ratio of the segment length to $d_H = (1+\nu') t$ is
\begin{equation}
\frac{l}{d_H} \propto \frac{a \epsilon (\sigma - \sigma')}{t} \propto \frac{\sigma - \sigma'} {\tau^{1 + 2 \nu'\bar v^2 }} \ .
\end{equation}
The logic of our earlier discussion is that we use simulations to determine the value of $h_{++}$ at $l/d_H$ somewhat less than one, and then evolve to smaller scales using the Nambu action.
That is,
\begin{equation}
h_{++}(\sigma - \sigma', \tau) = h_0 \ \mbox{when}\  \sigma - \sigma' = x_0 \tau^{1 + 2 \nu'\bar v^2 } \ ,
\label{match}
\end{equation}
for some constants $x_0$ and $h_0$.
We assume scaling behavior near horizon scale, so that $h_0$ is independent of time.
Using this as an initial condition for the solution~(\ref{soln}) gives
\begin{equation}
\langle h_{++}(\sigma - \sigma', \tau) \rangle
\approx h_0 \Biggl( \frac{\sigma - \sigma'}{x_0 \tau^{1 + 2 \nu'\bar v^2 }}\Biggr)^{2\chi}
\approx {\mathcal A} (l/t)^{2\chi}  \ ,
\label{hpp}
\end{equation}
where
\begin{equation}
\chi  =  \frac{\nu'\bar\alpha}{1 + 2 \nu'\bar v^2 }  =  \frac{\nu\bar\alpha}{1 - \nu\bar\alpha}  \ .
\end{equation}
In the last form of equation~(\ref{hpp}) we have expressed the correlator in terms of physical quantities, the segment length $l$ defined earlier and the FRW time $t$.\footnote{We have not yet needed to specify numerical normalizations for $\sigma$ and $\tau$, or equivalently for $\epsilon$ and $a$.  The value of $h_0$ depends on this choice, but the value of ${\mathcal A}$ does not.}

\begin{table}[t]
\begin{center}
 \begin{tabular}{| l || c | c | c | c |}
  \hline
	Cosmological epoch   & $\nu$ & $\bar v^2$ & $\bar\alpha$ & $\chi$ \\
	\hline
	\hline
	Radiation dominated  &  1/2  &    0.41    &     0.18     &  0.10  \\
	\hline
	Matter dominated     &  2/3  &    0.35    &     0.30     &  0.25  \\
	\hline
 \end{tabular}
\end{center}
\caption{Summary of relevant quantities for the evolution of small scale structure on cosmic strings in a scaling regime for both radiation and matter dominated eras.  All quantities are defined in the body of the text.  The RMS velocities $\bar v^2$ were taken from simulations by C.~Martins and E.~P.~S.~Shellard (2005).}
\label{values}
\end{table}

Eq.~(\ref{hpp}) is our main result.
Equivalently (and using $\sigma$ parity),
\begin{equation}
\langle {\bf p}_+ (\sigma,\tau) \cdot {\bf p}_+(\sigma',\tau) \rangle
= \langle {\bf p}_- (\sigma,\tau) \cdot {\bf p}_-(\sigma',\tau) \rangle
\approx 1 - {\mathcal A} (l/t)^{2\chi} \ .
\label{pppp}
\end{equation}
In the radiation era $\chi_r \sim 0.10$ and in the matter era $\chi_m \sim 0.25$.
The most relevant constants that determine the evolution of small scale structure on cosmic strings in a scaling regime are summarized in Table~\ref{values} for both the radiation- and matter-dominated eras.

There can be no short distance structure in the correlator ${\bf p}_+ \cdot {\bf p}_-$, because the left- and right-moving segments begin far separated, and the order $\dot a/a$ interaction between them is too small to produce significant non-smooth correlation.
Thus, from~(\ref{ppa}) we get
\begin{equation}
\langle {\bf p}_+ (\sigma,\tau) \cdot {\bf p}_-(\sigma',\tau) \rangle
= -\bar\alpha + O(\sigma - \sigma') \ .
\label{pppm}
\end{equation}
%

\subsection{Small fluctuation approximation\label{SMALLFLUC}}

Before interpreting these results, let us present the derivation in a slightly different way.
The exponent $\chi$ is positive, so for points close together the vectors $ {\bf p}_+ (\sigma,\tau)$ and ${\bf p}_+(\sigma',\tau) $ are nearly parallel.
Thus we can write the structure on a small segment as a large term that is constant along the segment and a small fluctuation:
\begin{equation}
{\bf p}_+ (\sigma,\tau) = {\bf P}_+ (\tau) + {\bf w}_+ (\tau,\sigma) - \frac{1}{2} {\bf P}_+ (\tau)
w_+^2  (\tau,\sigma) + \ldots \ ,
\label{smallf}
\end{equation}
with $P_+(\tau)^2 = 1$ and ${\bf P}_+ (\tau) \cdot {\bf w_+} (\tau,\sigma) = 0$.
Inserting this into the equation of motion~(\ref{CS_EOM2}) and expanding in powers of ${\bf w}_+$ gives
\begin{eqnarray}
\dot{\bf P}_+ &=& - \frac{\dot{a}}{a} \left[ {\bf P}_- - ({\bf P}_+ \cdot {\bf P}_-)\, {\bf P}_+ \right]\ ,
\\[6pt]
\dot{\bf w}_+ - \frac{1}{\epsilon} {\bf w}'_+ &=& \frac{\dot{a}}{a} \left[ ({\bf w}_+ \cdot {\bf P}_-)\, {\bf P}_+ + ({\bf P}_+ \cdot {\bf P}_-)\, {\bf w}_+ \right]   \nonumber\\
&=& - ({\bf w}_+ \cdot \dot{\bf P}_+)\, {\bf P}_+ +  \frac{\dot{a}}{a} ({\bf P}_+ \cdot {\bf P}_-)\, {\bf w}_+ \ .
\label{CS_EOM3}
\end{eqnarray}
Since the right-moving ${\bf p}_-$ is essentially constant during the period when it crosses the short left-moving segment, we have replaced it in the first line of (\ref{CS_EOM3}) with a $\sigma$-independent ${\bf P}_-$.
In the second line of (\ref{CS_EOM3}) we have used ${\bf w}_+ \cdot {\bf P}_+ = 0$.
In the final equation for ${\bf w}_+$, the first term is simply a precession: ${\bf P}_+$ rotates around an axis perpendicular to both ${\bf P}_+$ and ${\bf P}_-$, and this term implies an equal rotation of ${\bf w}_+$ so as to keep ${\bf w}_+$ perpendicular to ${\bf P}_+$.
Eq.~(\ref{CS_EOM3}) then implies that in a coordinate system that rotates with ${\bf P}_+$, ${\bf w}_+$ is simply proportional to $a^{-\alpha}$.

It follows that
\begin{equation}
\left\langle {\bf p}_+ (\sigma,\tau) \cdot {\bf p}_+(\sigma',\tau) \right\rangle - 1 = - \frac{1}{2} \left\langle [{\bf w}_+ (\sigma,\tau) - {\bf w}_+(\sigma',\tau)]^2 \right\rangle
\label{upup}
\end{equation}
scales as $a^{-2\bar\alpha}$ as found above (again we are approximating as in eq.~(\ref{avexp}), and again this statement would be exact if we instead took the average of the logarithm).
Similarly the four-point function of ${\bf w}_+$ scales as $a^{-4\bar\alpha}$.
We have not assumed that the field ${\bf w}_+$ is Gaussian; the $n$-point functions, just like the two-point function, can be matched to simulations near the horizon scale.
We can anticipate some degree of non-Gaussianity due to the kinked structure; this will be discussed further in Chapter~6.

\section{Discussion\label{DISCUSS}}

Now let us discuss our results for the two-point functions.
We can also write them as
\begin{eqnarray}
{\rm corr}_x(l,t) \equiv
\frac{ \left<{\bf x}'(\sigma,\tau)\cdot{\bf x}'(\sigma',\tau)\right> }{\left<{\bf x}'(\sigma,\tau)\cdot{\bf x}'(\sigma,\tau)\right>}  &\approx& 1 - \frac{{\mathcal A}}{2(1 - \bar v^2)} (l/t)^{2\chi} \ ,
\nonumber\\
{\rm corr}_t(l,t) \equiv
\frac{\left< \dot{\bf x}(\sigma,\tau)\cdot \dot{\bf x}(\sigma',\tau)\right>}
{\left< \dot{\bf x}(\sigma,\tau)\cdot \dot{\bf x}(\sigma,\tau)\right>}  & \approx&
1 - \frac{{\mathcal A}}{2 \bar v^2} (l/t)^{2\chi} \ .
\label{corrs}
\end{eqnarray}
These are determined up to two parameters $\bar v^2$ and ${\mathcal A}$ that must be obtained from simulations.
A first observation is that these expressions scale as they are functions only of the ratio of $l$ to the horizon scale.
This is simply a consequence of our assumptions that the horizon scale structure scales and that stretching is the only relevant effect at shorter scales.
We emphasize that these results are for segments on long strings; we will discuss loops in Chapter~4.

It is natural to characterize the distribution of long strings in terms of a fractal dimension.
The mean squared spatial distance between two points (also known as the extension) separated by coordinate distance $\sigma$ is
\begin{eqnarray}
\langle r^2(l,t) \rangle &=& \left<{\bf x}' \cdot{\bf x}' \right>  \int_0^l dl'  \int_0^l dl'' \,
{\rm corr}_x(l' - l'',t)    \nonumber\\
&\approx& (1 - \bar v^2) l^2 \left[ 1 - \frac{{\mathcal A} (l/t)^{2\chi}}{(2\chi+1)(2\chi+2)(1 - \bar v^2)} \right] \ .
\label{r2}
\end{eqnarray}
We can then define the fractal dimension $d_f$ (which is 1 for a straight line and 2 for a random walk),
\begin{equation}
d_f  =  \frac{2 \, d \ln l}{d\ln \langle r^2(l,t) \rangle}   
\approx  1 + \frac{{\mathcal A} \chi (l/t)^{2\chi}}{(2\chi+1)(2\chi+2)(1 - \bar v^2)} + O((l/t)^{4\chi})  \ .
\label{fracdim}
\end{equation}
The fractal dimension approaches 1 at small scales: the strings are rather smooth.
There is a nontrivial scaling property, not in the fractal dimension but rather in the deviation of the string from straightness,
\begin{equation}
1 - {\rm corr}_x(l,t) \propto (l/t)^{2\chi} \ , \quad
1 - {\rm corr}_t(l,t) \propto (l/t)^{2\chi} \ .
\label{crit}
\end{equation}
We define the scaling dimension $d_s = 2\chi$.
Note that $d_s$ is not large, roughly $0.2$ in the radiation era and $0.5$ in the matter era, so the approach to smoothness is rather slow.

One might also consider the fractal dimension of the ${\bf p}_\pm$ curves which live on the unit sphere.
To this end we consider the extension between two such vectors separated by an infinitesimal worldsheet coordinate $\delta\sigma$, i.e. we take ${\bf p}_+(\sigma)$ and ${\bf p}_+(\sigma+\delta\sigma) \simeq {\bf p}_+(\sigma) + \delta{\bf p}_+(\sigma)$.
Now using our result~(\ref{pppp}) we see that
\begin{equation}
\langle {\bf p}'_+(\sigma) \cdot {\bf p}'_+(\sigma + \delta\sigma) \rangle  \propto  \delta\sigma^{2\chi-2}  \ .
\end{equation}
Therefore, the extension is given by
\begin{equation}
\langle | \delta{\bf p}_+ | \rangle  =  \langle | {\bf p}'_+ | \rangle \delta\sigma  \propto  \delta\sigma^\chi  \ .
\end{equation}
It then follows that the fractal dimension of the ${\bf p}_\pm$ curves is $1/\chi$.
Note that this exponent is quite large, in contrast with the smoothness of the string configuration itself; it equals $10$ in the radiation dominated era and $4$ in the matter dominated era.

\begin{figure}
\center \includegraphics[width=20pc]{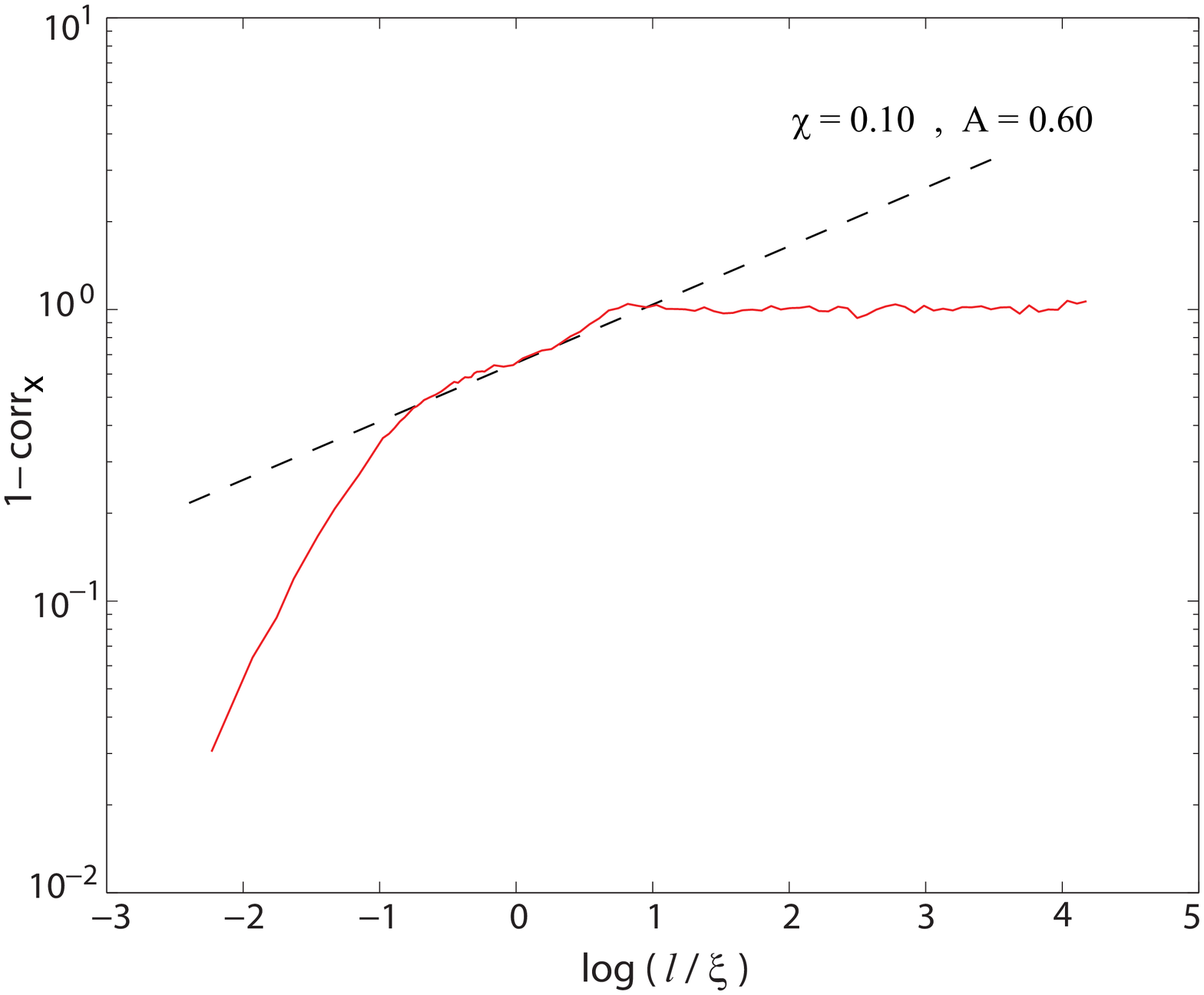}
\caption{Comparison of the model (dashed line) with the data provided by C.~Martins and E.~P.~S.~Shellard (solid red line) in the radiation-dominated era, for which the correlation length is $\xi \simeq 0.30 t$.}
\label{mvd_rad}
\end{figure}

Our general conclusions are in agreement with the simulations of Ref.~\cite{Martins:2005es}, in that the fractal dimension approaches 1 at short distance.
To make a more detailed comparison it is useful to consider a log-log plot of $1 - {\rm corr}_x(l,t)$ versus $l$, as suggested by the scaling behavior~(\ref{crit}).\footnote{We thank C. Martins for replotting the results of Ref.~\cite{Martins:2005es} in this form.}
The comparison is interesting.
At scales larger than $d_H$ the correlation goes to zero.\footnote{In terms of the correlation length $\xi$ which is used in reference~\cite{Martins:2005es} the horizon scale is $\sim 6.7 \xi$ in the radiation era and $\sim 4.3 \xi$ in the matter era.}
Rather abruptly below $d_H$ the slope changes and agrees reasonably well with our result.
It is surprising to find agreement at such long scales where our approximations do not seem very precise.
On the other hand, at shorter scales where our result should become more accurate, the model and the simulations diverge; this is especially clear at the shortest scales in the radiation-dominated era (Fig.~\ref{mvd_rad}).
Note that at these smaller scales the simulations seem to indicate a larger exponent $\chi\sim 0.5$, which corresponds to the functions ${\bf p}_\pm$ mapping out a random walk on the unit sphere.

\begin{figure}
\center \includegraphics[width=20pc]{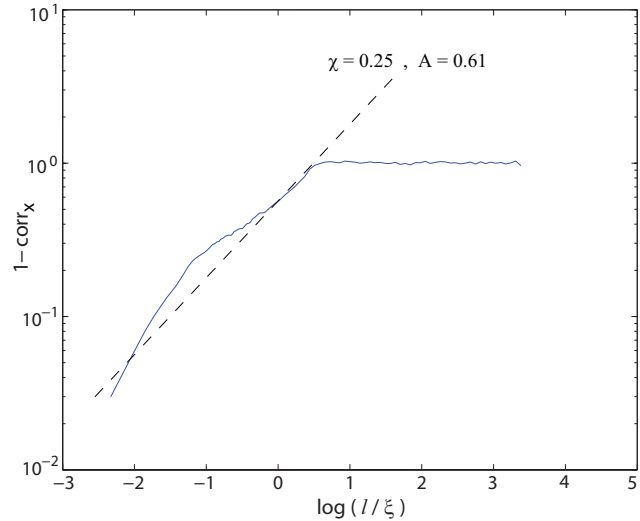}
\caption{Comparison of the model (dashed line) with the data provided by C.~Martins and E.~P.~S.~Shellard (solid blue line) in the matter-dominated era, for which the correlation length is $\xi \simeq 0.69t$.}
\label{mvd_mat}
\end{figure}

One possible explanation for the discrepancy would be transient behavior in the simulations.
We have argued that the structure on the string is formed at the horizon scale and `propagates' to smaller scales (in horizon units) as the universe expands.
In Ref.~\cite{Martins:2005es} the horizon size increases by a factor of order 3, and so even if the horizon-scale structure forms essentially at once, the maximum length scale over which it can have propagated is $3^{1 - \zeta}$, less than half an order of magnitude.
At smaller scales, the small scale structure seen numerically would be almost entirely determined by the initial conditions.
On the other hand, the authors of Ref.~\cite{Martins:2005es} (private communication) argue that their result appears to be an attractor, independent of the initial conditions, and that loop production may be the dominant effect.
Motivated by this we have examined loop production and these studies are presented in Chapter~4.
Indeed, we find that this is in some ways a large perturbation, and it is conceivable that the chopping-off of small loops caused by self-intersecting long strings could significantly smoothen the latter.
However, it still remains an open question whether this mechanism can resolve the discrepancy seen in the short distance structure of the two-point function.

Thus far we have discussed ${\rm corr}_x$.
Our result~(\ref{corrs}) implies a linear relation between ${\rm corr}_x$ and ${\rm corr}_t$.
In fact, this holds more generally from the argument that there is no short-distance correlation between $ {\bf p}_+$ and ${\bf p}_-$, Eq.~(\ref{pppm}):
\begin{eqnarray}
(1-\bar v^2 )  {\rm corr}_x(l,t) - \bar v^2  {\rm corr}_t(l,t) &=&     \nonumber \\
&& \hspace{-150pt} =  -\frac{1}{2} \Bigl\langle {\bf p}_+ (\sigma,\tau) \cdot {\bf p}_-(\sigma',\tau) +  {\bf p}_- (\sigma,\tau) \cdot {\bf p}_+(\sigma',\tau) \Bigr\rangle   \to   \bar\alpha \ .
\end{eqnarray}
Inspection of Fig.~2 of Ref.~\cite{Martins:2005es} indicates that this relation holds rather well at all scales below $\xi$.

The small scale structure on strings is sometimes parameterized in terms of an effective mass per unit length~\cite{Vilenkin:1990mz,Carter:1990nb}.
The basic idea is to consider a coarse-grained description of a cosmic string at some scale.
Any wiggles on it with wavelengths much smaller than this scale will appear as smooth at the cost of introducing an effective mass per unit length $\mu_{\rm eff}$ (and an effective tension as well).
For a segment of length $l$ the effective mass per unit length is given by
\begin{eqnarray}
\frac{ \mu_{\rm eff} }{ \mu }  =  \frac{\sqrt{1 - \overline{v}^2} l}{\langle r(l) \rangle } 
\approx 1 + \frac{{\mathcal A}}{2(1 - \overline{v}^2)(2\chi+1)(2\chi+2)}\left( \frac{l}{t} \right)^{2\chi} \ ,
\end{eqnarray}
where we have made use of result~(\ref{r2}).
Note that this is strongly dependent on the scale $l$ of the coarse-graining.

In conclusion, let us emphasize the usefulness of the log-log plot of $1 - {\rm corr}_x$.
In a plot of the fractal dimension, all the curves would approach one at short distance, though at slightly different rates.
The difference is much more evident in Figs.~\ref{mvd_rad} and~\ref{mvd_mat}, and gives a clear indication either of transient effects or of some physics omitted from the model.

\chapter{Gravitational radiation from cosmic strings}

The ``vanilla'' cosmic strings we consider interact only gravitationally.
A straight cosmic string leaves the surrounding space locally flat, merely introducing a deficit angle (giving rise to a conical space, globally)~\cite{Vilenkin:1981zs}. 
This is of course not the case once we allow for oscillating strings.
In particular, gravitational radiation is emitted when waves traveling in opposite directions along the string collide and this effect is proportional to the dimensionless string tension $G\mu$.
The resulting radiation depends on the power spectrum of the perturbations on the string~\cite{Siemens:2002dj} and so we can use the results obtained in Chapter~2 to address this matter.
This is precisely what we present in this chapter.

We begin by reviewing in Section~\ref{BACKREACTOLD} the computation of the power emitted in gravitational radiation from colliding small perturbations on long strings and the traditional picture of the gravitational back-reaction cutoff on the scales of fluctuations.
In Section~\ref{BACKREACTIMPROV} this calculation is improved by incorporating the fact that colliding modes on a string only radiate efficiently if their wavelengths have the same magnitude and by employing the spectrum of perturbations found in Chapter~2.
This gravitational back-reaction introduces a new scale in the network, below which the strings become smoother.
Therefore, the two-point function obtained in the previous chapter must be corrected at small scales and this is done in Section~\ref{2PFREVISIT}.
For later reference, we briefly discuss the interplay between small scale structure and cusp formation in Section~\ref{SSSandCUSPS}.

This chapter is mainly based on Ref.~\cite{Polchinski:2007rg}, with Joseph Polchinski.

\section{Radiation from long strings and back-re\-ac\-tion\label{BACKREACTOLD}}

As we mentioned in the introductory chapter, it is generally believed that, after their formation, cosmic string networks evolve into a scaling regime where all length scales grow linearly with the Hubble time.
This would imply that the typical size of cosmic string loops is a fixed fraction $\alpha$ of the horizon scale.\footnote{There is not complete agreement even on this point: see Ref.~\cite{Bevis:2006mj} and references therein.}
However, there is no consensus on the value of the proportionality constant $\alpha$; in fact, estimates range over tens of orders of magnitude.

A key question is whether the straightening of the string by gravitational radiation is necessary for scaling of the loop sizes.
If so, $\alpha$ will depend on the dimensionless string tension $G\mu$.
If not, then the evolution of the network will be purely geometric, and $\alpha$ could be a pure number, independent of $G\mu$.
In Chapter~5 we will demonstrate that gravitational radiation effects are not needed to obtain scaling of the loop length distribution.
On the other hand, in Chapter~4 we will find that employing the simple stretching model down to arbitrary scales results in a UV divergence in the loop production function.
Since smoothing by gravitational radiation is after all a real effect, it actually determines the cutoff and so we argue that $\alpha$ does in fact depend on $G\mu$.

For many years, simulations appeared to show loops forming at the short distance cutoff scale, and this was interpreted as implying the need to include gravitational radiation.
The necessity to consider it in order to achieve scaling of small scale structure was also suggested by the detailed study of Ref.~\cite{Austin:1993rg}.
However, the conclusions of~\cite{Allen:1990mp,Martins:2000cs} were opposite: those authors found that stretching by the expansion of the Universe alone is sufficient to guaranty scaling of the small scale structure.
Indeed, several recent simulations, indicate that the loops are actually forming above the cutoff scale.
Refs.~\cite{Vanchurin:2005pa,Olum:2006ix} find notably large loops, $\alpha \sim 0.1$, but this is superposed on a power law distribution that grows toward smaller scales.
Refs.~\cite{Ringeval:2005kr,Martins:2005es} find loops a few orders of magnitude smaller, but the distributions are still evolving in a non-scaling fashion.
None of these simulations include gravitational radiation directly.

When a perturbation on a string encounters another one traveling in the opposite direction, gravitational waves are generated.
The energy radiated away is stolen from the colliding fluctuations and as a result the string straightens.
However, this smoothing of the long strings by gravitational radiation is the dominant effect only below some critical length scale, usually called the gravitational radiation scale.
This (time-dependent) scale then acts as a lower cutoff on the sizes of loops produced at any given time.
It was long assumed that gravitational radiation smoothed the long strings on scales of order the horizon length times $G\mu$, so this would be the size of loops if they formed at the gravitational radiation cutoff~\cite{Bennett:1989ak,Quashnock:1990wv}.
However, gravitational radiation is an efficient energy loss channel only when the perturbations have comparable wavelengths~\cite{Siemens:2001dx}.
Indeed, those authors showed that the radiation from long strings had been overestimated, and so it becomes important only at a shorter length scale, proportional to a larger power of $G\mu$~\cite{Siemens:2002dj}; the exponent depends on the power spectrum of fluctuations on the long string.

The model developed in Chapter~2, which determined the form~(\ref{pppp}) for the two-point function at short distances in a scaling regime, should then provide some answer for the gravitational radiation scale.
In the small fluctuation approximation, the short distance structure can be equivalently expressed in terms of the fluctuations ${\bf w}_\pm$ as
\begin{equation}
1 - \langle {\bf p}_\pm (\sigma,t) \cdot {\bf p}_\pm (\sigma',t) \rangle
= \frac{1}{2} \langle \left[ {\bf w}_\pm (\sigma,t) - {\bf w}_\pm (\sigma',t) \right]^2 \rangle  + O(w^4_\pm)
\approx  {\mathcal A} (l/t)^{2\chi} \ .
\label{ww}
\end{equation}
The model also predicts that the exponent $\chi$ which controls the deviation of the string from straightness is determined by the RMS velocity $\bar{v}^2$ of the long string population.
We have noted that at small scales the simulations of~\cite{Martins:2005es} disagree from our model and seem to approach a larger value for the exponent, namely $\chi\sim 0.5$, corresponding to the functions ${\bf p}_\pm$ mapping out a random walk on the unit sphere.
In any event, our discussion below just uses the general power law form~(\ref{pppp}), and one can insert any assumed values for $\chi$ and ${\mathcal A}$.

Gravitational radiation is important on scales short compared to the Hubble time and so we can use the flat metric $g_{\mu\nu} = {\rm diag}(-1,+1,+1,+1)$, and also make the gauge choice $a\epsilon = 1$.\footnote{This choice is possible when we consider scales small compared to the Hubble time.  In Section~\ref{2PFREVISIT}, when we again consider cosmological evolution, we must reintroduce $a(t)$ and $\epsilon(t)$.}
The modes are then functions only of $u = t + \sigma$ or $v = t - \sigma$.
This is not surprising, given that the string equations of motion reduce to a simple wave equation in the coordinates $t$ and $\sigma$ for a flat spacetime.
The solution is a superposition of left- and a right-moving waves,
\begin{equation}
x^\mu(u,v) = \frac{1}{2} [ a^\mu(u) + b^\mu(v) ]  \ .
\end{equation}
Hence, we may write
\begin{eqnarray}
\langle {\bf p}_+ (u) \cdot {\bf p}_+ (u') \rangle  &\approx&  1 - {\mathcal A} [(u-u')/t]^{2\chi}  \ ,  \nonumber \\
\langle {\bf p}_- (v) \cdot {\bf p}_- (v') \rangle  &\approx&  1 - {\mathcal A} [(v-v')/t]^{2\chi}  \ ;
\end{eqnarray}
(the explicit $t$ on the right-hand side is effectively a fixed parameter, varying only over the Hubble time).

The power radiated in gravitational waves by an infinitely long source was first considered in~\cite{Sakellariadou:1990ne}.
By employing a weak-field approximation, the gravitational field can be easily related to the energy-momentum tensor of the string and then the energy flux through a large radius cylinder surrounding it gives the power radiated, proportional to $G\mu^2$.
A general formula for the gravitational radiation was obtained in~\cite{Hindmarsh:1990xi} under the assumption that the colliding waves traveling along the string have small amplitudes.
Essentially, gravitational radiation is only emitted when left- and right-moving perturbations interact.
This is manifest in the expression for the energy radiated per unit solid angle in the $\bf k$ direction, given below in the notation of~\cite{Siemens:2001dx},
\begin{equation}
\frac{d\Delta E}{d\Omega} = \frac{G\mu^2}{16\pi^2} \int_0^{\infty} d\omega \, \omega^2
\left\{ |A|^2 |B|^2 + |A^* \cdot B|^2 - |A \cdot B|^2 \right\}  \ ,
\label{energy}
\end{equation}
with the left- and right-moving contributions given respectively by
\begin{eqnarray}
A^\mu(k) &=& \int_{-\infty}^{+\infty} du \, p_+^\mu(u) 
\exp\left\{ { -\frac{i}{2} \int_0^u du' \, k \cdot p_+(u') } \right\} \ , \nonumber\\
B^\mu(k) &=& \int_{-\infty}^{+\infty} dv \, p_-^\mu(v) 
\exp\left\{  -\frac{i}{2} \int_0^v dv' \, k \cdot p_-(v')  \right\}  \ .
\label{AB}
\end{eqnarray}
Here, $k^\mu$ is the gravitational radiation wave vector, and $p^\mu_\pm$ are null 4-vectors whose time component is identically 1 and with ${\bf p}_\pm$ being the unit vectors above.\footnote{These are related to the standard notation via $p^\mu_+(u) = a^{\mu\prime}(u)$ and $ p^\mu_-(v) = b^{\mu\prime}(v)$.}
In the spirit of the small fluctuation approximation we orient a given segment of string mainly along the $z$-axis and consider fluctuations ${\bf w}_\pm$ in the perpendicular $xy$-plane.
Thus,
\begin{eqnarray}
p^\mu_+(u)  &\simeq&  \left( 1, {\bf w}_+(u), \sqrt{1-w_+^2(u)} \right)  \ ,  \nonumber\\
p^\mu_-(v)  &\simeq&  \left( 1, {\bf w}_-(v), - \sqrt{1-w_-^2(v)} \right)  \ .
\label{pw}
\end{eqnarray}
We can ignore the last two terms in equation~(\ref{energy}): they are equal when $B^\mu$ is real, and so cancel when ensemble-averaged over the short distance structure.

Let us first review the argument of Refs.~\cite{Siemens:2001dx,Siemens:2002dj}.
First linearize the modes~(\ref{AB}) in the oscillations, so the exponential factors become $e^{i k_a u}$ and 
$e^{i k_b v}$ respectively, with 
\begin{equation}
\omega = k_a + k_b \ , \quad k_z = k_b - k_a  \ .
\label{conservation}
\end{equation}
Here $\omega$ and $k_z$ are the frequency and wavenumber of the radiation.
Then the ensemble averages in the linearized approximation ($\approx$) are
\begin{eqnarray}
\langle |A(k)|^2 \rangle &\approx& l_u \int_{-\infty}^{+\infty} du \, e^{i k_a u - \varepsilon |u|}
\left( \langle {\bf p}_+ (u) \cdot {\bf p}_+ (0) \rangle - 1 \right) =  l_u {\mathcal A} \, c_\chi \,  k_a^{-(1+2\chi)} t^{-2\chi} \ ,  \nonumber\\
\langle |B(k)|^2 \rangle &\approx&  l_v {\mathcal A} \, c_\chi \, k_b^{-(1+2\chi)} t^{-2\chi}  \ .
\label{AandB}
\end{eqnarray}
Here $c_\chi =  2 \sin (\pi\chi) \, \Gamma(1+2\chi)$, taking the values 1.25 and 0.57 in the matter- and radiation-dominated eras respectively.
The convergence factor $e^{-\varepsilon |u|}$ accounts for the decay of the correlations on horizon scales.
Its detailed form is unimportant: it only affects the Fourier transform for $k_a$ of order the inverse horizon size, while for gravitational radiation the relevant $k_a$ is much larger.
The factors $l_u$ and $l_v$ are volume regulators: we artificially cut the oscillations off to produce finite trains, but these regulators of course drop out when we consider rates per unit time and length.
Using
$
d\omega \, d\cos\theta = 2 {dk_a dk_b}/({k_a + k_b}) ,
$
we can write the total energy radiated as
\begin{eqnarray}
\langle \Delta E \rangle &=& \frac{G\mu^2}{4 \pi} \int_0^\infty\!\!  dk_a \int_0^\infty\!\! dk_b \, (k_a + k_b)
\langle |A(k_a)|^2 \rangle  \langle |B(k_b)|^2 \rangle   \nonumber\\
&\approx&
 \frac{G\mu^2  {\mathcal A}^2 c_\chi^2}{ 4\pi } l_u l_v \int_0^\infty\!\! dk_a \int_0^\infty\!\! dk_b \, \frac{k_a + k_b}
{ (k_a k_b)^{1+2\chi} t^{ 4\chi} }  \ .
\label{power2}
\end{eqnarray}
The volume of the world-sheet is $V = \frac{1}{2}  l_u l_v$, so this translates into a power per unit length
\begin{equation}
\left\langle \frac{dP}{dz} \right\rangle \approx
 \frac{G\mu^2  {\mathcal A}^2 c_\chi^2}{ 2\pi } \int_0^\infty\!\! dk_a \int_0^\infty\!\! dk_b \, \frac{k_a + k_b}
{ (k_a k_b)^{1+2\chi} t^{ 4\chi} }  \ .
\label{power3}
\end{equation}
The total energy of the wavetrains is~\cite{Hindmarsh:1990xi}
\begin{equation}
E \approx {\mu} \int_0^\infty \frac{dk_a}{2\pi}\,|A(k_a)|^2  +{\mu} \int_0^\infty \frac{dk_b}{2\pi}\,|B (k_b)|^2  \ .
\end{equation}
Isolating a momentum range $dk_a$, we have
\begin{equation}
\Delta  |A(k_a)|^2 \approx - \frac{G\mu k_a}{2}  |A(k_a)|^2 \int_0^{\infty} dk_b \, 
\langle |B(k_b)|^2 \rangle  \ .
\label{Adecay}
\end{equation}
We have used the fact that momentum conservation determines that the energy coming from the $A$ and $B$ modes is in the proportion $k_a$ to $k_b$.
A given point $u$ interacts with the $v$ train for a time $\frac{1}{2} l_v$, so the rate of decay becomes
\begin{equation}
\frac{d}{dt}  |A(k_a)|^2 \approx
-{G\mu}  k_a  |A(k_a)|^2 {  {\mathcal A} c_\chi} \int_0^\infty
 \frac{ dk_b} {k_b^{1+2\chi} t^{ 2\chi} }  \ .
\label{decay1}
\end{equation}
The integral is dominated by the lower limit: most of the energy loss from the left-moving mode $k_a$ comes from its interaction with right-moving modes of much longer wavelength.
Cutting off the lower end of the $k_b$ integral at the horizon scale, $k_b \sim 1/t$, we find that the decay rate is of order $G\mu k_a$.
This is faster than the Hubble time for
\begin{equation}
k_a > O(G\mu t)^{-1} \ ,
\label{naive}
\end{equation}
meaning that modes with wavelengths $\lambda < O(G\mu t)$ are exponentially suppressed.
Thus, Eq.~(\ref{naive}) reproduces the usual (naive) estimate for the gravitational damping length, namely that fluctuations on scales smaller than $\sim G\mu t$ quickly decay due to gravitational radiation.

\section{Radiation and back-reaction improved\label{BACKREACTIMPROV}}

We have already mentioned that including the fact that a perturbation on a string does not interact with the same efficiency with all other modes leads to a shorter gravitational radiation length scale~\cite{Siemens:2001dx,Siemens:2002dj}.
Here we rederive these results in perhaps a simpler way.
Our result for the exponent of $G\mu$ differs somewhat from Ref.~\cite{Siemens:2002dj}.
This difference arises in going from the periodic wave train calculation of Ref.~\cite{Siemens:2001dx} to the random distribution on the actual network.
Also, we use input from the model developed in Chapter~2 and from simulations~\cite{Martins:2005es} to determine the actual power spectrum on the long string.

The calculation we carried out in the previous section implies that the energy loss from the left-moving modes $k_a$ comes primarily from their interaction with much longer right-moving modes at the horizon scale.
Ref.~\cite{Siemens:2001dx} argued that this was paradoxical, because a wave that encounters an oncoming perturbation with much longer wavelength is essentially traveling on a straight string, and such a wave will not radiate~\cite{Garfinkle:1990jq}.
They showed that this paradox arose due to the neglect of the exponential terms in the modes~(\ref{AB}).
With their inclusion, the radiation is exponentially suppressed when the ratio of $k_a$ to $k_b$ becomes sufficiently large.

Ref.~\cite{Siemens:2001dx} considered monochromatic wavetrains 
\begin{equation}
w_+^x(u) = \epsilon_a \cos (k_a u)\ , \quad w_-^x(v) = \epsilon_b \cos (k_b v) \ ,
\end{equation}
and found that keeping the full form~(\ref{AB}) of the amplitude gives a large suppression when
\begin{equation}
k_b/ \epsilon_b^2 \lsim  k_a\ \ {\rm or}\ \ k_b \gsim k_a/\epsilon_a^2 \ .
\label{range}
\end{equation}
The first of these relations will cut off the integral~(\ref{decay1}) below the horizon scale, but first we need to extend it to the incoherent spectrum on a long cosmic string.
We will do this in a systematic way below, but we can anticipate the answer.
For a continuous spectrum, of course, a single frequency makes a contribution of measure zero.
Indeed, the units are wrong: the Fourier transform of the fluctuation $ {\bf w}_-(v) $ obeys
\begin{equation}
\langle \tilde{\bf w}_-(k) \cdot  \tilde{\bf w}_-(k')^* \rangle = 2\pi \delta(k - k') \epsilon^2(k) \ ,
\end{equation}
where $\epsilon^2(k) \propto k^{-(1+2\chi)}t^{-2\chi}$ has units of length, whereas in eq.~(\ref{range}) the amplitudes must be dimensionless.
Thus, in going to the continuum, we must replace 
\begin{equation}
\epsilon_b^2 \to k_b \epsilon^2(k_b) \approx (k_b t)^{-2\chi} \ .
\end{equation}
The amplitude is then suppressed unless 
\begin{equation}
(k_b t)^{1 + 2\chi} \gsim k_a t \ .
\label{socut}
\end{equation}

Using this lower cutoff in the decay rate~(\ref{decay1}) gives
\begin{equation}
\frac{d}{dt}  |A(k_a)|^2 \sim
- {G\mu} (k_a t)^{1/(1+2\chi)} t^{-1}  |A(k_a)|^2 \equiv - \frac{|A(k_a)|^2}{\tau_{\rm GR}(k_a)} \ .
\end{equation}
This is faster than the Hubble rate for
\begin{equation}
k_a \gsim (G\mu)^{-(1+2\chi)} t^{-1} \ .
\label{true}
\end{equation}
The gravitational length scale is reduced from the naive $O(G\mu t)$ by a factor $(G\mu)^{2\chi}$.
The qualitative conclusion is the same as in Ref.~\cite{Siemens:2002dj}, but the exponents do not seem to agree.
The result there was $k_a > (G\mu)^{-(1 + 2\beta)/2} t^{-1}$.
It is not clear whether $\beta$ is to be identified with $\chi+ \frac{1}{2}$ (as suggested by Eq.~(21) of Ref.~\cite{Siemens:2002dj}, compared with Eq.~(\ref {decay1}) above) or with $\chi$ (as suggested by \cite{Siemens:2002dj} Eq.~(31)), but in either case the exponents differ.
This stems from a different method of converting from the single-mode result to the continuous spectrum; our exponent will be borne out by the more formal treatment below.  

Finally, our analytic model of Chapter~2 gives for the exponent $1+2\chi$ the values 1.2 in the radiation era and 1.5 in the matter era;  using instead the simulations~\cite{Martins:2005es} would yield an exponent around 2.0 at the shortest scales in both eras.\footnote{Numerically, there are two changes from Ref.~\cite{Siemens:2002dj}: the corrected expression for the exponent, and a more accurate estimate of the power spectrum.  Ref.~\cite{Siemens:2002dj} effectively estimated the latter using $\bar v^2 = 0$, so that their Eq.~(31) is $\beta = \chi|_{\bar v^2 = 0} = \nu/(1-\nu).$}

The necessary correction to the naive radiation formula comes from the previously neglected exponential factor in $B^\mu(k)$, Eq.~(\ref{AB})~\cite{Siemens:2001dx}.
Expanding the exponential to second order in the fluctuations gives
\begin{eqnarray}
|B(k) |^2 &=& \int_{-\infty}^{+\infty} dv \int_{-\infty}^{+\infty} dv' \,( {\bf p}_-(v) \cdot  {\bf p}_-(v') - 1 )
\nonumber\\
&& \hspace{-50pt} \exp\left\{  i k_b (v'-v)  - \frac{i {\bf k}_\perp}{2} \cdot \int_v^{v'} dv'' \,
  {\bf w}_-(v'') - \frac{i k_z}{4} \int_v^{v'} dv''\, w_-^2(v'')  \right\}  \ .
\label{Bexp}
\end{eqnarray}
Noting that $k_\perp^2 = 4k_a k_b$, when the first term in the exponent is of order one, the second and third terms in the exponent are respectively of order $(k_a t)^{1/2} (k_b t)^{-1/2- \chi}$ and $k_a t (k_b t)^{-1-2\chi}$.
Thus, as $k_b$ decreases they become important at precisely the scale~(\ref{socut}) where the small-amplitude calculation is expected to break down.\footnote{The terms compete at this scale because the small fluctuation ${\bf w}_-(k_b^{-1})$ and the small number $(k_b/k_a)^{1/2}$ become comparable.  Further terms in the exponent, which have been dropped, are of higher order in the small parameter. This is in keeping with Ref.~\cite{Siemens:2001dx}, which also found that only terms up to quadratic order in the exponent were relevant.}
The linearized form for $A(k)$ remains valid, so we can simply insert the corrected form~(\ref{Bexp}) into the decay rate~(\ref{Adecay}).

By completing the square, the exponent in Eq.~(\ref{Bexp}) can be written as
\begin{eqnarray}
&& i \frac{(v'-v)}{4 {k_a} }  \left( {{\bf k}_\perp} -  \frac{k_a}{\sqrt{ v'-v } } \int_v^{v'} dv'' \, {\bf w}_-(v'') \right)^2
\nonumber \\
&& \qquad \qquad - \frac{i k_z}{8 (v' - v)}  \int_v^{v'} dv'' \int_v^{v'} dv'''\, [{\bf w_-}(v'') - {\bf w_-}(v''')]^2  \ .
\end{eqnarray}
Converting $\int dk_b \to (4 k_a)^{-1} \int d k_\perp^2 \to (4\pi k_a)^{-1} \int d^2 k_\perp$, we have at fixed $k_a$ the integral
\begin{eqnarray}
\int_0^\infty dk_b \, |B(k) |^2 &=& \int_{-\infty}^{+\infty} dv \int_{-\infty}^{+\infty} dv' \,
\frac{i ( {\bf p}_-(v) \cdot  {\bf p}_-(v') - 1 )}{v'-v}
\nonumber\\
&& \hspace{-50pt}
\exp\left\{ - \frac{i k_z}{8 (v' - v)} \int_v^{v'} dv'' \int_v^{v'} dv'''\, [{\bf w_-}(v'') - {\bf w_-}(v''')]^2  \right\}  \ .
\label{Bexp2}
\end{eqnarray}
We have used $-k_z \approx k_a \gg k_b$, since this is the regime where the correction is important.

We now approximate, replacing both the exponent and the prefactor with their mean values from equation~(\ref{ww}).
Then
\begin{eqnarray}
\int_0^\infty dk_b \, \langle |B(k)|^2 \rangle &=& 2 {\mathcal A} l_v\,  {\rm Im} \int_{-\infty}^{+\infty} dv \,
\frac{v^{2\chi-1}}{ t^{2\chi}}
\nonumber\\
&& \qquad \exp\left\{ \frac{i k_a {\mathcal A}}{4 v} \int_0^{v} dv'' \int_0^{v} dv'''\,\frac{ |v''-v'''|^{2\chi}}{t^{2\chi}} \right\}
\nonumber\\
&=&  2 l_v  b_\chi  ( k_a t )^{-2\chi/(1+2\chi)}  \ .
\label{Bexp3}
\end{eqnarray}
Here
\begin{equation}
b_\chi =  2 \sin\left( \frac{\pi\chi}{1+2\chi} \right) \Gamma\left( \frac{ 2\chi }{1 + 2\chi} \right) [4 (1+\chi)]^{2\chi/(1 + 2\chi)} \left[ \frac{\mathcal A} {1+2\chi}\right]^{1/(1 + 2\chi)} 
\end{equation}
is 2.5 in the matter era and 2.1 in the radiation era.
We can improve this approximation if we make the assumption that the ensemble is Gaussian.
The first order correction, keeping contractions between the prefactor and the exponent, involves a straightforward calculation; its inclusion simply renormalizes the constant $b_\chi \to (1+ \epsilon_\chi)b_\chi$, with
\begin{equation}
\epsilon_\chi = \chi (1+\chi) \left[ \frac{\Gamma(1+2\chi)^2}{\Gamma(2+4\chi)} - \frac{1}{2(1+4\chi)} \right] \ .
\end{equation}
However, this represents a small correction since $\epsilon_\chi = 0.035$ in the radiation epoch and $\epsilon_\chi = 0.045$ in the matter epoch.
Thus, to good approximation,
\begin{equation}
\frac{1}{\tau_{\rm GR}(k_a)} = b_\chi G\mu ( k_a t )^{1/(1+2\chi)} t^{-1} \ .
\label{tauGR}
\end{equation}
This is faster than the Hubble rate for the scales~(\ref{true}), as deduced earlier.

\section{The long-string two-point function revisited\label{2PFREVISIT}}

We can now improve our earlier result for the two-point function found in Chapter~2 through the inclusion of gravitational radiation.
Define
\begin{equation}
H(\kappa,t) = a(t) \epsilon(t) \int_{-\infty}^{\infty} d\sigma\, e^{-i \kappa\sigma}  \langle {\bf w}_+(\sigma, t)
\cdot {\bf w}_+(0, t) \rangle \ ,
\label{Hkappa}
\end{equation}
where $\sigma$ is related to the actual length along the string by $l = a(t) \epsilon(t) \sigma$.
Recall that $\epsilon(t) = a(t)^{-2\bar{v}^2}$.
Combining the stretching given by Eq.~(\ref{CS_EOM3}) with the gravitational radiation found above gives
\begin{equation}
\frac{d}{dt} H(\kappa,t) = - \frac{1}{a} \frac{da}{dt} \bar\alpha H(\kappa,t) - \frac{1}{\tau_{\rm GR}(\kappa/a \epsilon)} H(\kappa,t) \ .
\end{equation}
The derivative of the prefactor $a(t)\epsilon(t)$ would give the first term with a plus sign but when the derivative acts on the correlator of the fluctuations the result is twice as large and negative.
This equation is readily integrated, to give
\begin{equation}
\ln H(\kappa,t) = - \bar\alpha \ln a(t) + \ln f(\kappa) - b_\chi (1+\chi)(1+2\chi) G\mu \kappa^{1/(2\chi+1)} t^{1/(1+\chi)(1+2\chi) } \ .
\end{equation}
The function $f(\kappa)$ is obtained by matching onto the known result~(\ref{ww}) at early times, giving $f(\kappa) = {\mathcal A} c_\chi \kappa^{-1-2\chi}$.
Expressing the result in terms of the physical momentum scale $k = \kappa /a(t)\epsilon(t)$ gives
\begin{equation}
J(k,t) = H(\kappa,t) =  {\mathcal A} c_\chi k^{-1} (kt)^{-2\chi} \exp\left[ -b_\chi (1+\chi)(1+2\chi) G\mu (kt)^{1/(1+2\chi)} \right] \ .
\label{2pf}
\end{equation}

\begin{figure}
\center \includegraphics[width=25pc]{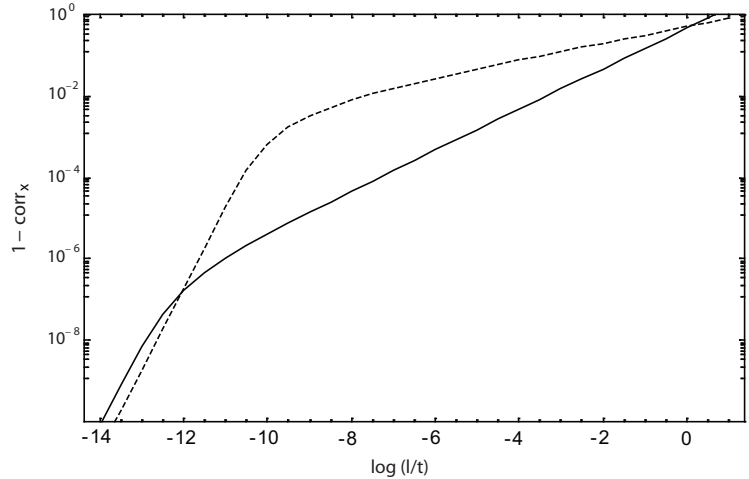}
\caption{The effect of the inclusion of gravitational radiation on the long-string two-point function is to smooth the string on scales below $\sim 20(G \mu)^{1+2\chi}t$.  The solid line corresponds to the matter epoch and the dotted line to the radiation epoch.  The value $G \mu = 10^{-9}$ has been adopted for illustrative purposes.}
\label{2pf plot}
\end{figure}

Thus, unless we consider modes with such short wavelength that the exponential suppression factor kicks in, we have $J(k) \sim k^{-(1+2\chi)}$.
At length scales smaller than $\sim (G \mu)^{1+2\chi} t$ this behavior is altered as a result of the inclusion of gravitational radiation, in such a way to make the two-point function smoother.
In Fig.~\ref{2pf plot} we show the numerical Fourier transform of $J(k,t)$, in terms of the correlation~\cite{Martins:2005es, Polchinski:2006ee}
\begin{equation}
1 - {\rm corr}_x  \approx  \frac{1 - \langle {\bf p}_+(l,t) \cdot {\bf p}_+(0,t) \rangle}{2(1-\overline{v}^2)}  =  \frac{ \langle w_+(l,t)^2 \rangle - \langle {\bf w}_+(l,t) \cdot {\bf w}_+(0,t) \rangle}{2(1-\overline{v}^2)} \ .
\end{equation}

The smoothing at short distance seen in Fig.~\ref{2pf plot} resembles that found in the numerical simulations of Ref.~\cite{Martins:2005es}, but here it is due to an additional physical effect, gravitational radiation, whereas in Ref.~\cite{Martins:2005es} such smoothing was found even without gravitational radiation.\footnote{As an aside, the correlator in Fig.~\ref{2pf plot} goes as $l^2$ at short distance, corresponding to an ensemble of smooth functions, whereas it appears to be close to $l^1$ in Ref.~\cite{Martins:2005es}, corresponding to a random walk in $ {\bf p}_\pm$.}
In Fig.~\ref{2pf plot} we have assumed that the enhanced straightness of the strings at the shortest scales seen in Figs.~\ref{mvd_rad} and~\ref{mvd_mat} is a transient effect, so that the two-point function follows a simple power law down to the gravitational radiation scale.

The exponential falloff in the world-sheet two-point function at large wavenumber implies that the strings are very smooth at short distance.
Inspection of Fig.~\ref{2pf plot} indicates that this smoothing sets in at
\begin{equation}
l_{\rm GW} \approx 20  (G \mu)^{1+2\chi} t
\label{lsize}
\end{equation}
in both the matter and radiation epochs.
The exponent we obtain is smaller than the estimates in~\cite{Siemens:2002dj}: $1.2$ in the radiation dominated era and $1.5$ in the matter dominated era here, compared to $3/2$ and $5/2$ respectively.
The difference is primarily from using a more accurate model of the effect of stretching, leading to a power spectrum with a slower falloff.
The length~$l_{\rm GW}$ is the approximate size of the smallest loops.
In particular, the divergent loop production found in Ref.~\cite{Polchinski:2006ee} is cut off at this scale.

The gravitational radiation back-reaction on the two-point function $\langle {\bf p}_\pm \cdot {\bf p}_\pm \rangle$ obviously carries over to the two-point function for the fluctuations $\langle {\bf w}_\pm \cdot {\bf w}_\pm \rangle$.
These are fixed by the relations~(\ref{pppp}) and~(\ref{upup}), or equivalently~(\ref{ww}), but only up to an undetermined function $\phi(u) \equiv \frac{1}{2} \langle w_+^2(u) \rangle$:
\begin{equation}
\langle {\bf w}_+(u) \cdot {\bf w}_+(u') \rangle = -{\mathcal A} \frac{|u-u'|^{2\chi}}{t^{2\chi}} + \phi(u) + \phi(u')  \ ,
\quad {\rm for} \; |u-u'| \gg l_{\rm GW} \ .
\end{equation}
To specify the function $\phi$ one needs to choose how to split ${\bf p}_\pm$ into the slowly-varying ${\bf P}_\pm$ and the fluctuation ${\bf w}_\pm$.
A natural choice is to let the self-correlation of the fluctuations to be independent of the point on the string, i.e. $\phi$ is a constant.
This constant is related to the behavior of the two-point function at large distances as can easily be seen by considering the Fourier transform of the above equation.
The transform of the first part was essentially computed in~(\ref{AandB}).
The constant piece $\phi$ can then be obtained by inserting a cutoff $L$ representing the separation between the long-distance and short-distance parts of the configuration, taken around the correlation length:
\begin{eqnarray}
\langle {\bf w_+}(u) \cdot {\bf w_+}(u') \rangle
&=&  {\mathcal A} c_{\chi} \int_{-\infty}^\infty \frac{dk}{2\pi} e^{i k (u-u')} \frac{1 }{ |k|^{1+ 2\chi} t^{2\chi}}
(e^{-|k| l_{\rm GW}} - e^{-|k| L})  \nonumber\\
&& \hspace{-60pt}  = \frac{\mathcal A}{t^{2\chi} \cos \pi\chi}  {\rm Re}\!\left[ (i |u - u'| + L)^{2\chi}  - (i |u - u'| +  l_{\rm GW})^{2\chi} \right]  \ ,
\label{2pfunc}
\end{eqnarray}
and similarly for ${\bf w}_-$; recall that $c_\chi = 2\sin(\pi\chi) \Gamma(1+2\chi)$.
We have also introduced a smooth short-distance cutoff at the scale $l_{\rm GW} \sim 20 (G\mu)^{1 + 2\chi} t$ to take account of the effects of gravitational radiation, as found above.
In the range $l_{\rm GW} \ll |u-u'| \ll  L$ expression~(\ref{2pfunc}) reduces to
\begin{equation}
\langle {\bf w_+}(u) \cdot {\bf w_+}(u') \rangle \approx {\mathcal A}\, (L'^{2\chi}  - |u - u'|^{2\chi} )/t^{2\chi}  \ ,
\label{2pfa}
\end{equation}
with $L'^{2\chi} = L^{2\chi}/{\cos \pi\chi}$.
So we see that our cutoff prescription corresponds to having $\phi = {\mathcal A}(L'/t)^{2\chi}$.
This expression for the two-point function will be used in the next chapter to compute the rate of small loop production.

\section{Small scale structure and cusps\label{SSSandCUSPS}}

Our work also allows us to address an old question, the interplay between the small scale structure and the cusps~\cite{Siemens:2001dx}.
A cusp is a point on a string that reaches momentarily the speed of light and where simultaneously the $\sigma$-parametrization becomes singular.
We shall discuss this in more depth in Chapter~4 but we can anticipate that cusps arise when the functions ${\bf p}_+(u)$ and ${\bf p}_-(v)$ cross on the unit sphere~\cite{Albrecht:1989mk}.
The {\it size} of the cusp is the inverse of the `velocity' of the functions ${\bf p}_{\pm}$ when they cross, their rate of change with respect to the length along the string.
Now, the average velocity of these functions on the long strings is (going again to the gauge $\epsilon(t) = a(t) = 1$)
\begin{equation}
V^2 = \langle  \partial_\sigma {\bf p}_+(\sigma,t) \cdot  \partial_\sigma {\bf p}_+(\sigma,t) \rangle
= \int \frac{d\kappa} {2\pi} \kappa^2 H(\kappa,t) = d_\chi(G\mu)^{-2(1-\chi)(1+2\chi)}  t^{-2}  \ ,
\label{speed}
\end{equation}
where
\begin{equation}
d_\chi = {\mathcal A} c_\chi \frac{1+2\chi}{2\pi} \Gamma[2(1-\chi)(1+2\chi)]\, [ b_\chi (1+\chi)(1+2\chi) ]^{2(\chi - 1)(1+2\chi)} \label{V2}
\end{equation}
takes the value 0.006 in the matter era and 0.008 in the radiation era.

On the long strings in the network, if the functions ${\bf p}_+(u)$ and ${\bf p}_-(v)$ were smooth and varied on the scale of the correlation length (which is not far below the horizon scale), then the typical velocity of any crossing would be the inverse correlation length, and this would be the size of cusps.
What we see from Eq.~(\ref{V2}) is that such slow crossings and large cusps are impossible.
Superimposed on the slow motion is the irregular motion from the short distance structure, so all crossings occur at the much larger velocity $V \sim 0.1(G\mu)^{-(1-\chi)(1+2\chi)} t^{-1}$, and all cusps have a size of order
\begin{equation}
V^{-1} \sim 10 (G\mu)^{(1-\chi)(1+2\chi)} t  \ ,
\label{cusps}
\end{equation}
which is very much smaller than the correlation length.
Effectively each large cusp breaks up into a large number of small cusps: the high fractal dimension of ${\bf p}_+(u)$ and ${\bf p}_-(v)$ above the gravitational radiation length, noted in Ref.~\cite{Polchinski:2006ee}, implies that these curves will cross many times near the would-be large cusp.
We can estimate the number: the total length of each curve ${\bf p}_+(u), {\bf p}_-(v)$ during one Hubble time is $Vt \sim 0.1(G\mu)^{-(1-\chi)(1+2\chi)}$.
The total number of crossings is roughly the total length of either path divided by the solid angle $4\pi$ of the sphere, and so up to numerical factors it is large to the same extent that the size of each cusp is small.

\chapter[Loop formation]{Loop formation}

The model presented in Chapter~2 assumes that stretching is the dominant mechanism governing the evolution of the small scale structure.
Recall that the output of this model is a power law describing the approach of the two-point function to straightness as we consider smaller separations along the string.
However, the simulations of Ref.~\cite{Martins:2005es} show strings even straighter than our stretching model suggests, at the smaller scales.
We have speculated in Section~\ref{DISCUSS} that this discrepancy could be due to loop production.
Given the smoothness of the strings on short distances, one might have thought that loop production on small scales would be suppressed, but we will see that this is not the case.

In this chapter we include the production of small loops in our model.
We will consider the stretching model as a leading approximation and add in the loop production as a perturbation.
Loop reconnection will still be ignored, based on the standard argument that this is rare for small loops~\cite{Bennett:1989yp}.

We shall begin by reviewing some useful results regarding loops, cusps and kinks in Section~\ref{FACTS}.
We then proceed to compute the rate of loop production.
This calculation is contained in Section~\ref{RATE}.
Section~\ref{DISCUSS2} is devoted to the discussion of the results and includes some speculative comments as well.

The treatment presented here is based on Refs.~\cite{Polchinski:2006ee,Dubath:2007mf}, with Joseph Polchinski and Florian Dubath.

\section{General results: loops, cusps and kinks\label{FACTS}}

Loops much smaller than the horizon size evolve in an essentially flat spacetime so we are able to use null coordinates $u = t + \sigma$ and $v = t - \sigma$.
In this case the left- and right-moving vectors ${\bf p}_\pm$ depend only on $u$ or $v$, respectively.
They are related to the spatial embedding of the string, ${\bf x}(t,\sigma)$, simply by
\begin{equation}
{\bf p}_+(u)  =  2 \partial_u {\bf x}(u,v)  \ ,  \quad\quad
{\bf p}_-(v)  =  2 \partial_v {\bf x}(u,v)  \ .
\label{pdx}
\end{equation}
Recall that these vectors are constrained to live on the unit sphere.

It is easy to derive the condition under which a loop will form by a self-intersecting long string.
Assuming that the reconnection probability is $P=1$, a loop of length $l$ will form if there exists a worldsheet point $(t,\sigma)$ for which\footnote{Here we choose to center the segment of the long string that originates the loop around $\sigma =0$.} ${\bf x}(t,\sigma - l/2) = {\bf x}(t,\sigma + l/2)$, or equivalently,
\begin{equation}
\int_{\sigma-l/2}^{\sigma+l/2} d\sigma' \, \partial_{\sigma'} {\bf x}(t,\sigma') = 0  \ .
\end{equation}
Thus, using $\partial_\sigma = \partial_u - \partial_v$ and the relations~(\ref{pdx}) we conclude that the condition for the formation of a loop of length $l$ is
\begin{eqnarray}
{\bf L}_+(u,l)  &=&  {\bf L}_-(v,l)  \ ,
\label{loopcondition} \\
{\bf L}_+(u,l)  &\equiv&  \int_{u-l/2}^{u+l/2} du'\, {\bf p}_+(u')  \ , \quad\quad
{\bf L}_-(v,l)  \equiv  \int_{v-l/2}^{v+l/2} dv'\, {\bf p}_-(v')  \ .   \nonumber
\end{eqnarray}
Stated differently, the unit vectors ${\bf p}_\pm$ must average to the same value in order to yield a self-intersection.

Note also that a loop of length $l$ actually has period $T = l/2$ in time.
The simplest way to see this is to recall that in flat space signals propagate {\em in both directions} at the speed of light.
Left- and right-moving signals originated at some point $\sigma_0$ on the loop will meet again at the `opposite end', $\sigma_0 + l/2$.
The period is equal to the distance traveled, which yields half the length of the loop.

Let us now consider the particular case when the two curves ${\bf p}_\pm$ cross.
Using $\partial_t{\bf x} = \frac{1}{2} [{\bf p}_+ + {\bf p}_-]$ it immediately follows that such points on a string (called cusps) move (instantaneously) with the speed of light~\cite{Turok:1984cn}, $(\partial_t{\bf x})^2 = 1$.
Correspondingly, the tangent along the string $\partial_\sigma{\bf x}$ momentarily vanishes at the tip of the cusp, signaling the singular behavior of the parametrization of the string by $\sigma$.
The highly relativistic velocities around these points translate into a strong emission of gravitational waves along the direction of motion of the cusp~\cite{Damour:2000wa,Damour:2001bk}.
This feature has been used by those authors to study the detectability of such gravitational waves.
In Chapter~6 we will consider instead gravitational waves originating from the lowest harmonics of small (and relativistic) loops.

When two smooth segments of string intersect and reconnect this results in two kinked segments, as shown in Fig.~\ref{reconnect}.
A kink is a discontinuity of the tangent vector $\partial_\sigma{\bf x}$.
In terms of the left- and right-moving unit vectors this is identified by a discontinuous jump in both ${\bf p}_\pm$.
As long as the approximation of a flat spacetime can be maintained, each discontinuity evolves into a pair of kinks traveling at the speed of light in opposite directions along the string.
A simple way to understand why this happens is to note that one of the kinks is associated with a discontinuity in ${\bf p}_+(u)$ and so is left-moving whereas its `twin' kink is related to a discontinuity in ${\bf p}_-(v)$ and therefore is right-moving.
More physically, such evolution shortens the length of a string between two points surrounding the initial kink and consequently reduces its tension.

\section{The rate of loop production\label{RATE}}

In this section we aim to determine the rate of loop production, which occurs when a string self-intersects.
More specifically, we want to compute the number of loops $d\mathcal N$ with invariant length between $l$ and $l+dl$ originating from a self-intersection at a string coordinate between $\sigma$ and $\sigma + d\sigma$ during a time interval $(t,t+dt)$.

\subsection{Preliminary considerations\label{CONSID}}

The calculation of the loop production rate requires separating the ${\bf p}_\pm$ into a long-distance piece and a short-distance piece.
In Ref.~\cite{Polchinski:2006ee} the long-distance part was simply averaged over the unit sphere, with a weight factor chosen to give the correct mean value of ${\bf p}_+ \cdot {\bf p}_-$.
Assuming the stretching model derived in Chapter~2 remains valid down to arbitrarily small scales, it was found that the loop production function diverged at short distance.
Clearly, this divergence must be cut off because the total rate at which string length goes into loops is fixed by energy conservation.
In the approach of Ref.~\cite{Polchinski:2006ee}, this condition saturated at a length scale just one order of magnitude below the horizon.
However, comparing with the numerical simulations of Refs.~\cite{Vanchurin:2005pa,Ringeval:2005kr} that treatment did seem to find the correct functional form for the loop production function.

An essential improvement was made in Ref.~\cite{Dubath:2007mf}, leading to a very different, and simpler, picture.
In that work the long-distance part was given a classical $(u,v)$-dependence, and then averaged over the short distance part with a weighting corresponding to Eq.~(\ref{fracdim}).
The inclusion of the time-dependence reveals that loop formation leads to a strong correction to the distribution of long-distance configurations, which was not accounted for in the earlier approach~\cite{Polchinski:2006ee}.\footnote{These same issues were faced in the pioneering work~\cite{Austin:1993rg}, but our framework differs.}

\begin{figure}[t]
\begin{center}
\ \includegraphics[width=15pc]{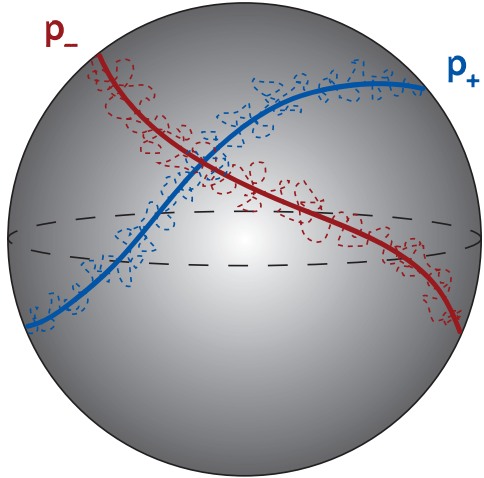}
\end{center}
\caption{The functions ${\bf p}_\pm$ are written as a fixed long-distance piece (bold) plus a random short-distance part.  The classical cusp has been defined to lie at $(u,v) = (0,0)$.}
\label{sphere}
\end{figure}

A typical configuration for the ${\bf p}_\pm$ vectors is shown in Fig.~\ref{sphere}.
The long-distance configuration is in bold, and the total configuration is dashed.
Recall from Section~\ref{DISCUSS} that, while the string configuration approaches fractal dimension 1 at short distance, the graph of its tangent has large fractal dimension $1/\chi$.
In the figure, we have shown a long distance configuration in which the curves ${\bf p}_+(u)$ and ${\bf p}_-(v)$ cross (at $u=0$ and $v=0$), meaning that there is a cusp in the spacetime evolution of the string.
It is evident that the effect of the short distance structure is to turn this one large cusp into many small cusps.
Therefore we see that the formation of small loops is closely associated with the cusps.

When condition~(\ref{loopcondition}) is satisfied the segment of length $l$, centered at $(u,v)$, breaks off as a loop.
For small $l$ the functions ${\bf p}_\pm$ will have small variation along the segment, so small loops will form where these are approximately equal~\cite{Albrecht:1989mu,Siemens:2003ra}.

Following this discussion, it is clear that at early times, when $u$ and/or $v$ are sufficiently negative, the curves will be outside the cusp region and the production of small loops will be negligible.
As the cusp begins to form, production of small loops begins to increase.
We will find that loops of all scales begin to form at the same time, extending all the way down to the gravitational radiation cutoff.
Thus we are led to conclude that the production of loops down to the gravitational radiation scale is inevitable, and not a transient effect.

The loop production cuts off when the probability for a given bit of string to be incorporated in a loop reaches 1.
We will see that this occurs long before the classical cusp point $(u,v)=(0,0)$ is reached.
In the approach of~\cite{Polchinski:2006ee} the region where the long-distance parts of ${\bf p}_+(u)$ and ${\bf p}_-(v)$ were parallel gave the dominant contribution, but in a correct treatment it is simply absent.

In the following section we will carry out the calculation just described.
As further support for our picture, we will obtain a prediction for the small loop production function that fits well with the simulations by two groups.
The fit actually works better than anticipated, as there is no sign that it is distorted by further fragmentation; we speculate as to why this might be.

\subsection{Calculation of the rate\label{RATELP}}

To perform the calculation of the rate of loop production from long strings, we separate the left- and right-moving unit tangent vectors into their long- and short-distance parts,
\begin{equation}
{\bf p}_\pm = {\bf l}_\pm + {\bf s}_\pm  \ .
\label{pls}
\end{equation}
Since the production of small loops takes place near the cusp, we take the simplest cusp form for the long-distance structure,
\begin{eqnarray}
{\bf l}_+(u) &=&  \hat {\bf z}  (1 - V_+^2 u^2)^{1/2}+ {\bf V}_+ u  \approx  \hat{\bf z} + {\bf V}_+ u -\frac{\hat{\bf z}}{2}
V_+^2 u^2  \ ,
\nonumber\\
{\bf l}_-(v) &=&  \hat {\bf z}  (1 - V_-^2 v^2)^{1/2}+ {\bf V}_- v  \approx  \hat{\bf z} + {\bf V}_- v -\frac{\hat{\bf z}}{2}
V_-^2 v^2  \ ,
\label{lplm}
\end{eqnarray}
with ${\bf V}_+ \cdot \hat {\bf z} = {\bf V}_- \cdot \hat {\bf z} = 0$.
We have chosen a coordinate frame in which the motion of the cusp is directed along the $z$-axis and we are expanding to order $u^2$, $v^2$.
The slopes ${\bf V}_{\pm}$ have units of inverse length, and are reciprocal to the size of the cusp.
They will be of order the correlation length, somewhat less than the horizon length.

The vectors ${\bf l}_\pm$ encoding the long-distance structure play the role of the vectors ${\bf P}_\pm$ defined in the small fluctuation approximation of Section~\ref{SMALLFLUC}.
In particular, they are normalized to unity.
The short-distance structure parts are thus expressed in terms of the small fluctuations as
\begin{equation}
{\bf s}_\pm = {\bf w}_\pm - \frac{1}{2} {\bf l}_\pm w_\pm^2  \ .
\label{spm}
\end{equation}
Since the fluctuations must be orthogonal to $\hat {\bf z}$, we can insert expressions~(\ref{lplm}) and~(\ref{spm}) into Eq.~(\ref{pls}) and obtain, again to quadratic order,
\begin{eqnarray}
{\bf p}_\pm &=& \hat {\bf z} + {\bf y}_\pm - \frac{1}{2} \hat {\bf z} y_\pm^2  \ , \nonumber\\
{\bf y}_+(u) &=& {\bf V}_+ u + {\bf w}_+(u)\ ,\quad {\bf y}_-(v) = {\bf V}_- v + {\bf w}_-(v)  \ .
\end{eqnarray}
For the fluctuation two-point function we shall use~(\ref{2pfa}).
Actually, the classical $(u,v)$-dependence given to the long-distance parts of ${\bf p}_\pm$ should add a term of the form $\frac{1}{2}V_+^2(u-u')^2$ to the right-hand side.
However, this is quadratic and we can ignore it when considering the production of loops as a result of the small scale structure.


The number of loops with length between $l$ and $l+dl$ formed by self-inter\-sec\-tions of a long string occurring within an infinitesimal worldsheet area $dudv$ is given by 
\begin{equation}
d{\mathcal N} = \left\langle \delta^{(3)}({\bf L}_+(u,l) - {\bf L}_-(v,l)) \left| \det {\bf J} \right| \right\rangle du\, dv\, dl \ ,
\label{dN}
\end{equation}
where ${\bf J}$ is the Jacobian for the transformation $(u,v,l) \rightarrow {\bf L}_+ - {\bf L}_-$:
\begin{equation}
{\bf J}(u,v,l) = \left[ \begin{array}{c} 
{\bf p}_+(u+l/2) - {\bf p}_+(u-l/2)\\
{\bf p}_-(v-l/2) - {\bf p}_-(v+l/2)\\
\frac{1}{2}{[{\bf p}_+(u+l/2) + {\bf p}_+(u-l/2) - {\bf p}_-(v+l/2) - {\bf p}_-(v-l/2)] }
\end{array}
\right]  \ .
\end{equation}
This formalism is as in Refs.~\cite{Embacher:1992zk,Austin:1993rg}.

In~\cite{Polchinski:2006ee} the computation proceeded with an estimation of the expectation value of the product~(\ref{dN}) by the product of expectation values, but here there are strong correlations: when the delta-function is nonzero, the third row of the Jacobian is much smaller than its mean value.
We therefore go to new variables, separating ${\bf w}_\pm$ into a piece constant on the segment and a piece with zero average on the segment,
\begin{eqnarray}
{\bf w}_+(u') &=& {\bf W}_+ + {\mbox{\boldmath$ \omega$}}_+(u')\ , \quad {\bf W}_+ = \frac{1}{l} \int_{u-l/2}^{u+l/2} du'\, {\bf w}_+(u') \ ,  \nonumber\\
{\bf w}_-(v') &=& {\bf W}_- + {\mbox{\boldmath$ \omega$}}_-(v')\ , \quad {\bf W}_- = \frac{1}{l} \int_{v-l/2}^{v+l/2} dv'\, {\bf w}_-(v') \ .
\end{eqnarray}
The variables ${\mbox{\boldmath$ \omega$}}_\pm$ and ${\bf W}_\pm$ depend on the parameters $u, v, l$ of the loop, but we leave this implicit.
We will be interested in the loop production in the range $l_{\rm GW} \ll l \ll  L$, so we use the form~(\ref{2pfa}).
One then finds the two-point functions
\begin{eqnarray}
t^{2\chi}\langle {{\mbox{\boldmath$ \omega$}}_+}(u') \cdot {{\mbox{\boldmath$ \omega$}}_+}(u'') \rangle
&=& - {\mathcal A} |u' - u''|^{2\chi} + f(u') + f(u'')  \ ,
\label{oo}  \\
t^{2\chi}\langle {{\mbox{\boldmath$ \omega$}}_+}(u') \cdot {\bf W_+} \rangle  &=&  - f(u')  \ ,
\label{cross}  \\
t^{2\chi}\langle {\bf W}_+ \cdot {\bf W_+} \rangle  &=&  {\mathcal A} L'^{2\chi} + O(l^{2\chi})  \ ,
\label{WW}
\end{eqnarray}
where
\begin{equation}
f(u') = \frac{{\mathcal A}}{(2\chi+1) l} \left[ (l/2 + u' - u)^{2\chi+1} + (l/2 - u' + u)^{2\chi+1} - \frac{l^{2\chi+1}}{2\chi+2} \right]
\label{fofu}
\end{equation}
is defined only for points on the segment forming the loop, i.e. $u - \frac{l}{2} \leq u' \leq u + \frac{l}{2}$.

Now let us express the rate in terms of these quantities.
First, for the transverse parts of ${\bf L}_\pm$ we have
\begin{equation}
{\bf L}^\perp_+ = l({\bf W}_+ + {\bf V}_+ u)  \ , \quad 
{\bf L}^\perp_- = l({\bf W}_- + {\bf V}_- v)  \ ;
\end{equation}
the transverse part of the delta-function sets these equal ${\bf L}^\perp_+ = {\bf L}^\perp_-  \equiv {\bf L}^\perp$, and we use this in evaluating the remaining terms.
For the $z$ component,
\begin{eqnarray}
{L}^z_+ &=& l - \frac{1}{2} \int_{u-l/2}^{u+l/2} du' \left[ {\bf V}_+ u'  + {\bf W}_+ + {\mbox{\boldmath$ \omega$}}_+ (u') \right]^2 \nonumber\\
&=& l - \frac{1}{2l} {{\bf L}^\perp}^2  - \frac{1}{2} \int_{u-l/2}^{u+l/2}  du' \left[ {\mbox{\boldmath$ \omega$}}_+ (u') + {\bf V}_+ (u' - u) \right]^2  \ ,
\end{eqnarray}
where we have used the fact that the mean value of ${\mbox{\boldmath$ \omega$}}_+$ is zero.
In the integrand, the fluctuations of ${\mbox{\boldmath$ \omega$}}_+$ are of order $l^\chi$, while the classical term is of order $l$, so we can drop the latter for small loops.
Thus,
\begin{equation}
{L}^z_+ - {L}^z_- = - \frac{1}{2} \int_{u-l/2}^{u+l/2} du' \left[ {\mbox{\boldmath$ \omega$}}_+ (u') \right]^2
+ \frac{1}{2} \int_{v-l/2}^{v+l/2} dv' \left[ {\mbox{\boldmath$ \omega$}}_- (v') \right]^2 = O(l^{1+2\chi})  \ .
\end{equation}

This takes care of the delta function in~(\ref{dN}).
Now we turn our attention to the Jacobian.
For the transverse part of the first row of ${\bf J}$,
\begin{eqnarray}
{\bf p}^\perp_+(u+l/2) - {\bf p}^\perp_+(u-l/2) &=& {\mbox{\boldmath$ \omega$}}_+(u+l/2) - {\mbox{\boldmath$ \omega$}}_+(u-l/2) + {\bf V}_+ l  \nonumber\\
 &=& {\mbox{\boldmath$ \omega$}}_+(u+l/2) - {\mbox{\boldmath$ \omega$}}_+(u-l/2) + O(l)  \ .
\end{eqnarray}
Similarly in the second and third rows, in the transverse terms we can replace ${\bf p}_\pm$ with $ {\mbox{\boldmath$ \omega$}}_\pm$, after imposing ${\bf L}^\perp_+ = {\bf L}^\perp_-$.
In the $z$ component of the first row,
\begin{eqnarray}
{p}^z _+(u+l/2) - {p}^z_+(u-l/2) &=& - \frac{1}{l} {\bf L}^\perp \cdot [ {\mbox{\boldmath$ \omega$}}_+(u+l/2) - {\mbox{\boldmath$ \omega$}}_+(u-l/2)] \nonumber\\
&& \hspace{-80pt}
- \frac{1}{2} \Bigl( [{\mbox{\boldmath$ \omega$}}_+(u+l/2)]^2 - [{\mbox{\boldmath$ \omega$}}_+(u-l/2)]^2 \Bigr) + O(l)  \ .
\label{jz1}
\end{eqnarray}
The first term is of order $l^\chi$ and the second of order $l^{2\chi}$, but the first actually drops out. 
In all three rows of ${\bf J}$ one finds the same pattern, $J^z = {\bf L}^\perp \cdot {\bf J}^\perp + J^z[{\mbox{\boldmath$ \omega$}}]$.
Thus the first term in the $z$-column is linearly dependent on the other two columns, and drops out of the determinant, leaving $J^z$ with ${\bf p}_\pm^\perp$ replaced by ${\mbox{\boldmath$ \omega$}}_\pm$.\footnote{Note this implies that $p_\pm^z$ is replaced by $1 - \frac{1}{2}{\mbox{\boldmath$ \omega$}}_\pm^2$, since the vectors ${\bf p}_\pm$ are normalized to unity.  Furthermore, the constant term may be dropped because it cancels out in all calculations.}

We have now expressed the terms multiplying $\delta^{(2)}({\bf L}^\perp_+ - {\bf L}^\perp_- )$ all in terms of ${\mbox{\boldmath$ \omega$}}$.
The cross-correlation~(\ref{cross}) is of order $l^{2\chi}$, smaller than the geometric mean $L'^\chi l^\chi$ of the diagonal correlators~(\ref{oo},$\,$\ref{WW}).
We can therefore ignore it, giving
\begin{equation}
\left\langle \delta^{(3)}({\bf L}_+ - {\bf L}_- ) \left| \det {\bf J} \right |  \right\rangle 
= \left\langle \delta^{(2)}({\bf L}^\perp_+ - {\bf L}^\perp_- ) \rangle  \times 
\langle \delta({L}^z_+ - {L}^z_-) \left| \det {\bf J} \right | \right\rangle_{ {\bf p}^\perp  \to {\mbox{\boldmath\scriptsize$ \omega$}}}  \ .
\label{dNw}
\end{equation}
Thus we have achieved our aim of factorizing the expectation value.
In the second term the delta-function scales as $l^{-1-2\chi}$, inversely to its argument, while the columns of ${\bf J}$ scale as $l^\chi, l^\chi, l^{2\chi}$, giving the overall $l^{-1+2\chi}$ scaling, as found in~\cite{Polchinski:2006ee}.
This second term is evaluated in Appendix A with the assumption that the short-distance structure is Gaussian.
One finds that
\begin{equation}
\langle \delta({L}^z_+ - {L}^z_-) \left| \det {\bf J} \right| \rangle_{ {\bf p}^\perp  \to {\mbox{\boldmath\scriptsize$ \omega$}}}
=  \: \eta \frac{{\mathcal A}}{t} \left( \frac{l}{t} \right)^{-1+2\chi}  \ ,
\end{equation}
where $\eta \approx 2.3$ in the radiation era and $\eta \approx 1.3$ in the matter era.
As we will discuss in Chapter~6, the Gaussian approximation may not be valid here, but this does not alter the scaling with $l$.

It remains to evaluate
\begin{eqnarray}
\left\langle \delta^{2}({\bf L}^\perp_+ - {\bf L}^\perp_- ) \right\rangle
&=&
\int \frac{d^2 q}{(2\pi)^2} \left\langle e^{i {\bf q} \cdot ({\bf L}^\perp_+ - {\bf L}^\perp_- ) }\right\rangle
\nonumber\\
&=& \int \frac{d^2 q}{(2\pi)^2} e^{- {\mathcal A} L'^{2\chi} l^2 q^2/2 t^{2\chi} + i l {\bf q}\cdot ({\bf V}_+ u - {\bf V}_- v)}
\nonumber\\
&=& \frac{t^{2\chi} e^{-({\bf V}_+ u - {\bf V}_- v)^2 t^{2\chi}/2 {\mathcal A} L'^{2\chi}}}{2 \pi {\mathcal A} L'^{2\chi} l^2}  \ .
\end{eqnarray}
In all,
\begin{equation}
d{\mathcal N} = C L'^{-2\chi} e^{-({\bf V}_+ u - {\bf V}_- v)^2 t^{2\chi}/2 {\mathcal A} L'^{2\chi}} \, du\, dv\, \frac{dl}{ l^{3 - 2\chi}}  \ ,
\label{dn} 
\end{equation}
where (with the Gaussian approximation) $C = 0.35$ in the radiation era and $C = 0.2$ in the matter era.
These constants are actually independent of ${\mathcal A}$ so the normalization of the two-point function only enters the calculation of loop production through the exponential suppression factor.

The result above indicates that the total rate of string loss $\int l \, d{\mathcal N}$ diverges at small $l$ for $\chi \leq 0.5$, as is the case in both the radiation and matter eras.
Thus, loop formation can have a significant effect on the evolution of small scale structure and could invalidate the assumptions of our model but we will argue against this in Section~\ref{LOOPEFFECTON2PF}.
Of course, the total rate of long string lost into loops is bounded.
Indeed, the small-$l$ divergence must be cut off at the gravitational radiation smoothing scale found in Chapter~3.
But, for now, let us verify the statement made in Section~\ref{CONSID} that segments on the long string with the potential to form cusps are carried away by loops before the cusp actually forms.

\subsection{The normalization\label{NORM}}

\begin{figure}[t]
\centering
\includegraphics[width=18pc]{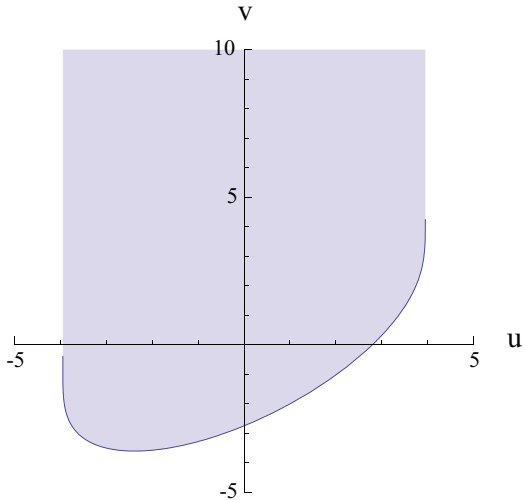}
\caption{The shaded region in the $(u,v)$-plane is excised by loop formation.  The cusp occurs at $u=v=0$ and the curve that delineates the excised region is $v_{max}(u)$.  Both $u$ and $v$ are expressed in units of the cosmological time $t$.  For this particular example the values $G\mu=10^{-9}$ and $\cos(\theta) = 2/\pi$ were used, where $\theta$ denotes the angle between ${\bf V}_+$ and ${\bf V}_-$.}
\label{excision}
\end{figure}

Consider a left-moving point with given $u$.
The total probability per unit $v$ that this point be incorporated into a loop is
\begin{equation}
\frac{d{\mathcal P}}{dv} = C L'^{-2\chi}e^{-({\bf V}_+ u - {\bf V}_- v)^2 t^{2\chi}/2 {\mathcal A} L'^{2\chi}} \int \frac{dl}{ l^{2 - 2\chi}} \ .
\end{equation}
Note that we have replaced $\int du \to l$ to count the loops containing the given point.
This diverges at small $l$ because we have not yet taken into account the smoothing due to gravitational radiation in the two-point function~(\ref{2pfunc}).
The smoothed two-point function is quadratic at $u' - u \to 0$, corresponding to the form~(\ref{2pfa}) at $\chi = 1$: at the shortest distances the divergence is gone.
Thus we cut off the integral at $l_{\rm GW}$ to get
\begin{equation}
\frac{d{\mathcal P}}{dv} \approx  l_{\rm GW}^{-1 + 2\chi}  L'^{-2\chi} e^{-({\bf V}_+ u - {\bf V}_- v)^2 t^{2\chi}/2 {\mathcal A} L'^{2\chi}} \ . \label{dpdv}
\end{equation}
At early times, where $v$ is large, the probability of a loop containing $u$ is small.
However, $l_{\rm GW}^{-1 + 2\chi}$ is large compared to the other dimensionful quantities, so as we integrate in $v$ we soon reach total probability $1$ for a range of values of $u$ near the cusp: this portion of the string is removed by loop production (see Fig.~\ref{excision}).
Therefore, for each value of the $u$-coordinate, the maximum value of $v$ that may still not be included in a loop is the solution (when it exists) of the equation ${\mathcal P}(v_{max})=1$, where
\begin{equation}
{\mathcal P}(v(u)) = \frac{C}{(1-2\chi)l_{GW}} \left(\frac{l_{GW}}{L'}\right)^{2\chi} 
  \int_{-\infty}^{v(u)} dv \, e^{-({\bf V}_+ u - {\bf V}_- v)^2 t^{2\chi}/2 {\mathcal A} L'^{2\chi}}  \ .
\end{equation}
We see that the excised region depends on the the angle $\theta$ between ${\bf V}_+$ and ${\bf V}_-$ for each cusp, not to mention the ubiquitous $G\mu$.
To obtain the normalization of the loop production function would require the exact knowledge of the probability distribution for the angles $\theta$ as well as the density of cusps.
One can certainly obtain normalizations in accordance with the simulations but there are too many free parameters.

Thus, it is beyond our present capabilities to determine the normalization of the loop production function directly from the knowledge of the two-point function.
Nevertheless, one can still make some progress by employing energy conservation.
The total rate of string loss is bounded: the long string network cannot loose more energy than it initially contains!
In a scaling solution, the total amount of long string in a comoving volume scales as $\ell_\infty \propto a^3/t^2$, while stretching alone would give $\ell_\infty \propto a^{\bar\alpha}$.
The rate of string lost to loops must be
\begin{equation}
\frac{1}{\ell_\infty} \Biggl( \frac{\partial \ell_\infty}{\partial t} \Biggr)_{\rm loops} = \frac{2 - 2 \nu(1 + \bar v^2)}{t}  \ .
\label{string2loops}
\end{equation}
Averaging expression~(\ref{dn}) over the ensemble of segments and noting that $dudv = 2dtd\sigma$ we must have an average rate of loop production of the form
\begin{equation}
\frac{\overline{d {\mathcal N}}}{dt d\sigma dl} = \frac{c}{t^3} \left( \frac{l}{t} \right)^{2\chi - 3}  \ .
\end{equation}
We were not able to compute the normalization $c$ from our stretching model but now we can fix it by equating the rate of string length going into loops to the rate of length lost by the long strings.
\begin{eqnarray}
\frac{1}{\ell_\infty} \left( \frac{\partial \ell_\infty}{\partial t} \right)_{loops} 
&=& \int_{l_{GW}}^\infty dl \, l \frac{\overline{d {\mathcal N}}}{dt d\sigma dl}
= \int_{l_{GW}}^\infty dl \, \frac{c l}{t^3} \left( \frac{l}{t} \right)^{2\chi-3}   \nonumber \\
&=& \frac{c}{(1-2\chi) t} \left( \frac{l_{GW}}{t} \right)^{2\chi - 1}  \ .
\end{eqnarray}
We have inserted a cutoff at the gravitational radiation scale $l_{GW}$.
Equating the two rates gives
\begin{equation}
c = (1-2\chi) \left[ 2 - 2\nu (1 + \bar{v}^2) \right] \left( \frac{l_{GW}}{t} \right)^{1-2\chi}  \ .
\label{c/l_GW}
\end{equation}
So we conclude that it is the gravitational radiation (or at least some UV cutoff on the sizes of loops produced) that determines the normalization of the loop production function.

\section{Discussion\label{DISCUSS2}}

These are the main results of this chapter.
First, loops of all sizes down to the UV cutoff form simultaneously, rather than in a cascade of fragmentation.
Therefore the production of the small loops is a robust physical result.
The distribution of the loop lengths obeys a simple power law with the exponent determined by $\chi$.

The production of very small loops takes place when ${\bf p_+} \sim {\bf p_-}$: that is, near cusps on the long strings as proposed in Refs.~\cite{Albrecht:1989mu,Siemens:2003ra}.
Indeed, all the loops that we consider are rather smaller than the correlation length, so the functions ${\bf p_+}$ and ${\bf p_-}$ are each somewhat localized on the unit sphere, and necessarily in the same region since ${\bf L}_+ = {\bf L}_-$.
In this sense all of our loops are produced near cusps.

The integral of the probability~(\ref{dpdv}) for a point on the string to break off becomes large long before the cusp $v=0$ is reached.
In Ref.~\cite{Polchinski:2006ee} this fact was not accounted for and this resulted in a large overcounting of loops formed by self-intersections.

It is possible that there is a second population of loops that form at much longer scales and we will speculate on this issue below.
But first let us compare our results with some of the most recent simulations.

\subsection{Comparison with simulations}

The results that we have found have some notable agreements, and disagreements, with simulations.
One success is an apparent agreement with the simulations of Ref.~\cite{Ringeval:2005kr} for the distribution of loop sizes: that reference finds a number density of loops per volume and per length $dn /dl \propto l^{-p}$ with $p = 2.41$ in the matter era and $p = 2.60$ in the radiation era.
We have exponents $3 - 2\chi = 2.5$ and $2.8$ respectively.
Our exponent is for the production rate rather than the density, but in this regime the density is dominated by recently produced loops and so these are the same.

Let us verify this, and also obtain the relative normalizations for the two quantities.
The number of loops per comoving volume of length between $l$ and $l+dl$ is unaffected by the expansion of the universe, and we are assuming that we are at scales where gravitational radiation can be neglected.
The number then changes only due to production:
\begin{equation}
\frac{d}{dt}\left( a^3 \frac{dn}{dl} \right) = \frac{a^3}{ \gamma^2 t^2} \frac{c}{l^3} \left( \frac{l}{t} \right)^{2\chi}  \ .
\end{equation}
(Recall that $\gamma$ is a dimensionless constant: the scaling value of the long string length per unit volume is $(\gamma t)^{-2}$).
This integral is dominated by recent times as long as $3\nu - 1 - 2\chi > 0$, as holds in both eras.
Integrating over time then gives
\begin{equation}
\frac{dn}{dl} = \frac{c}{(3\nu - 1 - 2\chi) \gamma^2 t^4} \left( \frac{l}{t} \right)^{-3 + 2\chi}  \ .
\end{equation}
In the notation of Ref.~\cite{Ringeval:2005kr} the normalization is defined as
\begin{equation}
C_\circ = \frac{c}{(3\nu - 1 - 2\chi) \gamma^2} (1 - \nu)^{-1 -2\chi}  \ ,
\end{equation}
where the last factor comes from the conversion from $d_H$ to $t$.

In Section~\ref{NORM} we determined the normalization $c$ in terms of the gravitational radiation cutoff.
Ref.~\cite{Ringeval:2005kr} does not explicitly consider gravitational radiation but their simulations have a minimum resolution: loops below a certain length are artificially removed from the network.
As we will see in the next chapter, this effectively introduces a gravitational radiation scale consistent with the values $l_{GW} \simeq 1.2 \times 10^{-4} t$ and $l_{GW} \simeq 6 \times 10^{-5} t$ for the matter and radiation eras, respectively.
Using $\gamma = 0.59, 0.30$ respectively in the matter and radiation eras~\cite{Martins:2005es}, this gives $C_\circ = 0.07, 0.03$.
These are smaller than found in Ref.~\cite{Ringeval:2005kr} by factors of 1.3 and 7.
Given the crudeness of our estimate, the disagreement by less than an order of magnitude is a pleasant surprise.

A different group~\cite{Olum:2006ix} has performed simulations that, due to a doubling trick, have been able to reach a larger dynamical range than in previous numerical works.
Their results relate more directly to ours since they consider explicitly the rate of loop production.
Those authors express this in terms of $x^2 f(x)$, where $x = l/t$ and the loop production function (LPF) $f(x)$ is related to the rate through
\begin{equation}
f(x) = 2 \frac{t^3}{\gamma^2} \frac{d{\mathcal N}}{du dv dl}  \ .
\end{equation}
These simulations show a power law distribution above the UV cutoff, with exponents that match rather well with our model in both the matter and radiation eras.
The issue of the normalization is a more speculative story.
First, the normalization of the loop production function in~\cite{Olum:2006ix} is still decreasing when the simulations stop running.
Second, these simulations do not include gravitational radiation and have no minimum resolution, so according to Eq.~(\ref{c/l_GW}) it seems that the normalization should approach zero at late times.
Indeed, again by conservation of energy, if the LPF retains the power law behavior down to smaller and smaller loop sizes as the simulation progresses then its normalization must decrease.
Since the smoothing by gravitational radiation is a physical effect, the evolution of the LPF must stop when the scale $l_{GW}$ is reached and in that case the normalization should be fixed by Eq.~(\ref{c/l_GW}).
On top of the power law distribution we have been discussing, Ref.~\cite{Olum:2006ix} also detects a peak at quite large loop lengths, $l \sim 0.1 t$ and we shall comment on this below.

As we have already mentioned, Ref.~\cite{Ringeval:2005kr} also finds power law distributions for $x^2 f(x)$.
In Table~\ref{exponents} we compare the values of the exponents obtained in the simulations by both groups with the exponents predicted by our model.
The simulations~\cite{Ringeval:2005kr} and~\cite{Olum:2006ix} thus agree reasonably well with each other and our model lies in between.

\begin{table}[t]
\begin{center}
 \begin{tabular}{| l || c | c | c |}
  \hline
	$x^2 f(x)$          & Ref.~\cite{Olum:2006ix} & Ref.~\cite{Ringeval:2005kr} &  our model  \\
	\hline
	\hline
	Radiation dominated &       $x^{-0.95}$       &         $x^{-0.60}$         &  $x^{-0.8}$ \\
	\hline
	Matter dominated    &       $x^{-0.55}$       &         $x^{-0.41}$         &  $x^{-0.5}$ \\
	\hline
 \end{tabular}
\end{center}
\caption{Comparison of the exponents in the loop production function power law obtained from simulations by two groups with those predicted by our stretching model.}
\label{exponents}
\end{table}

\subsection{Effect of loop production on the two-point function\label{LOOPEFFECTON2PF}}

The model developed in Chapter~2 ignored the production of small loops as being a sub-dominant effect to stretching.
However, we have found that, although the fractal dimension of the long strings approaches 1 at short scales, the rate of loop production becomes relevant in cusp regions.
Does this invalidate the assumptions of our model?

Recall there was a discrepancy between the two-point function as dictated by our model and the results of~\cite{Martins:2005es}.
At the shortest scales in Figs.~\ref{mvd_rad} and~\ref{mvd_mat} the slope of the curve approaches and possibly even exceeds unity, corresponding to the critical value $\chi = 0.5$: if the two-point function is of this form then the small-loop production converges.
Thus it is appealing to assume that it is feedback from the production of short loops that accounts for the break in the curve.\footnote{Note also that $\chi = 0.5$ corresponds to the functions ${\bf p}_\pm$ being random walks on the unit sphere.  This suggests this value might be some dynamical fixed point.}

We believe this is not the case.
If the loop formation were distributed on the long strings, there would be at least one additional effect, where the string shortens due to loop emission and more distant (and therefore less correlated) segments are brought closer together.
However, the loop production is not distributed: it occurs when the functions ${\bf p}_{\pm}(\sigma)$, as they wander on the unit sphere, come close together.
Our picture is that whole segments in cusp regions are then excised.
Thus, most of the segments that remain on a long string at a given time would have been little affected by loop production.

Consider a very short segment on a long string.
With time, its size as a fraction of the horizon size decreases, so it moves to the left on Figs.~\ref{mvd_rad} and~\ref{mvd_mat}.
If it experiences only stretching, it will follow the slope of the dashed line.
Thus we would expect the two-point function to be approximately given by the stretching power law down to arbitrarily small scales, until gravitational radiation enters.\footnote{One effect that may increase $\chi$ slightly is that segments that do not fall into the loop production `hole' at ${\bf p}_+ \sim {\bf p }_-$ may experience on average a larger value of $\alpha = -{\bf p}_+ \cdot {\bf p }_-$ and so have a reduced two-point function --- see Eq.~(\ref{hppint}).  However, it seems impossible that this could account for the factor of 5 discrepancy of slopes evident in Fig.~\ref{mvd_rad}.}
The loop production must have a strong bias toward removing segments with large fluctuations to account for the simulations.

There is a somewhat related puzzle regarding the loop velocities, first noted in~\cite{Polchinski:2006ee}.
For a loop of length $l$, the mean velocity is
\begin{equation}
{\bf v} = \frac{1}{2l} ({\bf L}_+(u,l) + {\bf L}_-(v,l)) =  \frac{1}{l} {\bf L}_+(u,l) \ .
\end{equation}
Then
\begin{eqnarray}
\langle v^2(l) \rangle &=& \frac{1}{l^2} \int_0^l \!\!\!\int_0^l d\sigma\, d\sigma'\, \langle  {\bf p}_+ (\sigma) \cdot {\bf p}_+(\sigma') \rangle   \nonumber\\
&=& 1 - \frac{2A}{(2\chi + 1)(2\chi + 2)} (l/t)^{2\chi}   \ .
\label{vsquared}
\end{eqnarray}
Looking for example at loops with $l = 10^{-2} t$, we obtain a typical velocity $0.985$ in the matter era and $0.90$ in the radiation era.
These are significantly larger than are generally expected; Ref.~\cite{Bennett:1989ch} gives values around $0.75$ and $0.81$ for loops of this size.
We emphasize that this is not a consequence of our dynamical assumptions, but can be seen directly by using the two-point functions of Ref.~\cite{Martins:2005es} (the discrepancy is then even greater in the radiation era).
Assuming that both simulations are correct, we must conclude that the two-point function on loops is very different from that on long strings, in fact less correlated.
One possible explanation is that loop formation is biased in this way, but this could also be an indication of the complicated nature of the fragmentation: the final non-self-intersecting loops must contain segments which began on the long string many times further separated than the final loop size.

\subsection{Fragmentation}

The formation of loops from self-intersections on long strings is just the beginning of the story.
After the primary loops are chopped off (and even before this happens) the process of loop production continues on smaller scales, unaffected at least initially by the separation of the loop from the infinite string.
Thus, the primary loops fragment into smaller daughter loops.

We have found that fragmentation is not necessary to produce the small loops, but it still raises a puzzle.
Given the small-$l$ divergence of the rate of loop formation in cusp regions one might have expected a rather dramatic fragmentation process.
Indeed, if we were to apply our same calculation to study fragmentation of these small loops, we would find again a strong tendency to fragment into loops at the gravitational radiation scale, giving a sharp peak there rather than a power law distribution.

In fact, we believe this is not precisely the situation.
Rather, a non-self-intersecting loop will occasionally form and this length of loop will then be lost to the fragmentation process on smaller scales.
The point is that the short distance structure on these small loops is not that typical of the long strings, but is dominated by two or more large kinks moving in each direction.
One kink in each direction forms with the loop, and the remainder are present in the pre-existing distribution.
From Fig.~\ref{sphere} we expect that if there are any cusps on these loops then the regions containing them will dissolve into small loops.
However, if there are parts of the loop where ${\bf p}_+$ is sufficiently far from the ${\bf p}_-$ curve, and vice versa, then these will survive.
Thus we expect that if the functions ${\bf p}_\pm$ are plotted, there will be significant gaps in each, corresponding to kinks which form simultaneously with intercommuting events, and the surviving curves will avoid each other.

It was noted in Ref.~\cite{Polchinski:2006ee} that the tails of the small scale structure distribution are not Gaussian but are dominated by a single large kink (again, this does not alter the scaling with $l$).
Thus these kinks are likely associated with the tail of the loop production function.
If so, the functions  ${\bf p}_\pm$ on the small loops are dominated by large jumps, from the kinks, and will be unlikely to have cusps where smaller loops can form.
This conforms to the conclusions of~\cite{Scherrer:1989ha}, where numerical simulations of the fragmentation of loops was studied.
There it was found that there is an end to the fragmentation process, meaning that all loops eventually reach non-intersecting trajectories.\footnote{However, we note that the spectrum of perturbations used to define the initial loops in this study does not resemble our distribution for cosmic strings in expanding spacetimes.}

Finally, the fragmentation need not change the scaling of the LPF as long as the long-string correlation function remains a power law, because the shapes of the produced loops, and the resulting fragmentation, all scale. 
Thus, we do not consider the agreement between the exponents characterizing the loop production obtained from our model and from the simulations to be accidental.

\subsection{Large loops}

We have already mentioned that Refs.~\cite{Vanchurin:2005pa,Olum:2006ix} find two populations of loops, one of which scales at $\alpha \equiv l/t \sim 0.1 $, and a second at much shorter scales which is still evolving toward smaller lengths.
We believe the latter corresponds to the power law distribution borne out of our small scale structure model.
The production of large loops is outside our model, but these also seem inevitable.
In the course of the evolution of the network it must happen at least occasionally that a large loop will form and survive reconnection.
We must leave it to the simulations to determine the relative weight and average size of these loops since at these large scales the cosmic string evolution is highly non-linear, but the argument given in the previous section explains why they do not fragment entirely into much smaller loops.
An account of these large loops has been given in Ref.~\cite{Vanchurin:2007ee}.

It seems that the interpretation of the small loops as being transients is implausible.
First, there is no sign in Ref.~\cite{Olum:2006ix} that the total fraction of string length going into the small loops is diminishing in time.
That was the case for Ref.~\cite{Vanchurin:2005pa} but those simulations were done in flat spacetime.
Second, if this were a transient effect then its natural time scale should be small in proportion to the size of the loops.
Instead, these loops are still being produced after several Hubble times, showing that some longer-distance process is continuing to feed the small loop production.
Indeed, this mechanism has already been identified: the small scale structure on long strings and large loops, which originates at the horizon length and then is carried down to smaller relative lengths by the expansion of the universe, leads to production of loops on arbitrarily small scales.

Thus we are led to suggest the following picture.
After a large loop is formed, production of smaller loops continues near its cusp regions.
In the end, there is a large loop without cusps or self-intersections, reduced in size by the excision of the cusp regions.
There is also a population of much smaller loops; based on the non-scaling seen thus far~\cite{Vanchurin:2005pa,Olum:2006ix}, we conjecture that these will be at the gravitational radiation length.\footnote{In the absence of gravitational wave smoothing, these could even be at the scale set by the thickness of the string~\cite{Bevis:2006mj}.}
Inspection of Fig.~4 of Ref.~\cite{Olum:2006ix} suggests that as much as 80\% of the total string length goes into the small loops, with 20\% remaining in the large loop.

In conclusion, it appears that numerical and analytic methods are converging on a firm picture of the loop production, with one peak near the horizon scale and one near the gravitational radiation scale, with perhaps the larger fraction of string in small loops.\footnote{We are following Refs.~\cite{Siemens:2001dx,Polchinski:2007rg} in putting the UV cutoff at the gravitational radiation scale.}

\chapter{Properties of the small loop population}

As mentioned in the Introduction, the large scale properties of cosmic string networks are well understood from simulations~\cite{Albrecht:1989mk,Bennett:1989yp,Allen:1990tv} and there are analytic models that accurately capture those features~\cite{Martins:1996jp,Martins:2000cs}.
However, this is not the case regarding the issue of small scale structure.
For example, the existence or not of a scaling regime for the smaller scales (for example, the typical size of the small loops) still remains an open question~\cite{Polchinski:2007qc}.

In this chapter we collect relevant properties of small loops and attempt to characterize such a population.
The phenomenology of small loops is addressed in Section~\ref{PHENOM}.
In Section~\ref{SCALINGLOOPS} we consider the number density of loops -- it is the small scale structure that is responsible for their production and so the number density provides a measure of the former -- and demonstrate that this quantity scales, both neglecting and including the effects of gravitational radiation.

Most of the material contained in this chapter is drawn from Refs.~\cite{Polchinski:2007rg,Dubath:2007wu,Rocha:2007ni}, with Joseph Polchinski and Florian Dubath.

\section{Phenomenology of small loops\label{PHENOM}}

\subsection{Gravitational radiation from small loops and their lifetimes\label{GRAVRADLOOPS}}

Let us start by briefly reviewing the issue of gravitational radiation emitted by small cosmic string loops.
For the ``vanilla'' cosmic strings, the rate at which the loops radiate energy in gravitational waves determines their lifetime and this will be used in Chapter~6.
In addition, when the radiation from the whole population of loops is added up, it gives rise to a stochastic gravitational radiation background~\cite{Vilenkin:1981bx} and this can lead to detection or better bounds on the dimensionless string tension, namely by accurate measurements of pulsar timing~\cite{Hogan:1984is,Caldwell:1991jj,DePies:2007bm}.

A heuristic derivation of the gravitational radiation power can be obtained by using the quadrupole formula~\cite{Vilenkin:1981bx}.
\begin{equation}
\frac{dE}{dt}  \sim  G \left( \frac{d^3 I}{dt^3} \right)^2  \sim  G m^2 \ell^4 \omega^6  \ ,
\end{equation}
where $I \sim m \ell^2$ is the quadrupole moment, $m \sim \mu \ell$ is the mass of the loop and $\omega \sim \ell^{-1}$ is the characteristic frequency for a loop of size $\ell$.
Therefore, the power radiated in gravitational waves is expected to be roughly
\begin{equation}
P  =  \Gamma G\mu^2  \ .
\label{powrad}
\end{equation}
Thus, the rate of energy radiated by small loops is proportional to $G\mu^2$, just like for the long strings.
The proportionality coefficient $\Gamma$ depends on the trajectory of the loop in question but not on its overall size.

The validity of this heuristic argument can be questioned since the quadrupole formula ceases to be accurate in the presence of relativistic motion of the source, which is the typical case for oscillating small loops.
However, more careful treatments confirm the result~(\ref{powrad}).
The values suggested by numerical~\cite{Vachaspati:1984gt,Burden:1985md} and analytic~\cite{Garfinkle:1987yw,Allen:1994bs} studies are $\Gamma \sim 50$.

Since a loop of length $l$ has energy $\mu l$, equation~(\ref{powrad}) implies that loops shrink at a rate
\begin{equation}
\frac{dl}{dt} = - \Gamma G\mu  \ ,
\label{gravrad}
\end{equation}
and so a loop which formed with an initial length $l_i$ has a lifetime given by $\tau = (\Gamma G\mu)^{-1} l_i$, if gravitational radiation is the principal decay mode.
It is customary to parameterize the size of small loops as $l = \varepsilon \Gamma G\mu t$.
Thus, the parameter $\varepsilon$ corresponds to the lifetime of the loop in units of the FRW time.
Given the power law distribution for the small loop production found in Chapter~4 we expect a large population of loops accumulating at the gravitational radiation scale.
Therefore, recalling equation~(\ref{lsize}) from Chapter~3, we have found 
\begin{eqnarray}
\epsilon  & \approx &  0.4 (G\mu)^{0.5}  \quad\quad {\mbox {\rm in the matter era}}  \ ,  \\
\epsilon  & \approx &  0.4 (G\mu)^{0.2}  \quad\quad {\mbox {\rm in the radiation era}}  \ .
\end{eqnarray}

For loops containing ideal cusps the angular distribution of the power radiated is divergent along the direction of motion of the cusp~\cite{Vachaspati:1984gt}.
Nevertheless, the divergence is integrable and the total power remains safely finite.
Gravitational radiation is expected to round off the tip of the cusp on very short scales and so it never really reaches the speed of light but comes quite close.
The angular distribution of the power radiated is then bounded from above but this effect can still lead to very strong bursts of gravitational waves, which can be observed, in principle, at GW detectors~\cite{Damour:2001bk,Damour:2004kw,Siemens:2006vk}.
A similar idea will be exploited in Chapter~6, but instead of the high frequencies typical of cusps we shall consider the lowest harmonics of the loops which get Lorentz boosted by the high velocities of small loops, a subject to which we now turn our attention.

\subsection{Loop velocities and cusps}

For the phenomenology of the small loops there is another important property, first noted in Ref.~\cite{Polchinski:2006ee}: they move {\it extremely} fast.
The point is that the functions ${\bf p}_+(u)$ and ${\bf p}_-(v)$ must have equal mean values on the loop (the condition for the loop to form).
If the loop is small, so that these functions have limited range, this implies that both remain near one point on the sphere, and the velocity (the average of ${\bf p}_+(u)$ and ${\bf p}_-(v)$ over the loop) will be very close to unity.
This is easy to understand since the whole loop is formed from a small region around a cusp and such points approach the speed of light.
From Eq.~(\ref{vsquared}), the Lorentz contraction factor is
\begin{equation}
\gamma = (l/t)^{-\chi} [( 1+\chi)(1+ 2\chi) / {\mathcal A} ]^{1/2} \ .
\end{equation}
This is $\approx 0.8 (G\mu)^{-0.37}$ in the matter era, and is of order $10^3$ for tensions of interest;  in the radiation era it is $\approx 1.0 (G\mu)^{-0.12}$ and is of order $10^1$.
Since this factor is so large, we must be careful to recall that the specific definition of length is
\begin{equation}
dl = a(t) \epsilon(t) d\sigma =  a(t) \sqrt{ {\bf x}'^2 / (1-\dot{\bf x}^2) }  d\sigma
\end{equation}
and so $l$ is $\gamma$ times the rest frame length $l_{\rm rest}$.
It is equal to twice the inverse period of the loop in the FRW rest frame.

To see one effect of the boost, in Ref.~\cite{Siemens:2006yp} the total stochastic gravitational wave spectrum from small loops was considered.
The frequency spectrum simply scales inversely as the size of the loops if they are at rest, $\omega \propto 1/l_{\rm rest}$.
In the FRW rest coordinate this frequency is reduced by a factor of $\gamma$.
However, the large boost pushes essentially all of the radiation into the forward direction, where the frequency is enhanced by a factor $(1-v)^{-1} \approx 2 \gamma^2$.
Thus the gravitational radiation spectrum corresponds to an effective loop size
\begin{equation}
\varepsilon_{\rm eff} \approx \gamma^{-1} \varepsilon_{\rm rest}/2  \approx \gamma^{-2} \varepsilon/2  \ .
\label{effeps}
\end{equation}
This is $\varepsilon_{\rm eff} \approx 0.3 (G\mu)^{1.25}$ in the matter era, meaning that $l_{\rm eff} \approx 15 (G\mu)^{2.25} t$.
In the radiation era $\varepsilon_{\rm eff} \approx 0.2 (G\mu)^{0.44}$, meaning that $l_{\rm eff} \approx 10 (G\mu)^{1.44} t$.
We should note that Ref.~\cite{Siemens:2006yp} considered the sensitivity in the whole $(G\mu , \varepsilon)$ plane, and we are simply highlighting one line in this plane.
It is notable that Advanced LIGO is more sensitive than LISA to these smallest loops, because their small size puts them above the LISA frequency range.
In fact, for interesting values of the tension the loops will produce a periodic signal in the LIGO frequency band, which may be observable at Advanced LIGO.
Such a study has been conducted in~\cite{Dubath:2007wu} and we reproduce it in Section~\ref{PERIODICGW}.

Recall further that the cusp region is the site of rapid loop production, so the many small cusps surrounding a crossing of the long distance parts of ${\bf p}_\pm$ will actually end up on small loops, disconnected from the long string.
Therefore, we expect a high tendency of the small loops to contain cusps.
In loop trajectories with no kinks, the curves ${\bf p}_\pm$ are smooth and must be periodic, of course.
Hence, if they intersect this must happen an even number of times, and for the simplest motions this gives two cusps per loop oscillation~\cite{Turok:1984cn}.
However, we have seen that small loops should be dominated by large kinks and we expect them typically to possess one cusp per period.
It might seem paradoxical that the cusp size~(\ref{cusps}) is parametrically larger than the loop size~(\ref{lsize}), but this is another Lorentz effect: in the loop rest frame these sizes are of the same order.

\subsection{A model loop\label{MODELLOOP}}

In this section we will display explicitly the trajectories of typical small loops as predicted by the arguments of Chapter~4, where loop formation was studied.
Recall that the picture is that, due to the high fractal dimension of the curves ${\bf p}_\pm$ on the unit sphere, large cusps break into many small cusps, each one having a tendency to form small loops around them.

For such a ``model'' loop we take the functions ${\bf p}_\pm$ to be short straight arcs on the unit sphere:
\begin{eqnarray}
p_+^\mu(u) & = & \left(1,\sin(Vu),0,\cos(Vu)\right) \simeq \left(1,Vu,0,1-V^2u^2/2\right) \ ,  \nonumber \\
p_-^\mu(v) & = & \left(1,0,\sin(Vv),\cos(Vv)\right) \simeq \left(1,0,Vv,1-V^2v^2/2\right) \ .
\label{modelloop}
\end{eqnarray}
These definitions hold for $-\beta < u,v <\beta$, where $2\beta$ is the length of the loop.
However, these functions can be extended to the whole $(u,v)$-plane by imposing $2\beta$-periodicity.
The jumps at $u,v = \beta + 2 \mathbb {Z} \beta$ correspond to kinks.
The velocity of the cusp, $V$, and the period of the small loops, $\beta$, were determined in Chapter~3 as average values over the ensemble of small loops.
For a matter-dominated era these quantities are given by~\cite{Polchinski:2007rg}
\begin{eqnarray}
V & = & 0.08 (G\mu)^{-1.12}t^{-1} \ , \\
\beta & = & 10 (G\mu)^{1.5}t \ ,
\end{eqnarray}
where $t$ is the FRW time.

The embedding of the loop worldsheet in target space is given by
\begin{equation}
x^\mu_{\rm w.s.}(u,v) = \frac{1}{2}\int p_+^\mu(u)du+\frac{1}{2}\int p_-^\mu(v)dv  \ ,
\label{x/p}
\end{equation}
and is therefore aperiodic as a function of $u$ or $v$ separately:
\begin{eqnarray}
x^\mu_{\rm w.s.}(u+2\beta,v) & = & x^\mu_{\rm w.s.}(u,v+2\beta) = x^\mu_{\rm w.s.}(u,v) + \left(\beta,0,0,\sin(V\beta)/V\right)  \nonumber \\
& \approx & x^\mu_{\rm w.s.}(u,v) + \beta \left(1,0,0,1-V^2\beta^2/6\right)  \ .
\label{periodicity}
\end{eqnarray}
However, from the above equation it follows immediately that $x^\mu_{\rm w.s.}$ is indeed periodic in $\sigma$, with periodicity $2\beta$.

One can use equations~(\ref{modelloop}) and~(\ref{x/p}) to obtain snapshots of the loop as it evolves in time.
For a fixed time $t_*$, considered to belong to the interval $\left[0,\beta\right]$ for concreteness, the $v$ coordinate may be written in terms of the $u$ coordinate as $v_u = 2t_* - u$ and then the string is parametrized by $u$ alone.
If $u \in \left[2t_*-\beta,\beta \right]$,
\begin{equation}
{\bf x}_{\rm w.s.}(u) = \frac{1}{2V} \left( -\cos(V u) , -\cos(V v_u) , \sin(Vu) + \sin(V v_u) \right)  \ .
\end{equation}
On the other hand, if $u \in \left[\beta,2t_*+\beta \right]$ we define $u'_u = u - 2\beta \in \left[-\beta,2t_*-\beta \right]$ so that
\begin{equation}
{\bf x}_{\rm w.s.}(u) = \frac{1}{2V} \left( -\cos(V u'_u) , -\cos(V v_u) , \sin(V u'_u) + \sin(V v_u) + 2\sin(V \beta) \right)  \ .
\end{equation}
\begin{figure}[t]
\begin{center}
\includegraphics[width=6cm]{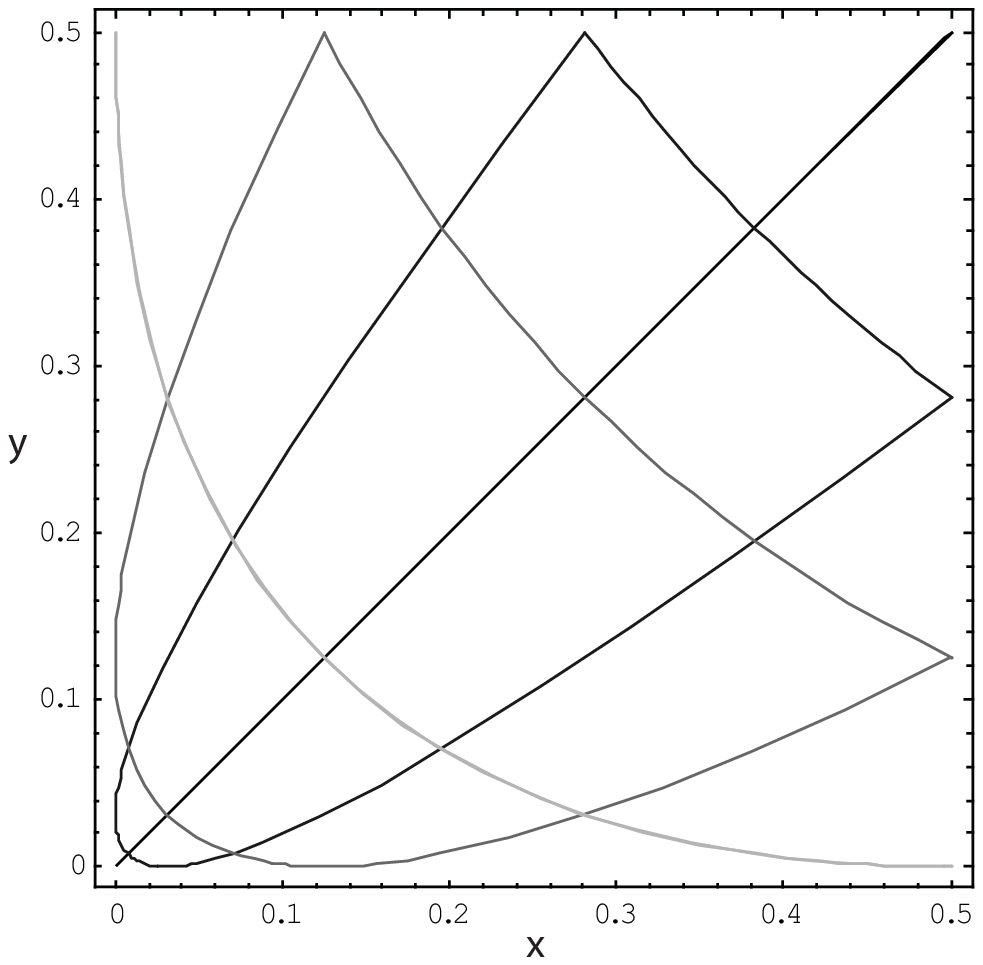}~\includegraphics[width=7cm]{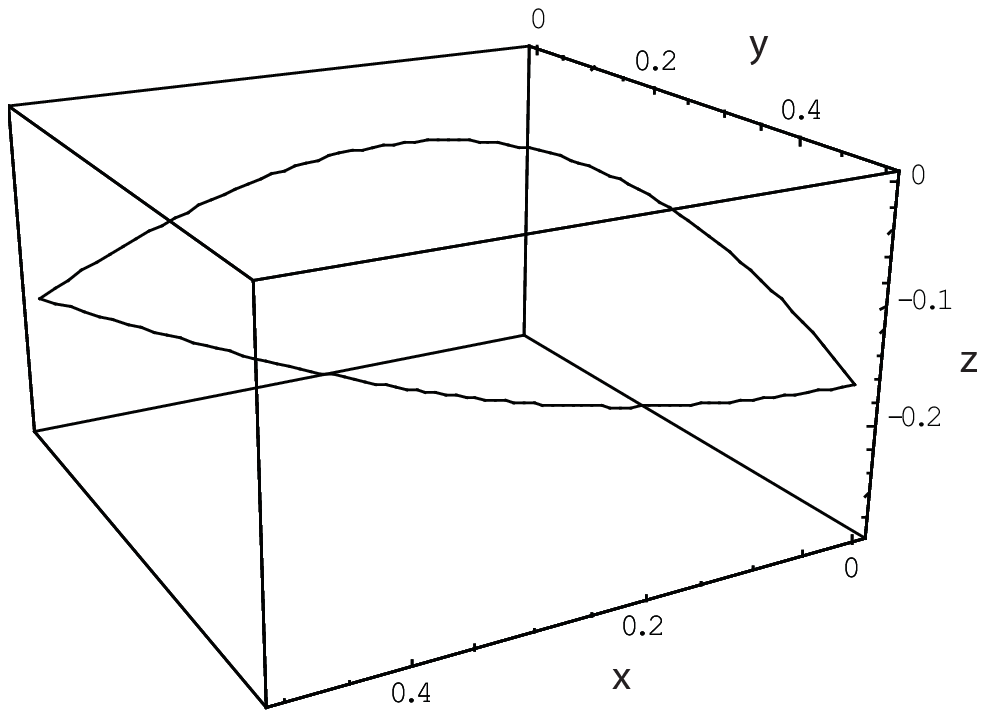}\\
\caption{The left panel shows the $xy$-projection of four superposed snapshots of the model loop considered.  For clarity a total translation by $(1,1)$ was performed and the $x$- and $y$-axis are in units of $(V\beta)^2$.  The right panel gives a 3D view of the model loop at one of the two moments (the lightest curve in the left panel) at which the $xy$-projection is degenerate.  The $z$-axis was translated by $-V\beta$ and is in units of $(V\beta)^3$.}
\label{loop}
\end{center}
\end{figure}
Figure~\ref{loop} shows several snapshots of the model loop~(\ref{modelloop}).
Strictly speaking, there is no self-intersection but at $t=0$ (or equivalently when $u+v=0$) the string actually folds back on itself.  The discontinuities of ${\bf p}_\pm$ at odd multiples of $\beta$ correspond to kinks on the loop traveling in opposite directions.
When they meet the loop becomes degenerate and at the same time a cusp develops at the other end.
However, adding a small (quadratic) perturbation generically renders it non-self-intersecting while the cusp is preserved.
Thus, small cosmic loops are expected to be generally stable against fragmentation.

Small loops with sizes of order $2\beta$ have very large Lorentz factors.
To show this, consider the impulsion of the loop, given by 
\begin{equation}
P^\mu(u,v) = \frac{1}{2}\left(p_+^\mu(u)+p_-^\mu(v)\right)  \ ,
\end{equation}
and the center-of-mass velocity, which is obtained by averaging the spatial part of $P^\mu(u,v)$ over the worldsheet coordinates,
\begin{equation}
{\bf v} \simeq \int_0^\beta \frac{dt}{\beta} \int_{0}^{2\beta} \frac{d\sigma}{2\beta} \dot{\bf x}_{\rm w.s.}(t,\sigma) 
= \int_{-\beta}^\beta \int_{-\beta}^\beta \frac{du dv}{(2\beta)^2} {\bf P}(u,v) 
= \left(0,0,1-\frac{(V\beta)^2}{6} \right)  \ .
\end{equation}
Therefore the boost factor is given by
\begin{equation}
\gamma \simeq \frac{\sqrt3}{V\beta}  \ .
\label{boost}
\end{equation}
For $G\mu = 10^{-9}$ the Lorentz factor~(\ref{boost}) is of order $10^{3}$, as we have already seen.
In its rest frame the $n$th harmonic will have frequency $f^{\rm rest}_n=\gamma\beta^{-1}n$, since $\beta$ is its time periodicity in the FRW frame.
Therefore, the observed frequency for a loop moving at an angle $\theta$ with respect to the line of sight is
\begin{equation}
f_n = \frac{f^{\rm rest}_n}{\gamma\left(1-|{\bf v}|\cos\theta \right)}  \simeq \frac{n}{\beta\left(1-\left(1-\frac{V^2\beta^2}{6}\right)\cos\theta \right)}  \ .
\label{harmonics}
\end{equation}
In the special case of a loop whose motion is exactly aligned with the line of sight we see that the observed frequency is boosted by a factor proportional to $\gamma^2$.
This effect can bring the lowest harmonics into the LIGO frequency band for $G\mu\lsim 10^{-9}$, i.e. values of the string tension not yet ruled out by observations, and so the potential for detection arises.
We shall consider such a possibility in the next chapter, but now let us address the scaling properties of the distribution of loops over their lengths.

\section{Scaling of the loop number density\label{SCALINGLOOPS}}

In this section we will study the evolution of the loop number density distribution.
The equation governing the evolution of the number density of loops in a cosmic string network is a detailed balance determined by conservation of energy.
In what follows we solve this equation, first ignoring gravitational radiation and then taking into account this effect.
But first let us make a few preliminary remarks and introduce some notation.

\subsection{Preliminaries\label{PRELIM2}}

Recall the many processes that can act to change the number density of loops with a given length $l$.
Expansion of the universe causes the strings to stretch, but only on scales larger than the horizon size~\cite{Vilenkin:1981kz}.
Therefore loops essentially do not grow in time.\footnote{Loops bigger than the horizon distance are regarded as long strings.}  Self-intersections of cosmic strings can produce loops from long strings and also fragment loops into smaller ones.
Of course, string intercommutation can lead to the absorption of loops back into the long string population but this process is strongly suppressed for small loops~\cite{Bennett:1989yp} and so can be neglected.
Finally, the coupling of matter in the form of cosmic strings to gravity means that the network radiates away part of its energy.
This has been amply discussed in the literature~\cite{Turok:1984cn,Vachaspati:1984gt,Garfinkle:1987yw,Siemens:2001dx} and as a consequence loops shrink as time progresses and eventually disappear from the network.

Obviously, the rate of loop production plays a crucial role in the detailed balance: it persistently feeds the loop population.
These get diluted by the expansion of spacetime.
On top of these effects, gravitational radiation continuously shifts the distribution toward smaller lengths as time elapses.
In Chapter~4 we found a power law for the loop production function.
The potential divergence at small distances is supposedly cut off at the gravitational smoothing length scale.
However, we will ignore the existence of this minimum length for the loops in what follows and see how far we can get.
In addition, it has been suggested in (a fraction of) the existing literature that gravitational radiation is unnecessary to achieve scaling~\cite{Ringeval:2005kr,Martins:2005es}.
The results of the following section confirm this hypothesis.

Let us now introduce the necessary notation for the rest of this chapter.
Denote by $dn(l,t)$ the number density of loops with size comprised between $l$ and $l+dl$ at a given time $t$.
Then the quantity $\frac{dn}{dl}$ has units of $({\rm length})^{-4}$.
Thus, if it ever reaches a scaling regime during the cosmological evolution, it must eventually approach the following form:
\begin{equation}
\frac{dn}{dl} = t^{-4} f(l/t)  \ .
\end{equation}
Borrowing notation from~\cite{Ringeval:2005kr}, define the length of loops in units of the horizon size, $\alpha\equiv l/d_{\rm h}$.
If the scale factor takes the form $a(t)\propto t^\nu$, the horizon size can be expressed as $d_{\rm h}=t/(1-\nu)$ and so the signature of a scaling regime in the loop number density is a solution of the form
\begin{equation}
\frac{dn}{d\alpha} = \frac{{\mathcal S}(\alpha)}{\alpha \, d_{\rm h}^3}  \ .
\label{scaling}
\end{equation}
In~\cite{Ringeval:2005kr} ${\mathcal S}(\alpha)$ was dubbed the {\em scaling function}.

\subsection{Evolution neglecting gravitational radiation}

As discussed in Section~\ref{DISCUSS2}, if gravitational radiation is neglected the number of loops within a comoving volume changes only due to loop production~\cite{Polchinski:2006ee}:
\begin{equation}
\frac{d}{dt} \left(a^3 \frac{dn}{dl}\right) = \frac{c\, a^3}{\gamma^2 t^2 l^3} \left(\frac{l}{t}\right)^{2\chi}  \ ,
\end{equation}
Defining $F(l,t)\equiv\frac{dn}{dl}(l,t)$ and inserting the power-law expression for the scale factor we obtain
\begin{equation}
t \dot F + 3\nu F = \frac{c}{\gamma^2 l^4} \left(\frac{l}{t}\right)^{1+2\chi}  \ ,
\label{evol}
\end{equation}
which has the general solution
\begin{equation}
F(l,t) = \frac{c}{\gamma^2(3\nu - 1 - 2\chi) t^4} \left(\frac{l}{t}\right)^{2\chi-3} + \frac{\mathcal{G}_0(l)}{t^{3\nu}}  \ .
\end{equation}
The function $\mathcal{G}_0$ depends only on the variable $l$.
Noting that the inequality $1+2\chi < 3\nu$ is satisfied both in the radiation- and matter-dominated eras, we conclude that at late times ($t\rightarrow\infty$) the loop number density approaches
\begin{equation}
\frac{dn}{dl}(l,t) \longrightarrow \frac{c}{\gamma^2(3\nu - 1 - 2\chi) t^4} \left(\frac{l}{t}\right)^{2\chi-3}  \ .
\end{equation}
Thus, expressing everything in terms of $\alpha$ and $d_{\rm h}$ we indeed find that the loop number density approaches a scaling regime, i.e. it takes the form~(\ref{scaling}) with 
\begin{equation}
{\mathcal S}(\alpha) = \frac{c\, (1-\nu)^{-1-2\chi}}{\gamma^2 (3\nu-1-2\chi)} \alpha^{2\chi-2}  \ .
\label{sol1}
\end{equation}
This confirms that the loop number density does approach a scaling regime without taking into account gravitational radiation.
The exponent $2\chi-2 \equiv -p$ takes the values $-1.5$ in the matter era and $-1.8$ in the radiation era.
As we have already mentioned in Chapter~4, this is in good agreement with the numerical results of~\cite{Ringeval:2005kr} who quote $p_{mat} = 1.41^{+0.08}_{-0.07}$ and $p_{rad} = 1.60^{+0.21}_{-0.15}$.
However, these simple power-laws become good fits to the data only above a physical length $\ell_{\rm c}$, which was identified in~\cite{Ringeval:2005kr} with the initial correlation length of the network.
Furthermore, if the solution~(\ref{sol1}) were valid over the full range of loop lengths the total energy density would diverge in the UV since
\begin{equation}
\int l \frac{dn}{dl} dl = d_{\rm h}^{-2} \int {\mathcal S}(\alpha) d\alpha  \ .
\label{total}
\end{equation}
We will now show that including the process of shrinkage of the loops due to gravitational radiation changes the power-law below the gravitational radiation scale, thus yielding a convergent total energy density in loops.

\subsection{Evolution including gravitational radiation}

Inclusion of gravitational radiation into the evolution equation for the loop number density introduces an extra term on the left-hand side of equation~(\ref{evol}) because loops shrink at a rate given by~(\ref{gravrad}).
Therefore, we now have
\begin{equation}
t \dot F + 3\nu F - \Gamma G \mu t F' = \frac{c}{\gamma^2 l^4} \left(\frac{l}{t}\right)^{1+2\chi}  \ .
\end{equation}
Defining for convenience $b \equiv \Gamma G \mu$, the solution of the above differential equation may be written as
\begin{eqnarray}
F(l,t) & = & \frac{c\, b}{\gamma^2} \frac{(l+bt)^{2\chi-4}}{t^{2\chi}} \left(\frac{bt}{l+bt}\right)^{2\chi-3\nu}
 B\left( \frac{bt}{l+bt}; 3\nu-1-2\chi, 2\chi-2 \right) \nonumber \\ [10pt]
&&  \hspace{30pt}  + \: \frac{\mathcal{G}(l+bt)}{t^{3\nu}}  \ ,
\label{sol2}
\end{eqnarray}
where $B$ represents the Euler incomplete beta function and $\mathcal{G}$ can be any function of the combination $l+bt$.

By employing the Taylor expansion
\begin{equation}
B\left( \epsilon; 3\nu-1-2\chi, 2\chi-2 \right) = \frac{\epsilon^{3\nu-1-2\chi}}{3\nu-1-2\chi} + O(\epsilon)  \ ,
\label{taylor}
\end{equation}
we find that the first term in~(\ref{sol2}) behaves, for large $l$ and fixed $t$, as $\sim l^{2\chi-3}$, whereas the second term goes like $\sim \mathcal{G}(l)$.
Requiring that the energy density in loops converges in the IR limit $l \rightarrow \infty$ imposes that the general function $\mathcal{G}(x)$ decays faster than $x^{-2}$.
Therefore, using the expansion
\begin{equation}
B \left( 1-\epsilon; 3\nu-1-2\chi, 2\chi-2 \right) =
 - \, \frac{ \pi\csc(2\pi\chi)\epsilon^{2\chi-2} }{ \Gamma(2\chi-1)\Gamma(3-2\chi) } 
 + \, O\left(\frac{1}{\epsilon}\right)  \ ,
\label{taylor2}
\end{equation}
the second term in~(\ref{sol2}) is dominated by the first term as $t \rightarrow \infty$:
\begin{equation}
\frac{dn}{dl}(l,t)  \longrightarrow  \frac{c\, b}{\gamma^2} \frac{(l+bt)^{2\chi-4}}{t^{2\chi}} \left(\frac{bt}{l+bt}\right)^{2\chi-3\nu}
 B\left( \frac{bt}{l+bt}; 3\nu-1-2\chi, 2\chi-2 \right)  \ .
\end{equation}
Once again converting to the variables $\alpha$ and $d_{\rm h}$, we obtain a solution of the form~(\ref{scaling}) with
\begin{equation}
{\mathcal S}(\alpha) = \frac{c\, b^{2\chi-3} \alpha}{\gamma^2 (1-\nu)^4} \left(\frac{\alpha+(1-\nu)b}{(1-\nu)b}\right)^{3\nu-4} 
 B\left( \frac{(1-\nu)b}{\alpha+(1-\nu)b}; 3\nu-1-2\chi, 2\chi-2 \right)  \ .
\end{equation}

Now we can use the series expansions~(\ref{taylor}) and~(\ref{taylor2}) to recover the limits for `small' and `large' loops.
The separation between these two regimes is set by the gravitational radiation scale $\Gamma G \mu$, and we find for $\alpha \gg \Gamma G \mu$ the same result~(\ref{sol1}), whereas for $\alpha \ll \Gamma G \mu$
\begin{equation}
{\mathcal S}(\alpha)  \simeq  \frac{ \pi(1-2\chi)\csc(2\pi\chi) }{ \Gamma(2\chi)\Gamma(3-2\chi) }
 \frac{c}{\gamma^2 \Gamma G \mu (1-\nu)^{2+2\chi}} \alpha^{2\chi-1}  \ .
\end{equation}
Since $0 < 2\chi < 1$ holds in both cosmological eras, the integral~(\ref{total}) is manifestly convergent and so the total energy density in loops is finite, as desired.

Thus, taking into account gravitational radiation has the effect of bending the curves for $\alpha \lsim \Gamma G \mu$ so that the exponent in the scaling function becomes $2\chi-1$.
Indeed, Figure~3 of that Ref.~\cite{Ringeval:2005kr} does appear to show a certain range of the parameter $\alpha$ in which the scaling function behaves as such a power-law.
At first sight this might seem intriguing since those simulations did not include the gravitational radiation process directly.
However, small loops behave like matter and so the expansion of the universe effectively shrinks the loops.
Because the simulations only keep loops with sizes greater than a fixed fraction of the horizon, they are eventually removed from the game, hence emulating gravitational radiation.
There, the minimum counting size was $\alpha_{min} = 10^{-5}$ so we should expect the sub-gravitational radiation regime to set in for comparable scales or smaller.
A more accurate estimate, equating the lifetime of the loops determined by gravitational decay to the lifetime set by the minimum counting size, yields $\alpha_{bend} = \frac{2-\nu}{1-\nu}\alpha_{min}$ for the scale at which the bending would occur.
Nonetheless, it is curious that this sub-gravitational radiation regime shows up also at early times in the simulations, when scaling is yet to be achieved.

In conclusion, we have shown that under the very mild assumption that the energy density in cosmic string loops converges in the infra-red (IR), then the loop distribution evolves toward a scaling configuration in which it is also free from divergences in the UV as long as the process of loop decay by GW emission is considered.
Note that the inclusion of gravitational radiation in our equations leads to a finite energy density of loops, even though the loop production function used as an input diverges at small scales.
This just means that the rate at which loops are removed from the network is sufficiently high to balance the diverging loop formation.

\chapter{Some implications for observation}

This chapter is concerned with observational signatures of cosmic strings.
The tangible effects of these networks often depend on their detailed properties and the methods developed in previous chapters allow us to address some of these issues.

We start by considering the implications of our model for the lensing by cosmic strings in Section~\ref{LENS}.
In this context we study both the typical distortion of the double images produced and the alignment of lensed pairs.
The latter is related to the non-Gaussianity of the string distribution.
We devote Section~\ref{PERIODICGW} to the possibility of detecting (quasi-)periodic GW at the Laser Interferometer Gravitational Wave Observatory (LIGO) from the highly boosted population of small loops.

This chapter is mostly drawn from Refs.~\cite{Polchinski:2006ee,Dubath:2007wu}, with Joseph Polchinski and Florian Dubath, respectively.

\section{Lensing\label{LENS}}

Let us now consider the effect of the small scale structure on the images produced by a cosmic string lens.
A well known fact is that a straight string of tension $\mu$ has the sole effect of introducing a deficit angle in the transverse dimensions, equal to $8 \pi G\mu$~\cite{Vilenkin:1981zs}, and this can induce identical images of sources in the background.
However, once we include the possibility of small fluctuations propagating along the cosmic string the two copies will become distorted.

Previous work has discussed possible dramatic effects~\cite{deLaix:1997dj,Bernardeau:2000xu}, including multiple images and large distortions.
We can anticipate that the rather smooth structure that we have found, which again we note is subject to our assumptions, will produce images with only small distortion.
We will use our stretching model of the two-point function.  If this proves incorrect one could apply the analysis using phenomenological values of $\chi$ and ${\mathcal A}$; for example, the extrapolation of the results of Ref.~\cite{Martins:2005es} give a smoother string, and even less distorted images.

It is worth mentioning a recent occurrence that led to some excitement in the field.
In 2003 a candidate for a cosmic string lens, CSL-1, was identified in the sky~\cite{Sazhin:2003cp}.
The two images of the putative lensed pair were identified with galaxies having apparently equal redshifts, morphologies and spectra.
Furthermore, their angular separation placed the cosmic string tension at the higher end of the permitted range and several other potential candidates were identified nearby.
Alas, a closer investigation with the Hubble Space Telescope revealed that the two images indeed corresponded to distinct galaxies~\cite{Sazhin:2006fe}, showing no sign of a sharp discontinuity and the two images appearing distorted relative to each other.

\subsection{Distortion of images}

\begin{figure}[t]
\center \includegraphics[width=20pc]{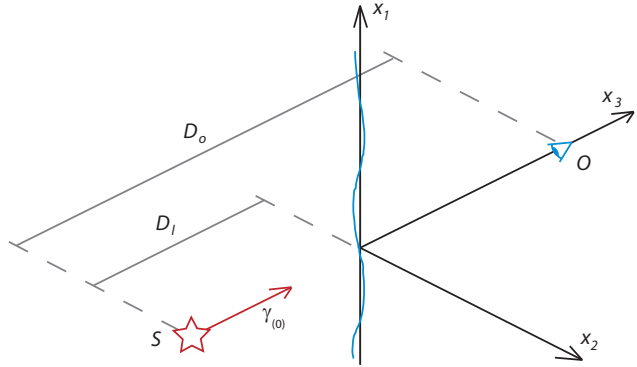}
\caption{Schematic representation of the system considered, with the string lens displayed along the $x_1$-axis and the distant source and the observer located at points $S$ and $O$ respectively.}
\label{schematic}
\end{figure}

We quote the result of ref.~\cite{deLaix:1996vc} for the angular deflection of a light ray by a string,
\begin{equation}
\mbox{\boldmath $\gamma$}_\perp({\bf y}_{\perp})
= 4G\mu \int d\sigma \Biggl[ \frac{F_{\mu\nu} \gamma^\mu_{(0)} \gamma^\nu_{(0)}}{1 - \dot x_\parallel} 
\frac{ {\bf x}_\perp - {\bf y}_{\perp} }{ |{\bf x}_\perp - {\bf y}_{\perp}|^2 } \Biggr]_{t = t_0(\sigma)}  \ .
\end{equation}
Here $\gamma_{(0)}^\mu$ is the four-velocity of the unperturbed light ray, which we take to be $(1,0,0,1)$ as shown in Fig.~\ref{schematic}, $y^\mu$ is a reference point on this ray, and subscripts $\perp$ and $\parallel$ are with respect to the spatial direction of the ray.
Also, $x^\mu(\sigma,t)$ is the string coordinate,\footnote{The region where the the light ray passes the string is small on a cosmological scale, so to use our earlier results we can locally set $a = \epsilon = 1$, $dl = d\sigma$, $\partial_t = \partial_\tau$.}
in terms of which
\begin{equation}
F^{\mu\nu} = \dot x^\mu \dot x^\nu - x^{\mu\prime} x^{\nu\prime} - \frac{1}{2}\eta^{\mu\nu}
(\dot x^\rho \dot x_\rho - x^{\rho\prime} x_\rho^{\prime})  \ ,
\end{equation}
and $t_0(\sigma)$ is defined by $t_0(\sigma) = x_3(\sigma,t_0) - y_3$.

In keeping with Section~\ref{SMALLFLUC} we separate the string locally into a straight part and a fluctuation; we will keep the deflection only to first order in the fluctuation.
We consider here only the simplest geometry, in which the straight string is perpendicular to the light ray and at rest, so that ${\bf P}_+ = -{\bf P}_- = (1,0,0)$.
To first order in the fluctuation, $x^\mu(\sigma,t) = (t,\sigma,x_2(\sigma,t),x_3(\sigma,t))$.
One then finds
\begin{eqnarray}
\gamma_1({\bf y}_{\perp}) &=& 4 G \mu \int d\sigma\, \frac{- \dot x^{\vphantom2}_3(\sigma,t_0) (\sigma - y_1) + x_2^{\prime}(\sigma,t_0) y_2}{(\sigma - y_1)^2 + y_2^2}  \ ,
\nonumber\\
\gamma_2({\bf y}_{\perp}) &=& - {\rm sgn}(y_2) \Delta + 4 G \mu \int d\sigma\, \frac{\dot x^{\vphantom2}_3(\sigma,t_0) y_2 + x_2^{\prime}(\sigma,t_0) (\sigma - y_1)}{(\sigma - y_1)^2 + y_2^2}  \ ,
\end{eqnarray}
with $\Delta = 4\pi G\mu$ being half the deficit angle of the string.
To the order that we work $t_0$ is a constant, corresponding to the time when the light ray is perpendicular to the straight string.

We can use our results for the small scale structure to calculate the two-point functions of the deflection.
Let us focus on the local magnifications parallel and perpendicular to the string, given by basic lensing theory as
\begin{eqnarray}
M_1 ({\bf y}_{\perp}) &=& 1 - \frac{D_l (D_o - D_l)}{D_o} \frac{\partial\gamma_1}{\partial y_1} ({\bf y}_{\perp})  \ , \nonumber\\
M_2 ({\bf y}_{\perp}) &=& 1 - \frac{D_l (D_o - D_l)}{D_o} \frac{\partial\gamma_2}{\partial y_2} ({\bf y}_{\perp})  \ .
\end{eqnarray}
Here $D_o$ and $D_l$ are the distances of the source from the observer and the lens respectively (these would be the angular diameter distances on cosmological scales).
It is particularly interesting to consider the differential magnifications for the two images produced by a string,
\begin{equation}
\delta M_1 = M_1 ({\bf y}_{\perp}) - M_1 ({\bf y}'_{\perp})  \ ,  \quad
\delta M_2 = M_2 ({\bf y}_{\perp}) - M_2 ({\bf y}'_{\perp})  \ .
\end{equation}
We take for simplicity ${\bf y}_{\perp} = (0, b)$ and ${\bf y}'_{\perp} = (0, -b)$ for $b = \Delta D_l (D_o - D_l)/D_o$; this corresponds to the symmetric images of an object directly behind the string.
Then
\begin{equation}
\delta M_1 = - \delta M_2 = -\frac{2b^2}{\pi} \int d\sigma\, x_2^{\prime}(\sigma,t_0) \partial_\sigma \left( \frac{1}{\sigma^2 + b^2} \right)  \ .
\end{equation}

From Section~\ref{SMALLFLUC} we obtain
\begin{eqnarray}
\left\langle {\bf w}_+ (\sigma,t) \cdot {\bf w}_+(\sigma',t) \right\rangle
&=& \left\langle {\bf w}_- (\sigma,t) \cdot {\bf w}_-(\sigma',t) \right\rangle
= 4  \left\langle x_2'(\sigma,t) x_2'(\sigma',t) \right\rangle          \nonumber\\
&=& {\mathcal A} \Bigl\{  (\sigma/t)^{2\chi} + (\sigma'/t)^{2\chi} - ([\sigma - \sigma']/t)^{2\chi} \Bigr\}  \ .
\label{u2p}
\end{eqnarray}
Recall that Eq.~(\ref{upup}) does not fully determine the two-point functions~(\ref{u2p}), and in fact that the latter cannot be translation invariant.
Here, we have fixed the ambiguity by defining the expectation value to vanish when $\sigma$ or $\sigma'$ vanishes (that is, at the point on the string nearest to the light ray); this amounts to a choice of how one splits ${\bf p}_\pm$ into ${\bf P}_\pm$ and ${\bf w}_\pm$.

As is stands, equation (\ref{u2p}) ignores the transition from a radiation dominated evolution to a matter dominated cosmology.
In effect, if we take a small segment of string (in the matter era) and evolve it backward in time it will grow relative to the horizon size and eventually reach it, as we have alluded to before.
Therefore, if the segment under consideration is short enough, it would have been comparable to the horizon distance before the radiation-to-matter transition, at a time $t_{\rm eq}$, where the string dynamics is determined by the radiation era values.
Hence, there is some critical length $l_{\rm c}$ and for $|\sigma -\sigma'|$ shorter than this one must take into account the transition.
The details of this effect are collected in Appendix~B. 
According to equation (\ref{radmat}), the right-hand side of (\ref{u2p}) just gets multiplied by a power of $t/t_{\rm eq}$.
From $\left\langle x_2' x_2' \right\rangle$ we obtain\footnote{The lensing scale, which is provided by $b$, is typically of the order of $10^{-7}d_H$, thus much smaller than the critical length $l_{\rm c}$.  We can then safely use the formula for $\left\langle x_2' x_2' \right\rangle$ valid for the smallest scales over the full range of integration.  Corrections from the longest scales will increase the final result, but not significantly.}
\begin{equation}
\left\langle \delta M_1^2 \right\rangle
= \left\langle \delta M_2^2 \right\rangle
= \frac{\chi_{\rm r}(1-2\chi_{\rm r})}{2 \cos (\pi\chi_{\rm r})} {\mathcal A}_{\rm r} (2b/t_0)^{2\chi_{\rm r}}(t_0/t_{\rm eq})^{-2\zeta_{\rm m}
- 2\chi_{\rm r}\zeta_{\rm m} + 2\chi_r}  \ .
\end{equation}
Plugging in the numeric values for $\chi$, ${\mathcal A}$ and $\zeta$ and using a representative value for the dimensionless parameter $G\mu \sim 10^{-7}$, we obtain a RMS differential magnification slightly below $1\%$:~\footnote{For this sample calculation the source was taken to be at a redshift of $z \simeq 2.5$ and the string was placed midway between the observer and the source.  The values $t_0 = 10 {\rm Gyr}$ and $t_{\rm eq} = 0.10 {\rm Myr}$ were used for the intervening cosmological times.}
\begin{equation}
\left\langle \delta M^2 \right\rangle^{1/2} \simeq 0.009  \ .
\end{equation}
Considering a smaller value of $G\mu$ reduces the final answer.
In any case, the take-away lesson is that the smoothness of the strings predicted by our model implies very small differential magnifications between the two images produced by a long cosmic string.

\subsection{Alignment of lenses and non-Gaussianity}

Another question related to short-distance structure is the alignment of lenses.
Suppose we see a lens due to a long string, with a certain alignment.
Where should  we look for additional lens candidates?
Previous discussions~\cite{Huterer:2003ze,Oguri:2005dt} have considered the two extreme cases of a string that is nearly straight, and a string that is a random walk on short scales; clearly the networks that we are considering are very close to the first case.

We keep the frame of the previous section, with the lens at the origin in ${\bf x}_\perp$ and aligned along the $x_1$-axis.
Then as we move along the string, the RMS transverse deviation is
\begin{eqnarray}
\langle x_2^2(l) \rangle &=& \int_0^l d\sigma'  \int_0^l d\sigma'' \, \langle x'_2(\sigma) x'_2(\sigma') \rangle  \nonumber\\
&=& \frac{{\mathcal A} l^2}{4(\chi + 1)} (l/t)^{2\chi}  \ .
\end{eqnarray}
The extension in the $x_1$ direction is just $l$, so the RMS angular deviation is
\begin{equation}
\delta\varphi  \sim  \sqrt{{\mathcal A} / 4(\chi + 1)} (l/t)^{\chi} \equiv {\overline{\delta\varphi}}  \ .
\end{equation}
If we put in representative numbers, looking at an apparent separation on the scale of arc-minutes for a lens at a redshift of order $0.1$,  we obtain with the matter era parameters a deviation ${\delta\varphi} \sim 0.05$ radians.
That is, any additional lenses should be rather well aligned with the axis of the first.
If the string is tilted by an angle $\psi$ to the line of sight, then projection effects increase $\delta\theta$ and $\delta\varphi$ by a factor $1/\cos\psi$.
Of course, for lensing by a {\it loop}, the bending will be large at lengths comparable to the size of the loop.

Lens alignment provides an interesting setting for discussing the non-Gaussia\-ni\-ty of the structure on the string.
If the fluctuations of $x_2'$ were Gaussian, then the probability of finding a second lens at an angle $\delta\varphi$ to the axis of the first would be proportional to $e^{-\delta\varphi^2/2\overline{\delta\varphi}^2}$, and therefore very small at large angles.
However, we have considered thus far a typical string segment, which undergoes only stretching.
There will be a small fraction of segments that contain a large kink, and one might expect that it is these that dominate the tail of the distribution of bending angles.

Let us work this out explicitly.
Consider a left-moving segment of coordinate length $\sigma$, and let $P(\sigma,\tau,k)\, dk$ be the probability that it contain a kink for which the discontinuity $|{\bf p}_+ - {\bf p}_+'|$ lies between $k$ and $k+dk$ ($0 < k < 2$).
There are two main contributions to the evolution of $P$.
Intercommutations introduce kinks at a rate that we assume to scale, so that it is proportional to the world-sheet volume in horizon units, $a^2 \epsilon \sigma \tau^{-2\nu'/\nu} d\tau$, and to some unknown function $g(k)$.
Also, the expansion of the universe straightens the kink, $k \propto a^{-\bar\alpha}$~\cite{Bennett:1989yq}.
Then\footnote{Other effects are often included in the discussion of kink density, such as the removal of kinky regions by loop formation~\cite{Allen:1990mp}, but these have a small effect.}
\begin{equation}
\frac{\partial P}{\partial \tau} = \tau^{2\nu' (1 - \bar v^2 - 1/\nu)}
 \sigma g(k) + \bar\alpha \frac{\dot a}{a} \frac{\partial}{\partial k}(kP)  \ .
\label{pevol}
\end{equation}
We set $P$ to zero at the matching time $\tau_0$ defined by
\begin{equation}
\sigma= x_0 \tau_0^{1 + 2 \nu'\bar v^2 }  \ ,
\end{equation}
as in Eq.~(\ref{match}): earlier kinks are treated as part of the typical distribution, while $P$ identifies kinks that form later.
For simplicity we assume that $x_0$ is small enough that the probability of more than one kink can be neglected.

To solve this, define
\begin{equation}
\kappa = k \tau^{\zeta'}  \ , \quad
Q(\sigma,\tau,\kappa) = k P(\sigma,\tau,\kappa)  \ .
\end{equation}
Then
\begin{equation}
\tau \partial_\tau Q = \sigma g(\kappa \tau^{-\zeta'}) \kappa \tau^{-1-\nu'}  \ .
\end{equation}
This can now be integrated to give
\begin{eqnarray}
P(\sigma,\tau,\kappa) &=&
\sigma \tau^{\zeta'} \int_{\tau_0}^\tau \frac{d\tau'}{\tau'} \tau'^{-1-\nu'}
g(k \tau^{\zeta'} /\tau'^{\zeta'})    \nonumber\\
&=& \frac{\tilde\sigma}{\zeta'} k^{-1/\zeta}
\int_k^{k_0} \frac{dk'}{k'} k'^{1/\zeta} g(k')  \ .
\label{psol}
\end{eqnarray}
Here $k_0 = k (x_0 /\tilde\sigma)^{\zeta'/(1+2\nu'\bar v^2)}$ and $\tilde\sigma = \sigma/\tau^{1 + 2 \nu'\bar v^2}$.
Note that $\tilde\sigma$ is just a constant times $l/t$, so the probability distribution scales.
The source $g(k)$ vanishes by definition for $k > 2$, so $k_0 > 2$ is equivalent to $k_0 = 2$.

Rather than the angle $\delta\varphi$ between the axis of the first lens and the position of the second lens, it is slightly simpler to consider the angle $\delta\theta$ between the two axes.
In the small fluctuation approximation this is just $x_2'(\sigma)$, and so the RMS fluctuation is
\begin{equation}
\overline{\delta\theta} = \sqrt{{\mathcal A}/2}(l/t)^\chi  \ .
\end{equation}
In the Gaussian approximation, the probability distribution is $e^{-\delta\theta^2/2\overline{\delta\theta}^2}$ and so is very small for large angles.
On the other hand, a large angle might also arise from a segment that happens to contain a single recent kink.
Treating the segment as straight on each side of the kink, the probability density is then precisely the function $P(\sigma,\tau,k)\, dk$ just obtained, with $k = \delta\theta$.
If we consider angles that are large compared to $\overline{\delta\theta}$ but still small compared to 1, the range of integration in the solution~(\ref{psol}) extends essentially to the full range $0$ to $2$ and so the integral gives a constant.
Then
\begin{equation}
P(\sigma,\tau,k) \, dk \propto k^{-1/\zeta} dk = \delta\theta^{-1/\zeta} d\delta\theta  \ .
\end{equation}
Thus the tail of the distribution is not Gaussian but a power law, dominated by segments with a `recent' kink.
One finds the same for the distribution of $\delta\varphi$.
It is notable however that the exponent in the distribution is rather large, roughly 4 in the matter era and 10 in the radiation era.
Thus the earlier conclusion that the string is rather straight still holds.
The sharp falloff of the distribution suggests that a Gaussian model might be a good approximation.

\section{Periodic gravitational waves from small cosmic string loops\label{PERIODICGW}}

We have seen in the previous chapter that the small loops which our stretching model predicts to be abundant are characterized by very high velocities.
Such a strong Lorentz boost can push the natural oscillation frequency of the loops into the frequency band of GW interferometers.
The small loops may then be potential sources for these experiments and if so they should give rise to somewhat special GW signatures.
In this section we study this possibility.

One of the most promising ways to test the presence of a cosmic string network in our Universe is through its gravitational wave (GW) emission~\cite{Vachaspati:1984gt,Damour:2004kw,Siemens:2006vk,Hogan:2006we}.
One can expect a GW background from the network~\cite{Economou:1991bc,Battye:1997ji,Siemens:2006yp} and on top of this the loops typically form cusps which emit strong bursts of radiation~\cite{Damour:2000wa,Damour:2001bk}. 
The presence of a population of highly boosted loops leads to a peculiar GW signature and therefore deserves a specific study as these may be potential sources for present and planned gravitational wave detectors.
The large Lorentz factor has some consequences on the analysis of GW detection which were not previously considered in~\cite{Damour:2004kw,Siemens:2006vk,Hogan:2006we,Siemens:2006yp,Garfinkle:1987yw}. 

All cosmic string loops are expected to have (quasi-)periodic behavior, but one of the key features of small high-velocity loops is the fact that their observed period may be as short as a few mili-seconds.
One may even expect that the first few harmonics of the loop enter the frequency band of the GW interferometers.
Such a case is very promising since, on one hand, the lowest harmonics emit the strongest signal and furthermore it has little dependence on the exact loop trajectory: we can expect more robust wave-forms than those originated from the single cusp case (whose behavior under back-reaction or in the presence of small-scale structure has been questioned~\cite{Thompson:1988yj,Quashnock:1990wv,Siemens:2001dx,Siemens:2003ra,Chialva:2006ak}).
On the other hand, if this frequency is short enough in order to enter the GW interferometer band, this allows to search for periodic GW signals accumulating a large number of loop periods.
As a result the strain sensitivity will be improved by a huge factor corresponding to the square root of the ratio between the observation time and the lifetime of a single cusp.

The fraction of the small loop population moving directly toward us sees its GW strength boosted.
However, the large Lorentz factor also pushes most of the emission of gravitational radiation into the direction of the loop motion.
The fact that the signal is potentially strong enough is not sufficient to guaranty observations; we need the fraction of sources meeting the observability requirements to be numerous enough in order to get a reasonable probability for detection.

\subsection{Emission of GW}

We now compute the GW spectrum emitted by a typical small loop.
The model loop described in Section~\ref{MODELLOOP} will be considered as a representative of this class.
To distinguish between the location of the cosmic string and a generic point in spacetime we shall use coordinates $x_{\rm w.s.}^\mu$ for the former and coordinates $x^\mu$ for the latter.

A cosmic string acts as a source term for the gravitational field through its energy-momentum tensor.
For a classical string it is given by~\cite{1994csot.book.....V}
\begin{equation}
T^{\mu\nu}({\bf x},t)=\mu\int\left(p^\mu_+p^\nu_-+p^\mu_-p^\nu_+\right)\delta^4\left(x-x_{\rm w.s.}(u,v)\right)dudv  \ ,
\end{equation}
with $x=(t,{\bf x})$, while the four-vector $x_{\rm w.s.}(u,v)$ gives the spacetime location of the world-sheet point with coordinates $(u,v)$. Since the loop has $2\beta$-periodicity we choose the world-sheet to be a strip $(u,v) \in (-\infty,\infty)\times[-\beta,\beta]$.

Furthermore, for a source of size $\sim d$ localized around the origin, the trace-reversed metric perturbation in the local wave zone ($r\equiv|{\bf x}|>>d$ ) is given, in the time domain, by~\cite{1994csot.book.....V,Misner:1974qy,Weinberg:1972book}
\begin{eqnarray}
\bar{h}^{\mu\nu}(t,{\bf x})&=&4G\int \frac{1}{\vert {\bf x}-{\bf x}'\vert}T^{\mu\nu}(t-\vert {\bf x}-{\bf x}'\vert,{\bf x}') \, d^3{\bf x}'   \nonumber \\
&& \hspace{-50pt}  \simeq  \frac{4G}{r}\int T^{\mu\nu}(t-r+{\bf x}' \cdot {\bf n},{\bf x}')\, d^3{\bf x}'  \nonumber \\
&& \hspace{-80pt}  = \frac{4G\mu}{r}\int\left({p}^\mu_+{p}^\nu_-+{p}^\mu_-{p}^\nu_+\right)\delta\left(t-r+{\bf x}_{\rm w.s.}(u,v)\cdot{\bf n}-x_{\rm w.s.}^0(u,v) \right)dudv \ ,
\end{eqnarray}
where ${\bf n} \equiv {\bf x}/r$.
When analyzing high frequencies originated by cusps and kinks one must be careful with the fact that the dominant contributions to $p^\mu_\pm$ in the series expansion is a pure gauge term, as was first noted in~\cite{Damour:2001bk}.
However, converting to transverse traceless (TT) gauge eliminates this term.\footnote{One can also eliminate the gauge term explicitly by replacing $p^\mu_\pm$ by  $\check{p}^\mu_\pm =p^\mu_\pm-(1,-{\bf n})$.}
In any case, we will be interested in the lowest harmonics since these are the frequencies which contribute the most to the observation rate.
Performing a temporal Fourier transform one obtains
\begin{equation}
\tilde{\bar{h}}^{\mu\nu}(\omega,{\bf x}) = \frac{4G\mu}{r}e^{i\omega r}\int \left({p}^\mu_+{p}^\nu_-+{p}^\mu_-{p}^\nu_+\right)e^{i\omega (\frac{u+v}{2}-{\bf x}_{\rm w.s.}(u,v)\cdot{\bf n})} dudv  \ .
\end{equation}

Now, decomposing the integral over $u$ into
\begin{equation}
\int_{-\infty}^{\infty}du=\sum_{m\in \mathbb{Z}}\int_{(2m-1)\beta}^{(2m+1)\beta}du  \ ,
\end{equation}
using property~(\ref{periodicity}) and defining $\theta$ as the angle subtended between ${\bf n}$ and the $z$-axis (along which the loop is traveling) we get
\begin{eqnarray}
\tilde{\bar{h}}^{\mu\nu}(\omega,{\bf x}) & = & \frac{4G\mu}{r}e^{i\omega r}\sum_{m\in \mathbb{Z}}
e^{ i\omega m\beta\left(1-|{\bf v}|\cos\theta \right) }  \nonumber \\
&& \hspace{60pt}  \times \underbrace{ \int_{-\beta}^\beta\int_{-\beta}^\beta \left({p}^\mu_+{p}^\nu_-+{p}^\mu_-{p}^\nu_+\right)
e^{i\omega (\frac{u+v}{2}-{\bf x}_{\rm w.s.}(u,v)\cdot{\bf n})} dudv }_{\mathcal{I}^{\mu\nu}(\theta,\phi)}   \nonumber \\
& = & \frac{4G\mu}{r}e^{i\omega r}\frac{2\pi}{\beta\left(1-|{\bf v}|\cos\theta  \right)}\sum_n\delta\left(\omega-\frac{2\pi n}{\beta\left(1-|{\bf v}|\cos\theta  \right)}\right)\mathcal{I}^{\mu\nu}(\theta,\phi) \nonumber \\
& = & \frac{8\pi G\mu}{r}e^{i\omega r} f_1(\theta)\sum_n\delta\left(\omega-2\pi f_n(\theta)\right)\mathcal{I}^{\mu\nu}(\theta,\phi)\label{h01}  \ .
\label{htilde}
\end{eqnarray}
As a check, note that we have recovered the discrete set of frequencies obtained in~(\ref{harmonics}).
The computation of $\mathcal{I}^{\mu\nu}(u,v)$ involves the following integrals:
\begin{eqnarray}
I^n_{u,j}
&=&\int_{-\beta}^\beta u^je^{2\pi i f_n(\theta)\left(\frac{u}{2}\left(1-\cos\theta\right)-\sin\theta\cos\phi\frac{Vu^2}{4}+\cos\theta\frac{V^2u^3}{12} \right)} du \ , \\
I^n_{v,j}
&=&\int_{-\beta}^\beta v^je^{2\pi i f_n(\theta)\left(\frac{v}{2}\left(1-\cos\theta\right)-\sin\theta\sin\phi\frac{Vv^2}{4}+\cos\theta\frac{V^2v^3}{12} \right)} dv \ ,
\  j=0,1,2  \ .
\end{eqnarray}

Before we move on to computing~(\ref{htilde}) one more consideration is in order.
The interaction between a GW and a detector is usually described in the TT gauge and it is convenient to rotate the coordinate frame in order to match the observer description in which the GW arrives along the $z$-axis, so that $\hat{z}^{\rm new} \equiv {\bf n}$.
The rotation matrix that converts between the source frame and the observer frame is given by
\begin{equation}
R=\left(\begin{array}{ccc}\cos\theta \cos\phi & \cos\theta \sin\phi & -\sin\theta \\
-\sin\phi & \cos\phi & 0 \\
\sin\theta \cos\phi & \sin\theta \sin\phi & \cos\theta \end{array}\right)  \ .
\end{equation}
Rewriting $\tilde{\bar{h}}_{\mu\nu}$ in the TT gauge is performed by means of the projector $\Lambda_{ij,kl}$ defined by (see section 10.4.15 in~\cite{Weinberg:1972book})
\begin{equation}
\Lambda_{ij,kl}=P_{ik}P_{jl}-\frac{1}{2}P_{ij}P_{kl}  \ ,
\end{equation}
where $P_{ij}=\delta_{ij}-n_in_j$.
Then,
\begin{equation}
\tilde{h}_{ij}^{TT}=\Lambda_{ij,kl}R_{kk'}R_{ll'}\tilde{\bar{h}}_{k'l'}=\left(\begin{array}{ccc}\tilde{h}_+ & \tilde{h}_\times & 0 \\\tilde{h}_\times & -\tilde{h}_+ & 0 \\0 & 0 & 0\end{array}\right)  \ ,
\end{equation}
and we can define the observed strength as
\begin{equation}
\tilde{h}=\sqrt{|\tilde{h}_+|^2+|\tilde{h}_\times|^2}  \ .
\end{equation}

\begin{figure}
\begin{center}
\includegraphics[width=7cm]{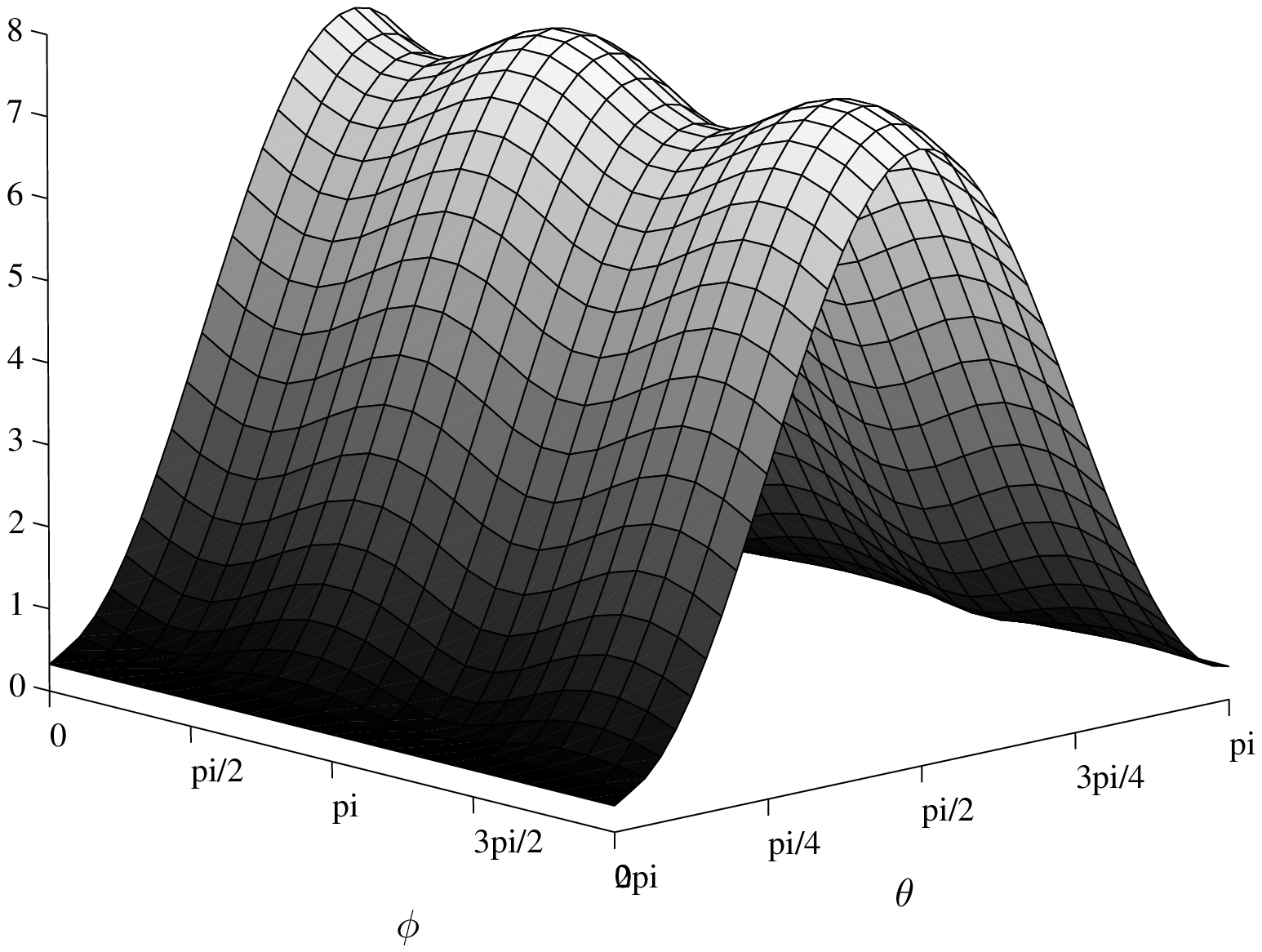}~\includegraphics[width=7cm]{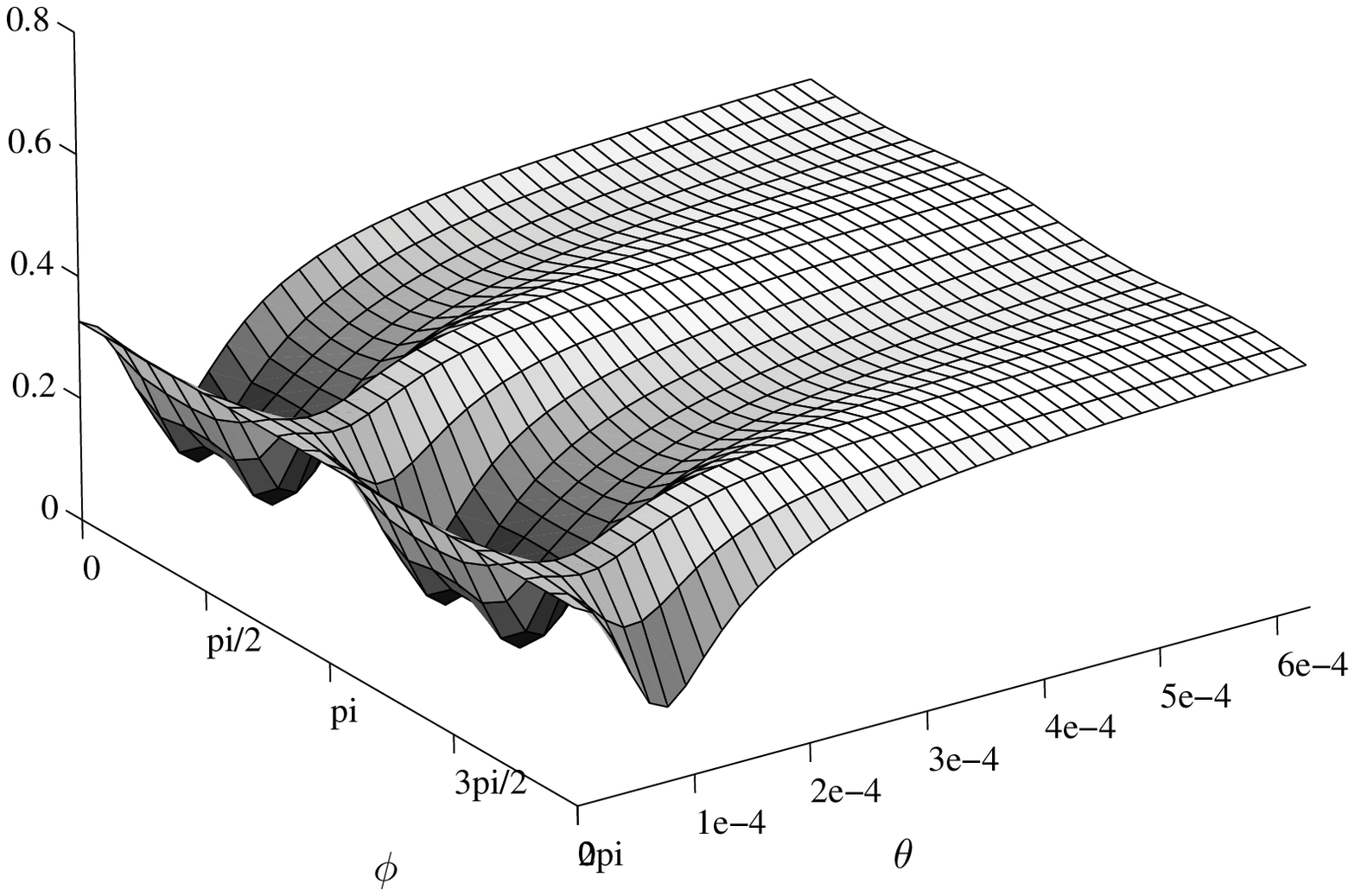}\\
\caption{The angular dependence of the first harmonic ($n=1$). The plots show the quantity  $\tilde{h}\cdot\frac{r}{2 \pi G\mu}\frac{\left(1-{\bf v}\cos\theta  \right)}{V^2\beta^3}$. We have set $G\mu=10^{-9}$. The right panel is a closeup on the value of $\theta$ of order $V\beta\simeq1.5\cdot10^{-4}$.}
\label{FIG_II_1}
\end{center}
\end{figure}

\begin{figure}
\begin{center}
\includegraphics[width=8cm]{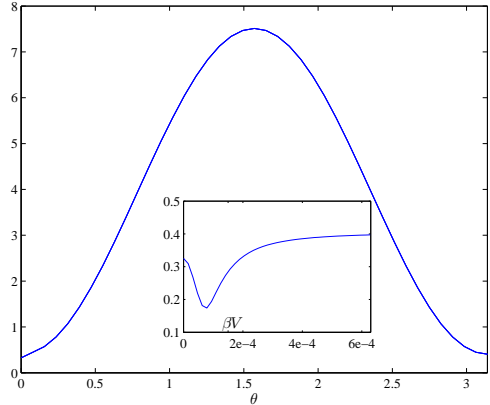}
\caption{Same as Figure~\ref{FIG_II_1} but after averaging over $\phi$. Note that for $V\beta < \theta < \pi-V\beta$ the curve is approximately proportional to $\sin^2\theta$. This dependence arises from the rotation and posterior conversion to TT gauge.}
\label{FIG_II_2}
\end{center}
\end{figure}

Note that the main dependency of $\tilde{h}$ on the direction of observation comes from the $f_1(\theta)$ factor in Eq.~(\ref{h01}) which, for small values of $\theta$, causes an enhancement by a factor $(V\beta)^{-2}$ ($\sim10^7$ for $G\mu=10^{-9}$ ), as was discussed in Section~\ref{MODELLOOP}.
The remaining angular dependence arising from $\mathcal{I}^{\mu\nu}(\theta,\phi)$, for the first harmonic, is shown in Figures~\ref{FIG_II_1} and~\ref{FIG_II_2}.
One can observe that this gives some enhancement\footnote{This enhancement is a consequence of the rotation between the source frame and the observer frame.  This is more clearly seen by taking $\phi=0$, in which case the dominant contribution to $\tilde{h}/f_1(\theta)$ is proportional to $\sin^2(\theta)$.} for $\theta\sim\pi/2$.
However, this corresponds to the range in the parameter $\theta$ where the large Lorentz boost enhancement is lost.
Let us anticipate that in order to be able to detect the loop we need a big enhancement of the emitted GW (in other terms, we are able to `see' only loops moving nearly in our direction).
Therefore any sub-leading effect at large $\theta$ may be neglected and in the remainder of this study we will discard this sub-leading dependence.
This leads to the following form (omitting the higher harmonics) for $\tilde{h}$:
\begin{eqnarray}
\tilde{h}&\simeq&\frac{8 \pi G\mu}{r}\frac{V^2\beta^3}{\left(1-|{\bf v}|\cos\theta \right)}\left\vert \int_{-1}^1xe^{i\pi x^3} dx \right\vert^2\cdot0.5   \nonumber \\
&\simeq&1.3\frac{\pi}{r}\frac{G\mu V^2\beta^3}{\left(1-\cos\theta+\frac{V^2\beta^2}{6}\cos{\theta}\right)}  \ ,
\label{e:h}
\end{eqnarray}
where the final factor of 0.5 shifts the value of the integral at $\theta=0$ to its actual minimal value (see Figure~\ref{FIG_II_2}) in order to obtain a lower estimate for the strain~$\tilde{h}$.

\subsection{Visibility of the small loops\label{VISIB}}

Having computed the GW spectrum, we now want to determine whether or not current or planed GW detectors can observe the small relativistic loops under consideration and, if so, for what range of the parameters.
This will depend on the cosmology and therefore we must fix the dependence of $t$ and $r$ on the redshift $z$.
To this end, we consider a flat universe with the cosmological constant contribution included in the cosmology.

Any GW detector is characterized by its frequency window $[f_{-};f_{+}]$ as well as the minimal amplitude needed for a detection $h_{\rm det}(f)$.
For LIGO's second science run (LIGO S2) the frequency window is $[160;728.8]Hz$.
Furthermore, the sensitivity to a continuous signal from a 10-hour search can be modeled by
\begin{equation}
h_{\rm det}(f) = h_{160}\left(\frac{f}{160 Hz}\right)^q  \ ,
\end{equation}
where $q \sim 0.6$ and $h_{160}\sim10^{-22}$, see~\cite{Abbott:2006vg}.
LIGO is expected to be 10 times more sensitive and Advanced LIGO to bring another factor of 10 (see for example~\cite{Creighton:2003nm}).

In order to detect a given loop there are 3 conditions which must be met:
first, the frequency of the incoming GW produced by the loop has to lie within the detector window;
second, its amplitude has to be above the limit $h_{\rm det}(f)$;
finally, the observed lifetime of the loop, $\tau_{\rm obs}$, must be large compared to the time span of the experiment, $T_{\rm obs}$, to insure that its periodicity does not change significantly during observation, in which case it would quickly drop out of the frequency window anyway.

\begin{figure}[t]
\begin{center}
\includegraphics[width=9cm]{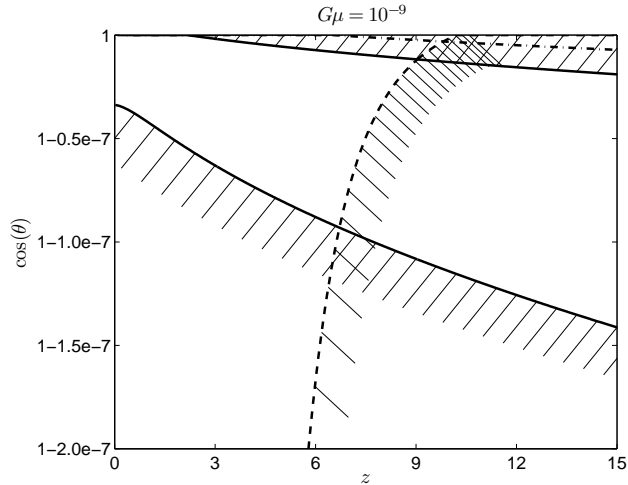}
\caption{Observational window for LIGO and for $G\mu=10^{-9}$.  We plot $\cos(\theta)$ as a function of the red-shift.  The solid curves represent the frequency window, the dashed curve is the sensitivity bound and the dash-dotted curve corresponds to the lifetime constraint.  Note that, for such a value of $G\mu$, we can observe loops up to $z\simeq9$ when the direction of the motion of the loop lies on a cone of aperture $\sim2\cdot10^{-4}rad\simeq0.7'$ around the line of sight.}
\label{FIG_wnd}
\end{center}
\end{figure}

Inspection of equation~(\ref{harmonics}) shows that the frequency of the received GW depends on the angle $\theta$ subtended between the line of sight and the velocity of the loop.
Thus, the strategy is to first fix $G\mu$ and determine the range of angles under which observation is possible, for a given $z$.
We shall give the different bounds in terms of the cosine of the corresponding angle.
Then, for each value of $G\mu$, a combination of the above constraints determines an observational window.
An example with $G\mu=10^{-9}$ is shown in Figure~\ref{FIG_wnd}.
Let us point out that, if our assumptions are correct, one can potentially observe these loops at LIGO if $10^{-10} \lsim G\mu \lsim 7 \times 10^{-9}$.
Finally, the integration over the redshift can be performed to determine the expected number of detections.
The capacity of observing such loops (or placing constraints on the string tension) will then depend on their number density in the universe, and this quantity will be worked out in Section~\ref{DETECTRATE}.
But before we do so, let us turn to the analysis of the several constraints mentioned above.

\subsubsection*{Constraint from the frequency window}
  
In order to determine the observed frequency of a GW produced by a small loop we must correct equation~(\ref{harmonics}) for the cosmological redshift, thus obtaining
\begin{equation}
f_n\vert_{\rm observed}=\frac{1}{\beta(1+z)}\frac{n}{(1-\cos\theta(1-V^2\beta^2/6))}  \ .
\end{equation}
The bounds $f_\pm$ on the observed frequency lead to corresponding constraints on the angle $\theta$ which can be expressed as $\cos\theta_- \leq \cos\theta \leq \cos\theta_+$, with
\begin{equation}
\cos\theta_\pm = \frac{1-\frac{n}{f_{\pm} \beta (1+z)}}{1-\frac{1}{6}V^2\beta^2}  \ .
\label{bfreq}
\end{equation}
%

\subsubsection*{Constraint from the sensitivity}

The amplitude $\tilde{h}$ of the GW wave produced by a small cosmic string loop was obtained in~(\ref{e:h}).
Taking into account the cosmology and converting to the time domain, the strain caused by the lowest harmonic takes the following form:
\begin{equation}
h(t,\theta) = \frac{1.3\ G\mu V^2\beta^3}{(1-\cos\theta(1-V^2\beta^2/6)) \,r(z)} 
\cos(2\pi f_1(\theta)t-\psi_0)  \ ,
\end{equation}
where $\psi_0$ is an (arbitrary) phase.
The condition to be satisfied in order to be able to detect the GW is then $h(t,\theta) \geq h_{\rm det}(f_1\vert_{\rm observed}(\theta))$.
Solving for the angle yields a lower bound, $\cos\theta \geq \cos\theta_h$, where
\begin{equation}
\cos\theta_h = \frac{1 - \left( \frac{1.3\ G\mu V^2\beta^{3+q} (1+z)^q (160Hz)^q}{h_{160} \, r(z)} \right)^{1/(1-q)}}{1 - \frac{1}{6} V^2\beta^2}  \ .
\label{c:s}
\end{equation}
%

\subsubsection*{Constraint from the lifetime}

Another feature that must be taken into account is that cosmic string loops shrink with time due to the loss of energy in the form of gravitational waves themselves.
We have already seen in Section~\ref{GRAVRADLOOPS} that the lifetime of a loop of size $2\beta$ in the FRW frame is $\tau=(2\beta)(\Gamma G\mu)^{-1}$.
Since the loops are moving with a relative velocity which makes an angle $\theta$ with the line of sight their observed lifetime, including the cosmological redshift, is
\begin{equation}
\tau_{\rm obs} = (1-|{\bf v}|\cos\theta)(1+z)\tau = (1-|{\bf v}|\cos\theta)(1+z) \frac{2\beta}{\Gamma G\mu}  \ .
\label{lifetime}
\end{equation}
Given the extremely high velocities of the loops, one finds that for very small angles $\theta$ this apparent lifetime can be very short since it is suppressed by a factor of $\gamma^2$.
In that case, if we want to consider loops whose frequency does not change significantly during an observation time $T_{\rm obs}$ we need to guaranty that $\tau_{\rm obs} \geq T_{\rm obs}$, which translates into the constraint $\cos\theta \leq \cos\theta_L$, where
\begin{equation}
\cos\theta_L = \frac{1-\frac{\Gamma G\mu T_{\rm obs}}{2\beta(1+z)}}{1-\frac{1}{6}(V\beta)^2}  \ .
\label{blife}
\end{equation}
This bound has the same form as the bound~(\ref{bfreq}) coming from the frequency window.
The lifetime constraint can thus be recast as a constraint on the frequency window instead.
Combining equation~(\ref{bfreq}) and~(\ref{blife}) we obtain an effective maximal frequency given by $f^{\rm eff}_+ = \min\left\{f_+,f_L\right\}$, with
\begin{equation}
f_L = \frac{2n}{\Gamma G\mu T_{\rm obs}}  \ .
\end{equation}

The lifetime constraint has the effect of introducing a $G\mu$-dependent maximal frequency.
For the first harmonic ($n=1$) and a time span of $T_{obs}=10$ hours in the settings of LIGO, this does not affect the original frequency window for $G\mu \leq 1.5 \times 10^{-9}$.
Beyond that, the maximal frequency is given by $f_L$, which decreases as $G\mu$ is increased, until it finally closes the frequency window for $G\mu \simeq 6.9 \times 10^{-9}$.
If we want to observe or constrain higher values of $G\mu$ we have to look at higher harmonics or make a shorter run.
Both possibilities help relax the bound coming from the lifetime but on the other hand they also tighten the sensitivity constraint.

\subsection{Loop number density and expected number of detections\label{DETECTRATE}}

For the purposes of computing the expected detection rate we shall assume that the string network is in a scaling regime.
The rate at which the long string is converted into small loops can be obtained from equation~(\ref{string2loops}).
Inserting the numbers for a matter-dominated era, where the energy density in long strings is $\rho_\infty \simeq 4 \mu t^{-2}$, one finds that
\begin{equation}
\left(\frac{\partial\ell_\infty}{\partial t}\right)_{\rm loops} \simeq \frac{4{\rm Vol}}{5\,t^3}  \ ,
\end{equation}
where ${\rm Vol}$ is the Hubble volume.
We obtain the number density in small loops by multiplying the above equation by the lifetime and dividing by their length as well as the volume factor.
Since the lifetime of the loop~(\ref{lifetime}) introduces a dependency on the angle of observation $\theta$, the apparent number density is given by
\begin{equation}
n(\theta) = (1-|{\bf v}|\cos\theta)n_{FRW} \equiv (1-|{\bf v}|\cos\theta) \frac{4}{5}\frac{1}{\Gamma G\mu}\frac{1}{t^3}  \ .
\end{equation}

Now we can include the constraints from the Section~\ref{VISIB}.
Only loops moving along certain directions and within certain distances emit GW which enter the frequency window with an amplitude high enough to be detected.
The range of $\theta$ for which the detector can `see' the loop is given by $\cos\theta_{min} \leq \cos\theta \leq \cos\theta_{max}$, where
\begin{eqnarray}
\cos\theta_{max} &=& \max\left\{-1,\min\left\{\cos\theta_+,\cos\theta_L\right\}\right\}  \ ,  \nonumber \\
\cos\theta_{min} &=& \min\left\{1,\max\left\{-1,\cos\theta_-,\cos\theta_h\right\}\right\}  \ .
\end{eqnarray}
Therefore, averaging over the sphere gives
\begin{eqnarray}
\frac{1}{2}\int_{\theta_{min}}^{\theta_{max}} n(\theta) \sin\theta \, d\theta &=&   \nonumber \\
&&  \hspace{-110pt}  
= \underbrace{\frac{1}{2}\left(\cos\theta_{max}-\cos\theta_{min}-\frac{|{\bf v}|}{2}\cos^2\theta_{max} + \frac{|{\bf v}|}{2}\cos^2\theta_{min}\right)}_{\zeta(\theta_{max},\,\theta_{min})} n_{FRW}  \ .
\end{eqnarray}

Finally, to compute the expected number of observations we must account for the non-trivial evolution of the Universe.
To this end we will follow Ref.~\cite{Siemens:2006vk}, where the cosmology is encoded in the function $h(z) \equiv H(z)/H_0$ which expresses the time-dependent Hubble factor in terms of the redshift:
\begin{equation}
h(z)  =  \sqrt{ \Omega_m (1+z)^3 + \Omega_r (1+z)^4 + \Omega_\Lambda }  \ .
\end{equation}
The coefficients $\Omega_m$, $\Omega_r$ and $\Omega_\Lambda$ are the present energy density contributions from matter, radiation and cosmological constant, respectively.
These add up to unity in a flat universe and $\Omega_m = 0.25$, $\Omega_r = 4.6 \times 10^{-5}$.
The comoving variables are expressed as $t = H_0^{-1} \varphi_t(z)$ and $r = H_0^{-1} \varphi_r(z)$, where
\begin{eqnarray}
\varphi_t(z)  &=&  \int_z^\infty \frac{dz'}{(1+z')h(z')}  \ ,  \\
\varphi_r(z)  &=&  \int_0^z \frac{dz'}{h(z')}  \ .
\end{eqnarray}

Then, the number of observed loops is simply
\begin{equation}
N=\int \zeta(\theta_{max},\theta_{min}) \, n_{FRW} \, dV  \ ,
\end{equation}
where the comoving volume is given by
\begin{equation}
dV=\frac{4\pi}{H_0^3}\frac{\varphi^2_r(z)}{\left(1+z\right)^3h(z)} dz \ .
\end{equation}
and $H_0$ represents the current value of the Hubble parameter.
Therefore, we obtain the following expression for the expected number of observations:
\begin{equation}
N= \frac{16\pi}{5\Gamma G\mu}\int_0^\infty \frac{\varphi_r^2(z)}{\varphi_t^3(z)}\frac{ \zeta(\theta_{max},\theta_{min})}{(1+z)^3h(z)} \, dz  \ .
\end{equation}

\begin{figure}[t]
\begin{center}
\includegraphics[width=9cm]{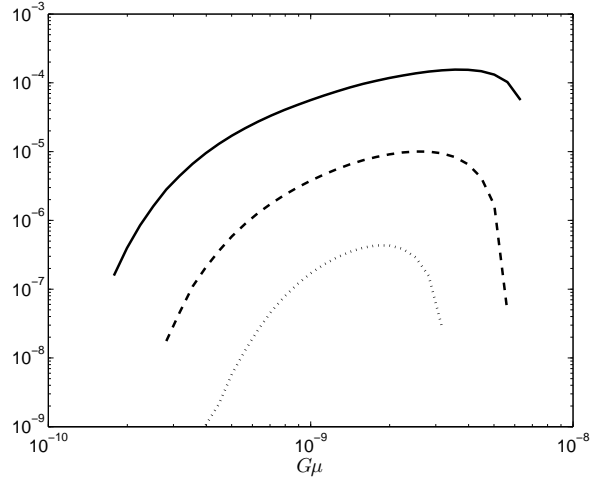}
\caption{Expected number of observations for a run of $T_{obs}=10$ hours.  The dotted, dashed and continuous curves represent LIGO S2, LIGO and Advanced LIGO, respectively.}
\label{FIG_res}
\end{center}
\end{figure}

As a result, we report in Figure~\ref{FIG_res} the expected number of detections for a run of 10 hours with the different versions of LIGO.
We can understand heuristically some qualitative features of the curves.
The rise of the curves for increasing $G\mu$ comes mainly from the dependence of the GW strength on this parameter (see equation~(\ref{e:h})) which allows to observe a larger volume,\footnote{Equation~(\ref{c:s}) shows how the sensitivity constraint becomes less restraining with increasing $G\mu$} while the abrupt cutoff at large values of $G\mu$ has its origin in the closing of the frequency window by the upper bound from the lifetime constraint.
Another effect that contributes to this fall-off is the fact that the regions allowed by the frequency window and the sensitivity bound do not overlap anymore for high $G\mu$.

Note that reducing the observation time has 2 effects: the first one is to drop the GW strength relative to the detector sensitivity since $h_{160}$ is proportional to $T_{\rm obs}^{-1/2}$~\cite{Abbott:2006vg}.
On the other hand, this allows the detection of short-lived loops at higher redshifts by relaxing the lifetime constraint and consequently we can scan larger values of $G\mu$.
However, this leaves the expected number of observations far from being of order 1 or larger.

Note also that we have only taken into account the first harmonic.
Adding higher harmonics has the effect of expanding the curves in Figure~\ref{FIG_res} to higher values of $G\mu$ but this does not cause the curves to rise.
For specific combinations of the parameters $G\mu$ and $z$ it is possible to observe more than one harmonic in the detector frequency window.
This may provide (through a specific search for direction correlated signals) better sensitivity.
However, this is certainly not enough to get an expected number of observations of order 1.

\subsection{Summary and discussion}

In conclusion, we have shown that the observation of boosted cosmic string loops is, in principle, possible for very interesting values of $G\mu$ in the range $10^{-10}-10^{-8}$.
However, the chances of such an observation during a 10 hour period are slim due to the low effective number density of such loops.
Even though a large boost factor can bring the typical frequency of the GW produced into the LIGO frequency band, this will only occur if the loop in question is moving along a trajectory very close to the line of sight ($\theta \lsim (V\beta)^2$), which effectively cuts down the apparent number density of loops with these characteristics.
An improvement can be obtained by increasing the duration of observation, thereby gaining in sensitivity of the detector.
If one keeps increasing the observation time, the apparent lifetime of the loops will eventually be reached.
We point out that, due to the production mechanism of the small loops (most of the loops that are produced around a large cusp move in coincident directions), we do not expect a Poisson distribution.
As a result, while longer times of observation lead to higher probabilities of detection, the expected number $N$ of loops observed should grow less than linearly with $T_{\rm obs}$.

We should stress that the possibility of observing GW from cosmic strings we have considered concerns only the continuous emission of GW by the lower harmonics of the loops.
This is in contrast with~\cite{Damour:2004kw,Siemens:2006vk,Damour:2000wa,Damour:2001bk} which focus on the burst of GW produced by the cusps and kinks.
The results we have presented here depend both on the exact characteristics of the population of small loops (namely, their size) and also on the type of cosmic string network.
We have assumed the distribution of loop size to be sharply peaked at the gravitational radiation scale but a decaying power law would reproduce a more realistic population.
Also, cosmic superstrings can have reconnection probabilities as small as $P \sim 10^{-3}$~\cite{Jackson:2004zg} and the number density of loops is proportional to a negative power of $P$.
Numerical studies~\cite{Sakellariadou:2004wq,Avgoustidis:2005nv} suggest values between $-1$ and $-0.6$ for the exponent.
This would lead to an enhancement of the probability of detection but not quite enough to obtain $N \sim 1$.
On the other hand, our conclusions are independent of the presence or absence of high frequency features on the loops, i.e. cusps and kinks.

Note also that the discrimination between a signal from a boosted loop and other periodic signals is not difficult since the former is expected to have a rising frequency and a decreasing amplitude.
This is in contrast with both spinning stars, which are expected to spin down, and with mergers, which see their GW amplitude growing with time.
However, a complete understanding of the waveform is not possible without a description of the higher harmonics.
Obtaining a template for such waveforms would be a valuable achievement which would allow direct searches of GW emission from cosmic string loops.

Finally, we have restricted our analysis to the LIGO frequency band.
For the lower frequencies of the LISA band the potential sources are loops with high $G\mu$ at very large $z$.
However, in order to get a sufficient signal-to-noise ratio we have to consider longer observation times and then the corresponding lifetime bound closes the observational window.

\chapter{Final Remarks}

The evolution of the interest in the subject of cosmic strings has shifted since their original proposal.
In the early eighties cosmic strings were regarded as a possible explanation for structure formation in the Universe, but more recently the interest has been revived due to the connection with String Theory.
In particular, such objects could provide an observational window into this promising unifying framework.
If this is the case and cosmic strings are ever detected, detailed knowledge of their networks is crucial to extract the real physics contained in the data.
The work presented in this dissertation represents a step toward determining properties and effects of cosmic strings.

Over 25 years of investigations have not led yet to a complete understanding of these networks.
This is a clear indication of their high degree of complexity: the non-linearity of the equations of motion, the presence of a large hierarchy of scales and the interplay between the several mechanisms that take place.
The issue of the build up of small scale structure on such systems stands out among these uncertainties.

Thus, our approach has been to focus on a range of scales for which one mechanism (stretching due to the expansion of the Universe) becomes dominant and the equations of motion become approximately linear.
The analytic model presented here therefore addresses only the short distance properties of the networks, avoiding the complex dynamics at scales comparable to the horizon size.
The numerical simulations are needed to address this highly non-linear evolution at large scales but analytic methods are important to separate real effects from transients and numerical inaccuracies.
Purely analytic or numerical treatments are frustrated by the complexity of these systems but the conjunction of both methods has led to some advances, namely on our understanding of the distribution of cosmic string loops.

How successful have we been with our efforts?
Using the stretching model we have been able to fix the functional form of the two-point function between points on a long string, confirming their smoothness at short distances.
The expression we have found fits the data from numerical simulations on a range of length-scales below the horizon but there is some discrepancy at smaller scales which still begs for a unanimous explanation.
The model developed also enables a more accurate determination of the scale of string smoothing by gravitational radiation.
This is parametrically smaller than the horizon size by a power of the dimensionless string tension, yielding a large hierarchy of scales.
The exponent is related to the two-point function.
Furthermore, a novel picture for the distribution of loop sizes has arisen from both numerical and analytic investigations.
Traditionally, loops were considered to be produced at a single length-scale.
More realistically there is a power law distribution accumulating loops at the gravitational radiation smoothing scale and a second peak roughly an order of magnitude below the horizon distance.
Unfortunately, our methods have not allowed us to determine the normalization of the distribution to confront with the simulations.
Furthermore, the issue of loop fragmentation still remains a little obscure.
We have also demonstrated that the loop distribution approaches a scaling regime in matter- and radiation-dominated eras, a point closely related to the long-standing question of the presence or not of scaling in the small scale structure.

More generally, our work has unveiled several correlations among the two-point function, loop production, fragmentation, and loop velocity.
These connections have led us to examine some observational consequences of cosmic string networks.
Regarding the lensing phenomenon by cosmic strings we now have a better sense for the expected magnitude of the distortion of lensed pairs.
On the other hand, detection at LIGO of gravitational waves generated from the low harmonics of small cosmic string loops is an exciting possibility for an interesting range of the string tension.

In conclusion, despite the simplicity of the equations of motion, the evolution of cosmic string networks turns out to be inherently (and persistently!) complicated.
Our efforts have been directed toward the study of the simplest possible kind of network and even so we still fall short of a complete understanding.
Nonetheless, with the stretching model we have made some progress in relating properties of the system at different scales and in identifying subtle relations between disparate quantities characterizing the network.

Gradually we are evolving toward a clearer understanding of the properties of cosmic string networks.
Hopefully, the desired level of knowledge is an attractor.

\begin{appendix}
\pagebreak

\chapter{Gaussian averages}

In this appendix we include the calculations that support the estimation of the second term in the expectation value~(\ref{dNw}) obtained in Section~\ref{RATELP}, namely
\begin{equation}
\langle \delta({L}^z_+ - {L}^z_-) \left| \det {\bf J} \right| \rangle_{ {\bf p}^\perp  \to {\mbox{\boldmath\scriptsize$ \omega$}}}
\ .
\end{equation}
We shall assume that the functional probability distribution for the fluctuations is Gaussian with variance given by the two-point function.
Furthermore, we see no reason for a strong correlation between the $\delta$-function and Jacobian factors in the expression above\footnote{The whole point of the separation of variables in Section~\ref{RATELP} was to accomplish this factorization.}, so we take the product of their averages.

For the $\delta$-function, a Gaussian average gives
\begin{eqnarray}
\left< \delta(L_+^z - L_-^z) \, \right> 
&=&  \int_{-\infty}^\infty \frac{dy}{2\pi} \,  \left< e^{iy(L_+^z - L_-^z)}\, \right>   \nonumber\\
&=& \int_{-\infty}^\infty \frac{dy}{2\pi} \, 
    e^{ iy \left< (L_+^z - L_-^z) \right>_c - \frac{y^2}{2} \left< (L_+^z - L_-^z)^2 \right>_c
    - i \frac{y^3}{6} \left< (L_+^z - L_-^z)^3 \right>_c + \ldots }   \nonumber\\
&=& \int_{-\infty}^\infty \frac{dy}{2\pi} \,  e^{-y^2 R(\chi) + O(y^3) }   \nonumber\\
&\approx&  \frac{1}{\sqrt{4\pi R(\chi)}} \ .
\label{deltaz}
\end{eqnarray}
The subscripts $c$ in the second line refer to the connected expectation values, obtained by contracting the Gaussian fields ${\mbox{\boldmath$ \omega$}}_\pm$ with the propagator~(\ref{oo}).
We have defined
\begin{eqnarray}
R(\chi) &=& \frac{1}{4} \int_{u-l/2}^{u+l/2} du' \int_{u-l/2}^{u+l/2} du'' \,
           \langle {\mbox{\boldmath$ \omega$}}_+(u') \cdot {\mbox{\boldmath$ \omega$}}_+(u'') \rangle^2
  =  C(\chi) {\mathcal A}^2 \frac{l^{2 + 4\chi}}{t^{4\chi}}  \ ,  \\
C(\chi) &=& \textstyle \frac{1}{4 (1 + 2\chi)^2}
\left( {\frac{1 + 2\chi}{1 + 4\chi} + \frac{1}{(1 + \chi)^2} - \frac{4}{3 + 4\chi} 
- \frac{ 4 \Gamma^2(2 + 2\chi)}{\Gamma(4 + 4\chi)} } \right)  \ .
\end{eqnarray}
Numerically $C(\chi) = (0.0049, 0.0106, 0.0111)$ for $\chi = (0.1, 0.25, 0.5)$.


We now consider the expectation value of the Jacobian, $\left< |\det {\bf J}| \right>_{ {\bf p}^\perp  \to {\mbox{\boldmath\scriptsize$ \omega$}}}$.
The matrix itself can be written more concisely if we shift the coordinates $u$ and $v$ by $l/2$.
In that case the quantities ${\bf L}_\pm$ depend on $l$ only through one of the integration limits.
Consequently, the third row (obtained by differentiating ${\bf L}_+ - {\bf L}_-$ with respect to $l$) simplifies and we arrive at
\begin{equation}
{\bf J}_{ {\bf p}^\perp  \to {\mbox{\boldmath\scriptsize$ \omega$}}} = \left(
\begin{array}[tcb]{ccc}
\omega^x_+(l) - \omega^x_+(0) & \; \omega^y_+(l) - \omega^y_+(0) \; & \frac{1}{2}\!\left[ \omega^2_+(0) - \omega^2_+(l) \right] \\ 
\omega^x_-(0) - \omega^x_-(l) & \; \omega^y_-(0) - \omega^y_-(l) \; & \frac{1}{2}\!\left[ \omega^2_-(l) - \omega^2_-(0) \right] \\ 
\omega^x_+(l) - \omega^x_-(0) & \; \omega^y_+(l) - \omega^y_-(0) \; & \frac{1}{2}\!\left[ \omega^2_-(0) - \omega^2_+(l) \right]
\end{array}
          \right)  \ .
\end{equation}
For notational simplicity we have set $u=0$ and $v=l$ at the end.

In the Gaussian approximation the relevant probability distribution is
\begin{equation}
{\mathcal P}_\pm({\mbox{\boldmath$ \omega$}}_\pm(l) , {\mbox{\boldmath$ \omega$}}_\pm(0)) = \frac{\det {\bf M}_\pm}{(2\pi)^2} \,
\exp \left[ -\frac{1}{2} ({\bf B}_\pm^i)^T {\bf M}_\pm {\bf B}^i_\pm \right] \ ,
\end{equation}
where the index $i$ is summed over the two coordinates $x$ and $y$ since the fluctuations obey the orthogonality condition $\hat{\bf z} \cdot {\mbox{\boldmath$ \omega$}}_\pm = 0$, and
\begin{equation}
{\bf B}^i_\pm \equiv \left( \begin{array}{c} \omega^i_\pm(l) \\ \omega^i_\pm(0) \end{array} \right) \ .
\end{equation}
(That is, the columns and rows of the $2\times 2$ matrix ${\bf M}$ correspond to the points $0$ and $l$, not the index $i$.)
As usual, the whole distribution is determined solely by the two-point functions which have already been determined in~(\ref{oo}) and~(\ref{fofu}).
Thus, one finds that
\begin{equation}
{\bf M}_\pm = \frac{2(t/l)^{2\chi}}{{\mathcal A}(1-\chi)}
\left( \begin{array}{cc} 1 & \chi \\ \chi & 1 \end{array} \right) 
\;\; , \;\;
\det {\bf M}_\pm = \left( \frac{2}{{\mathcal A}} \right)^2 \left(\frac{t}{l}\right)^{4\chi} \frac{1+\chi}{1-\chi}  \ .
\end{equation}

We now have all we need to write down an expression for $\left< |\det {\bf J}| \right>$ .
Before we do so, let us perform a simplifying change of variables:
\begin{equation}
\begin{array}{lllllll} 
	X_\pm \equiv \sqrt{\frac{1+\chi}{1-\chi} \frac{1}{2{\mathcal A}}} \left({t}/{l}\right)^\chi 
							 \left[ \omega^x_\pm(l) + \omega^x_\pm(0) \right]  \ ,     \\ [7pt]
	Y_\pm \equiv \sqrt{\frac{1}{2{\mathcal A}}} \left({t}/{l}\right)^\chi
							 \left[ \omega^x_\pm(l) - \omega^x_\pm(0) \right]  \ ,	   \\ [7pt]
	Z_\pm \equiv \sqrt{\frac{1+\chi}{1-\chi} \frac{1}{2{\mathcal A}}} \left({t}/{l}\right)^\chi
							 \left[ \omega^y_\pm(l) + \omega^y_\pm(0) \right]  \ ,     \\ [7pt]
	W_\pm \equiv \sqrt{\frac{1}{2{\mathcal A}}} \left({t}/{l}\right)^\chi
							 \left[ \omega^y_\pm(l) - \omega^y_\pm(0) \right]  \ ,
				\end{array}
\end{equation}
under which the expectation value of the loop Jacobian takes a compact form:
\begin{equation}
\left< |\det {\bf J}| \right> =  \frac{{\mathcal A}^2}{2\pi^4} \left(\frac{l}{t}\right)^{4\chi}
\int d^8{\bf X} \: e^{-{\bf X}^2} \: \left| F({\bf X}) + \frac{1-\chi}{1+\chi} \, G({\bf X}) \right|  \ .
\label{integral}
\end{equation}
Here we have defined an 8-dimensional vector
\begin{equation}
{\bf X} \equiv (X_+,X_-,Y_+,Y_-,Z_+,Z_-,W_+,W_-)
\end{equation}
and two functions of it:
\begin{eqnarray}
F({\bf X}) &\equiv& ( Y_+ W_- - Y_- W_+ ) \left( Y_-^2 + W_-^2 - Y_+^2 - W_+^2 \right)\ , \nonumber \\ 
G({\bf X}) &\equiv&  2 ( W_+ W_- - Y_+ Y_- ) (X_+ - X_-) (Z_+ - Z_-) +															\\
					 & & + \: (Y_+ W_- + Y_- W_+) \left[ (X_+ - X_-)^2 - (Z_+ - Z_-)^2 \right] \ .  \nonumber
\end{eqnarray}
It is challenging to proceed further analytically but one can find upper and lower bounds for the value of the integral in~(\ref{integral}) by noting that
\begin{equation}
\left| \int |A| - \int |B| \, \right|  
\leq  \int |A + B|
\leq  \int |A| + \int |B|
\label{bounds}
\end{equation}
Indeed we can compute analytically the integrals of the functions $F$ and $G$ separately.
Here we show how to accomplish this for the integral of $F({\bf X})$.
Under the following change of variables
\begin{equation}
\left\{ \begin{array}{llll} 
	Y_+ = r \cos \theta  \\
	W_+ = r \sin \theta	\\
	Y_- = \rho \cos \varphi  \\
	W_- = \rho \sin \varphi
				\end{array} \right.
\end{equation}
that integral becomes
\begin{eqnarray}
\int d^8{\bf X} \: e^{-{\bf X}^2} \: |F| &=& \pi^2 \int_0^{2\pi} d\theta d\varphi |\sin(\varphi-\theta)|
    \int_0^\infty dr d\rho \: r^2 \rho^2 |\rho^2 - r^2| \: e^{-r^2-\rho^2}  \nonumber \\
&=& \pi^2 \int_0^{2\pi} d\theta \: 2 \int_\theta^{\theta + \pi} d\varphi \: \sin(\varphi-\theta) \; \frac{1}{4} 
    = 2\pi^3   \ .
\end{eqnarray}
A similar (nevertheless, more cumbersome) game can be played with the integral of $G({\bf X})$.
It involves more complicated changes of variables but the result is just as simple:
\begin{equation}
\int d^8{\bf X} \: e^{-{\bf X}^2} \: |G|  =  \pi^4   \ .
\end{equation}
Thus, the expectation value of the Jacobian is bounded by
\begin{equation}
\left| \frac{1}{\pi} - \frac{1-\chi}{2+2\chi} \right| 
\leq  {\mathcal A}^{-2} (t/l)^{4\chi} \left< |\det {\bf J}| \right>  \leq
\frac{1}{\pi} + \frac{1-\chi}{2+2\chi}
\end{equation}
Numerically, we find that
\begin{equation}
{\mathcal A}^{-2} (t/l)^{4\chi} \left< |\det {\bf J}| \right>  
\simeq  \left\{ \begin{array}{ll} 
                 0.57 \quad {\mbox{during the radiation era}}  \\
                 0.48 \quad {\mbox{during the matter era}}
				        \end{array}  \right.
\label{expdetJ}
\end{equation}

Finally, we can combine results~(\ref{deltaz}) and~(\ref{expdetJ}):
\begin{equation}
\langle \delta({L}^z_+ - {L}^z_-) \left| \det {\bf J} \right| \rangle_{ {\bf p}^\perp  \to {\mbox{\boldmath\scriptsize$ \omega$}}}
=  \: \eta \frac{{\mathcal A}}{t} \left( \frac{l}{t} \right)^{-1+2\chi}  \ ,
\end{equation}
where $\eta \approx 2.3$ in the radiation era and $\eta \approx 1.3$ in the matter era.

\pagebreak

\chapter{The matter-radiation transition}

In this subsection we will assume that our stretching model is actually valid down to the scale where gravitational radiation sets in.  If loop production or other relatively rapid processes are actually determining the small scale structure then this appendix is moot. 

We have noted that at very short scales we see structure that actually emerged from the horizon dynamics during the radiation era.
Thus we should take the radiation-to-matter transition into account in our calculation of the small-scale structure.
At the time of equal matter and radiation densities,
\begin{equation}
1 - \langle {\bf p}_+ (\sigma,\tau_{\rm eq}) \cdot {\bf p}_+(\sigma',\tau_{\rm eq}) \rangle
\approx {\mathcal A}_{\rm r} (l_{\rm eq}/t_{\rm eq})^{2\chi_{\rm r}}
\ , \label{teq}
\end{equation}
where $l_{\rm eq}$ is the length of the segment between $\sigma$ and $\sigma'$ at $t_{\rm eq}$.
Assuming that the transition from radiation-dominated to matter-dominated behavior is sharp (which is certainly an oversimplification), we evolve forward to today using the result~(\ref{soln}).
The right-hand side of Eq.~(\ref{teq}) is then multiplied by a factor $(t/t_{\rm eq})^{-2\nu_{\rm m}\bar\alpha_{\rm m}}$.
In terms of the length today, $l = l_{\rm eq} (t/t_{\rm eq})^{\zeta_{\rm m}}$, we have
\begin{equation}
1 - \langle {\bf p}_+ (\sigma,\tau) \cdot {\bf p}_+(\sigma',\tau) \rangle
\approx {\mathcal A}_{\rm r} (l/t)^{2\chi_{\rm r}} (t/t_{\rm eq})^{-2\zeta_{\rm m}
- 2\chi_{\rm r}\zeta_{\rm m} + 2\chi_r}
\ . \label{radmat}
\end{equation}
This expression applies to scales $l(t)$ that, evolved backward in time, reached the horizon scale $d_H$ before the transition occurred, i.e. at a time $t_*$ defined by $l(t_*) \sim d_H$ such that $t_* < t_{\rm eq}$.
For longer scales, which formed during the matter era (for which $t_* > t_{\rm eq}$), we have simply
\begin{equation}
1 - \langle {\bf p}_+ (\sigma,\tau) \cdot {\bf p}_+(\sigma',\tau) \rangle
\approx {\mathcal A}_{\rm m} (l/t)^{2\chi_{\rm m}}
\ . \label{radmat2}
\end{equation}

\begin{figure}
\center \includegraphics[width=22pc]{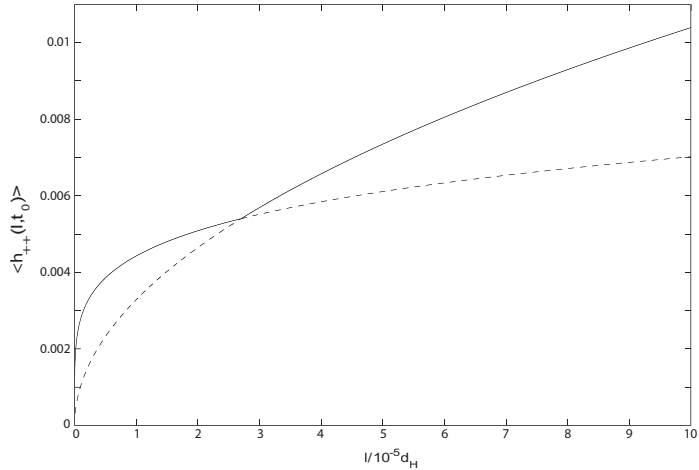}
\caption[]{Structure on the string, $\langle h_{++} \rangle$, as a function of the length $l$ at present time (solid curve).  On scales larger than the critical length $l_{\rm c} \sim 3\times 10^{-5} d_H$ the structure is determined by the matter era expression.  On scales below $l_{\rm c}$ the transition result~(\ref{radmat}) gives an enhanced effect.  The dashed curves show the extrapolations of the two relevant expressions: on small scales the actual structure is enhanced relative to the pure matter era result.} \label{transition}
\end{figure}

The transition between the two forms occurs along the curve determined by the intersection of the two surfaces~(\ref{radmat}) and~(\ref{radmat2}).
This determines the critical length at the time of equal matter and radiation densities, $l_{\rm c}(t_{\rm eq}) = ({\mathcal A}_{\rm r}/{\mathcal A}_{\rm m})^{1/(2\chi_{\rm m}-2\chi_{\rm r})}t_{\rm eq}$.
In terms of the length at some later time $t$, the transition occurs at
\begin{equation}
\frac{l_{\rm c}(t)}{t} \approx \left(\frac{{\mathcal A}_{\rm r}}{{\mathcal A}_{\rm m}}\right)^{1/(2\chi_{\rm m}-2\chi_{\rm r})} \left(\frac{t}{t_{\rm eq}}\right)^{\zeta_{\rm m}-1}
\ , \label{radmat3}
\end{equation}
so that the transition scale at the present time is $l_{\rm c} \sim 3\times 10^{-5} d_H$ (Fig.~\ref{transition}).
The transition result~(\ref{radmat}) implies more structure at the smallest scales than would be obtained from the matter era result, by a factor $(l_{\rm c}/l)^{2(\chi_{\rm m} - \chi_{\rm r})} \sim (l_{\rm c}/l)^{0.3}$.

Of course, precise studies of the small scale structure must include also the effect of the recent transition to vacuum domination; this period has been brief so the effect should be rather small.
\end{appendix}

\appendix

\clearpage
\ssp
\bibliographystyle{utphys}
\bibliography{ROCHA_thesis}

\end{document}